 \documentclass[manuscript]{emulateapj}

 \usepackage{apjfonts}

\slugcomment{to appear in the Astrophysical Journal}

\shorttitle{Supersoft X-Ray Phase of Novae}
\shortauthors{Hachisu \& Kato}

\begin{document}

\title{A Prediction Formula of Supersoft X-ray Phase of Classical Novae}


\author{Izumi Hachisu}
\affil{Department of Earth Science and Astronomy,
College of Arts and Sciences, University of Tokyo,
Komaba, Meguro-ku, Tokyo 153-8902, Japan}
\email{hachisu@ea.c.u-tokyo.ac.jp}

\and

\author{Mariko Kato}
\affil{Department of Astronomy, Keio University,
Hiyoshi, Kouhoku-ku, Yokohama 223-8521, Japan}
\email{mariko@educ.cc.keio.ac.jp}

%\and
%
%\author{Angelo Cassatella}
%\affil{INAF, Istituto di Fisica dello Spazio Interplanetario,
%Via del Fosso del Cavaliere 100, 00133 Rome, Italy}
%\email{cassatella@fis.uniroma3.it}


%\altaffiltext{1}{Visiting Astronomer, Cerro Tololo Inter-American Observatory.
%CTIO is operated by AURA, Inc.\ under contract to the National Science
%\altaffiltext{2}{Society of Fellows, Harvard University.}
%\altaffiltext{3}{present address: Center for Astrophysics,
%    60 Garden Street, Cambridge, MA 02138}
%\altaffiltext{4}{Visiting Programmer, Space Telescope Science Institute}
%\altaffiltext{5}{Patron, Alonso's Bar and Grill}


\begin{abstract}
  On the basis of the recently developed universal decline law of 
classical novae, we propose prediction formulae for
supersoft X-ray on and off times,
i.e., $t_{\rm X-on} = ( 10 \pm 1.8 ) t_3$
days and $t_{\rm X-off} = ( 5.3 \pm 1.4 ) (t_3)^{1.5}$ days
for $8 \lesssim t_3 \lesssim 80$ days.  Here $t_3$ is the
newly proposed ``intrinsic'' decay time during which the brightness
drops by 3 magnitudes from optical maximum
along our universal decline law fitted with
observation.  We have determined the absolute magnitude of
our free-free emission model light curves and derived 
maximum magnitude vs. rate of decline (MMRD) relations.
Our theoretical MMRD relations are governed by two parameters,
one is the white dwarf (WD) mass and the other is the initial
envelope mass at a nova outburst; this second parameter explains
the scatter of MMRD points of individual novae.
Our theoretical MMRD relations are also in good
agreement with the well-known empirical formulae.
We also show another empirical relation
of $M_V(15) \sim -5.7 \pm 0.3$ based on the absolute magnitude of 
our model light curves,
i.e., the absolute magnitude at 15 days after optical maximum is almost
common among various novae.
We analyzed ten nova light curves, in which a supersoft X-ray phase
was detected, and estimated their WD masses.  The models best
reproducing simultaneously the optical and supersoft X-ray
observations are ONeMg WDs with
$1.28 \pm 0.04~M_\sun$ (V598 Pup),
$1.23 \pm 0.05~M_\sun$ (V382 Vel),
$1.15 \pm 0.06~M_\sun$ (V4743 Sgr),
$1.13 \pm 0.06~M_\sun$ (V1281 Sco),
$1.2 \pm 0.05~M_\sun$ (V597 Pup),
$1.06 \pm 0.07~M_\sun$ (V1494 Aql),
$1.04 \pm 0.07~M_\sun$ (V2467 Cyg),
$1.07 \pm 0.07~M_\sun$ (V5116 Sgr),
$1.05 \pm 0.05~M_\sun$ (V574 Pup), and a CO WD with
$0.93 \pm 0.08~M_\sun$ (V458 Vul).
The newly proposed relationships are consistent with
the emergence or decay epoch of the supersoft X-ray phase
of these ten novae.
Finally, we discuss the mechanism of shock-origin hard X-ray component
in relation to the emergence of companion star from the WD envelope.
\end{abstract}




\section{Introduction}
     Classical novae are a thermonuclear runaway event on
a white dwarf (WD) in a binary system, in which the WD
accretes hydrogen-rich matter from the companion star.
When the accreted matter reaches a critical value, hydrogen
at the bottom of the WD envelope ignites to trigger a shell flash
(denoted by time $t_{\rm OB}$).
Just after a nova outburst, the envelope on the WD
rapidly expands to a giant size and optically thick winds
blow.  The WD envelope settles in a steady-state.
The optical magnitude attains the peak at the maximum
expansion of the photosphere (time $t_0$).
Then the photosphere gradually
shrinks whereas the total luminosity is almost constant during
the outburst.  Thus, the photospheric temperature $T_{\rm ph}$
increases with time.  The main emitting wavelength region
moves from optical to ultraviolet and
finally to supersoft X-ray, corresponding to from
$T_{\rm ph} \sim 10^4$~K through $\sim 10^6$~K.
The optically thick winds blow continuously from the very beginning
of the outburst until the photospheric temperature increases to
$\log T_{\rm ph}$~(K)$\sim 5.4$.  Just when the optically thick wind
stops (time $t_{\rm wind}$),
the temperature quickly rises and the supersoft X-ray phase
starts \citep[e.g.,][]{kat94h}.

Thus, classical novae become a transient supersoft
X-ray source in a later phase of the outburst,
but their X-ray detection is rather rare mainly
because of sparse observing time of X-ray satellites
\citep[e.g.,][]{kra96, ori01a, nes07a}.
If there is a formula to predict
a supersoft X-ray phase of novae, we can plan more efficient
observation.  The aim of this paper is to propose such a formula.

Novae show a rich variety of light curves; the timescales
of decline, shapes, and total durations of outbursts are very
different from nova to nova.  
Therefore, it is difficult to predict a later supersoft X-ray phase
only from the early optical light curve.  Recently we have proposed
the universal decline law of classical novae 
\citep{hac06kb, hac07k, hac08kc}, i.e., there is a basic light
curve which can basically follow (or overlap with) individual nova
light curves by squeezing or stretching it in the direction of time.
We call this ``a universal decline law.''
The universal shape of light curves is almost
independent of the WD mass and of the chemical
composition of its hydrogen-rich envelope.
This implies that each nova light curve can be uniquely
specified by one parameter (timescaling factor -- squeezing/stretching
factor against a universal shape of nova light curves).

We have reproduced a number of light curves of novae from early
phase to supersoft X-ray phase (turn-on/turnoff times)
\citep{hac06kb, hac07k, hac06kk, hac07kl, hac08kc}.  Based on these
experiences we are now able to construct empirical formulae
that tell us the emergence/decay times of a supersoft X-ray phase 
of individual novae from observational information of the early phase.
We hope that these formulae are useful for future X-ray observations
of classical novae. 

The next section (Section \ref{modeling_of_nova})
describes our numerical method for obtaining
X-ray, ultraviolet (UV), optical, and infrared (IR) light curves.  
In Section \ref{properties_of_light_curve}, we show homologous 
shapes of our model light curves.  Using characteristic
properties of free-free emission model light curves,
we derive the absolute magnitude of each optical light curve,
which provides a unique standard candle of novae.
We further propose another method for obtaining 
the nova absolute brightness by fitting a nova light
curve with the other the brightness of which is known.
After various nova timescales and relations between these
timescales are presented, we propose our prediction formulae of
a supersoft X-ray phase in Section \ref{various_timescales}.
Then, we present our theoretical
understanding on the so-called maximum magnitude vs. rate of decline
(MMRD) relation of classical novae in Section \ref{mmrd_relation}.
The light-curve analyses of ten classical novae are presented
in Section \ref{light_curve_of_ten_novae}.
The X-ray turn-on/turnoff times of these ten novae,
i.e., V598 Pup, V382 Vel, V4743 Sgr, V1281 Sco, V597 Pup,
V1494 Aql, V2467 Cyg, V5116 Sgr, V574 Pup, and V458 Vul,
are consistent with the above relations (formulae).
Conclusions follow in Sections \ref{conclusions}.

\section{Modeling of Nova Light Curves}
\label{modeling_of_nova}

The decay phase of novae can be
followed by a sequence of steady state solutions \citep[e.g.,][]{kat94h}.
Using the same method and numerical techniques as in \citet{kat94h},
we have followed evolutions of novae by connecting steady state solutions
along the decreasing envelope mass sequence.  The mass of the
hydrogen-rich envelope is decreasing due to wind mass-loss
and nuclear burning.  We solve a set of equations, consisting of
the continuity, equation of motion, radiative diffusion,
and conservation of energy, from the bottom of the hydrogen-rich
envelope through the photosphere assuming spherical symmetry.
Winds are accelerated deep inside the photosphere so that
they are called ``optically thick winds.''

     We have presented a unified model for IR, optical, UV,
and supersoft X-ray light curves of the decay phase of 
classical novae \citep{hac06kb, hac07k, hac07kl, hac08kc, kat07h}.
The optical and IR luminosity are reproduced by free-free emission
from optically thin plasma \citep[e.g.,][]{gal76, kra84}.
This free-free emission flux at the frequency $\nu$ is
estimated from Equation (9) of
\citet{hac06kb}, i.e.,
\begin{equation}
F_\nu \propto {{\dot M_{\rm wind}^2} \over {{v_{\rm ph}^2 R_{\rm ph}}}},
\label{wind-free-free-emission}
\end{equation}
where $\dot M_{\rm wind}$ is the wind mass-loss rate,
$v_{\rm ph}$ the wind velocity at the photosphere, and
$R_{\rm ph}$ the photospheric radius, all of which are taken
from our wind solutions.
It should be noted that the shapes of free-free emission
light curves are independent of the frequency (wavelength). 
After the optically thick wind stops, the total mass of the
ejecta remains constant in time.  The flux from such homologously
expanding ejecta is estimated from Equation (19) of \citet{hac06kb},
i.e., 
\begin{equation}
F_\nu \propto t^{-3},
\label{expansion-free-free-emission}
\end{equation}
where $t$ is the time after the outburst.

Assuming a blackbody spectrum with the photospheric temperature of
$T_{\rm ph}$, we have calculated the UV 1455 \AA~ light curve for
a narrow energy band of 1445--1465 \AA~  \citep[see, e.g.,][for
more details]{cas02, hac06kb} and have also estimated the supersoft
X-ray flux \citep{kat94h, hac06kb} throughout the present paper
for the energy range of 0.2--0.6 keV \citep[e.g.,][]{hac09k}. 
Even if we adopt a different energy range of 0.2--2.0 keV,
shapes of supersoft X-ray light curves hardly change
because the blackbody photosphere emits photons with energy
of $< 0.6$ keV for $M_{\rm WD} \lesssim 1.2~M_\sun$.

     The light curves of our optically thick wind model are
parameterized by the WD mass ($M_{\rm WD}$), chemical composition
of the WD envelope ($X$, $Y$, $X_{\rm CNO}$,
$X_{\rm Ne}$, and $Z$), and envelope mass
($M_{\rm env, 0}$) at the time of the outburst ($t_{\rm OB}$).
Here $X$ is the hydrogen, $Y$ the helium, $X_{\rm CNO}$ 
the carbon-nitrogen-oxygen, $X_{\rm Ne}$ the neon, and 
$Z$ the metallicity content of the envelope.  We adopt $Z=0.02$,
which also includes carbon, nitrogen, oxygen, and neon
with solar composition ratios.  Table \ref{chemical_composition}
shows nine sets of chemical compositions assumed in our models.
The details of our model light curves and their parameter 
dependence are described in \citet{hac06kb}.

\section{Universal Decline Law}
\label{properties_of_light_curve}

% Table 1
%%%\placetable{chemical_composition}

\begin{deluxetable}{llllll}
\tabletypesize{\scriptsize}
\tablecaption{Chemical Composition of White Dwarf Envelope
\label{chemical_composition}}
\tablewidth{0pt}
\tablehead{
\colhead{nova case} &
\colhead{$X$} &
\colhead{$Y$} &
\colhead{$X_{\rm CNO}$} &
\colhead{$X_{\rm Ne}$} &
\colhead{$Z$\tablenotemark{a}}
}
\startdata
CO nova 1 & 0.35 & 0.13 & 0.50 & 0.0 & 0.02 \\
CO nova 2 & 0.35 & 0.33 & 0.30 & 0.0 & 0.02 \\
CO nova 3 & 0.45 & 0.18 & 0.35 & 0.0 & 0.02 \\
CO nova 4 & 0.55 & 0.23 & 0.20 & 0.0 & 0.02 \\
CO nova 5 & 0.65 & 0.27 & 0.06 & 0.0 & 0.02 \\
Ne nova 1 & 0.35 & 0.33 & 0.20 & 0.10 & 0.02 \\
Ne nova 2 & 0.55 & 0.30 & 0.10 & 0.03 & 0.02 \\
Ne nova 3 & 0.65 & 0.27 & 0.03 & 0.03 & 0.02 \\
Solar & 0.70 & 0.28 & 0.0 & 0.0 & 0.02 
\enddata
\tablenotetext{a}{carbon, nitrogen, oxygen, and neon are also
included in $Z=0.02$ with the same ratio as the solar
abundance}
\end{deluxetable}

% Table 2
%%%\placetable{light_curves_of_novae_co}

\begin{deluxetable*}{lllllllllllllll}
\tabletypesize{\scriptsize}
%%%\rotate
\tablecaption{Light Curves of CO Novae\tablenotemark{a}
\label{light_curves_of_novae_co}}
\tablewidth{0pt}
\tablehead{
\colhead{$m_{\rm ff}$} &
\colhead{0.55$M_\sun$} &
\colhead{0.6$M_\sun$} &
\colhead{0.65$M_\sun$} &
\colhead{0.7$M_\sun$} &
\colhead{0.75$M_\sun$} &
\colhead{0.8$M_\sun$} &
\colhead{0.85$M_\sun$} &
\colhead{0.9$M_\sun$} &
\colhead{0.95$M_\sun$} &
\colhead{1.0$M_\sun$} &
\colhead{1.05$M_\sun$} &
\colhead{1.1$M_\sun$} &
\colhead{1.15$M_\sun$} &
\colhead{1.2$M_\sun$} \\
\colhead{(mag)} &
\colhead{(day)} &
\colhead{(day)} &
\colhead{(day)} &
\colhead{(day)} &
\colhead{(day)} &
\colhead{(day)} &
\colhead{(day)} &
\colhead{(day)} &
\colhead{(day)} &
\colhead{(day)} &
\colhead{(day)} &
\colhead{(day)} &
\colhead{(day)} &
\colhead{(day)} 
}
\startdata
  1.50 & 0.000 & 0.000 & 0.000 & 0.000 & 0.000 & 0.000 & 0.000 & 0.000 & 0.000 & 0.000 & 0.000 & 0.000 & 0.000 & 0.000 \\
  1.75 & 1.319 & 1.421 & 1.286 & 1.319 & 1.231 & 1.187 & 1.123 & 1.120 & 1.032 & 0.970 & 0.922 & 0.870 & 0.772 & 0.717 \\
  2.00 & 2.921 & 2.835 & 2.615 & 2.705 & 2.442 & 2.367 & 2.243 & 2.210 & 2.027 & 1.910 & 1.811 & 1.660 & 1.528 & 1.428 \\
  2.25 & 4.634 & 4.393 & 4.162 & 4.102 & 3.737 & 3.548 & 3.343 & 3.220 & 2.997 & 2.850 & 2.660 & 2.440 & 2.247 & 2.140 \\
  2.50 & 6.346 & 6.132 & 5.758 & 5.482 & 5.069 & 4.750 & 4.453 & 4.220 & 3.917 & 3.780 & 3.485 & 3.200 & 2.940 & 2.820 \\
  2.75 & 8.187 & 7.872 & 7.358 & 6.842 & 6.398 & 5.960 & 5.543 & 5.220 & 4.837 & 4.700 & 4.280 & 3.940 & 3.622 & 3.470 \\
  3.00 & 10.46 & 9.822 & 9.138 & 8.272 & 7.738 & 7.187 & 6.623 & 6.220 & 5.747 & 5.620 & 5.057 & 4.650 & 4.276 & 4.100 \\
  3.25 & 12.72 & 12.05 & 11.06 & 9.992 & 9.088 & 8.409 & 7.713 & 7.170 & 6.627 & 6.480 & 5.797 & 5.340 & 4.926 & 4.710 \\
  3.50 & 16.25 & 14.32 & 12.99 & 11.71 & 10.55 & 9.659 & 8.793 & 8.130 & 7.517 & 7.260 & 6.527 & 6.010 & 5.548 & 5.310 \\
  3.75 & 21.22 & 17.04 & 15.39 & 13.50 & 12.15 & 10.98 & 9.913 & 9.100 & 8.407 & 8.060 & 7.257 & 6.670 & 6.169 & 5.890 \\
  4.00 & 27.67 & 19.96 & 17.88 & 15.66 & 13.80 & 12.32 & 11.05 & 10.07 & 9.257 & 8.850 & 7.967 & 7.320 & 6.774 & 6.470 \\
  4.25 & 34.30 & 23.75 & 20.45 & 17.95 & 15.47 & 13.69 & 12.21 & 11.06 & 10.13 & 9.650 & 8.677 & 7.970 & 7.369 & 7.040 \\
  4.50 & 41.31 & 28.68 & 23.06 & 20.29 & 17.18 & 15.20 & 13.42 & 12.08 & 11.02 & 10.44 & 9.387 & 8.620 & 7.969 & 7.610 \\
  4.75 & 48.32 & 34.46 & 27.56 & 22.96 & 19.69 & 16.79 & 14.77 & 13.13 & 11.97 & 11.24 & 10.11 & 9.280 & 8.579 & 8.160 \\
  5.00 & 55.67 & 42.32 & 33.29 & 26.67 & 22.73 & 18.44 & 16.29 & 14.27 & 12.95 & 12.09 & 10.87 & 9.950 & 9.209 & 8.740 \\
  5.25 & 65.40 & 50.31 & 39.71 & 32.68 & 25.92 & 21.12 & 17.85 & 15.78 & 14.08 & 12.99 & 11.67 & 10.66 & 9.899 & 9.340 \\
  5.50 & 75.12 & 58.89 & 46.20 & 38.40 & 29.35 & 24.37 & 20.06 & 17.41 & 15.58 & 13.98 & 12.56 & 11.44 & 10.73 & 10.11 \\
  5.75 & 87.91 & 68.64 & 52.69 & 43.44 & 33.25 & 27.64 & 23.20 & 19.35 & 17.32 & 15.21 & 13.65 & 12.51 & 11.60 & 10.99 \\
  6.00 & 101.8 & 78.86 & 60.29 & 48.43 & 37.47 & 30.94 & 25.89 & 21.74 & 19.04 & 16.97 & 15.15 & 13.97 & 12.77 & 11.93 \\
  6.25 & 117.5 & 90.97 & 68.44 & 54.61 & 41.95 & 33.84 & 28.14 & 23.89 & 20.85 & 18.59 & 16.54 & 14.98 & 13.83 & 12.90 \\
  6.50 & 134.6 & 103.7 & 76.86 & 61.70 & 46.86 & 37.24 & 30.55 & 25.75 & 22.50 & 19.92 & 17.56 & 15.92 & 14.53 & 13.59 \\
  6.75 & 152.9 & 117.3 & 87.51 & 69.39 & 52.11 & 41.14 & 33.37 & 27.75 & 24.15 & 21.28 & 18.61 & 16.71 & 15.21 & 14.14 \\
  7.00 & 171.3 & 131.5 & 98.25 & 77.79 & 57.84 & 45.35 & 36.66 & 30.11 & 25.92 & 22.68 & 19.72 & 17.57 & 15.94 & 14.71 \\
  7.25 & 190.6 & 146.1 & 109.6 & 86.81 & 64.22 & 49.91 & 40.23 & 32.84 & 27.97 & 24.18 & 20.91 & 18.58 & 16.74 & 15.34 \\
  7.50 & 211.6 & 161.1 & 121.3 & 96.18 & 71.09 & 54.93 & 44.08 & 35.79 & 30.29 & 25.87 & 22.23 & 19.67 & 17.58 & 16.03 \\
  7.75 & 234.6 & 177.5 & 133.6 & 105.6 & 78.50 & 60.32 & 48.19 & 38.77 & 32.77 & 27.76 & 23.70 & 20.82 & 18.49 & 16.75 \\
  8.00 & 260.1 & 196.0 & 146.7 & 115.7 & 86.05 & 66.19 & 52.32 & 41.90 & 35.42 & 29.80 & 25.23 & 22.03 & 19.46 & 17.49 \\
  8.25 & 288.9 & 217.2 & 160.5 & 127.1 & 94.01 & 72.60 & 56.72 & 45.29 & 38.05 & 31.91 & 26.86 & 23.28 & 20.44 & 18.30 \\
  8.50 & 322.8 & 241.2 & 177.7 & 139.7 & 102.9 & 79.60 & 61.60 & 48.95 & 40.89 & 34.14 & 28.56 & 24.53 & 21.47 & 19.12 \\
  8.75 & 363.7 & 268.6 & 197.2 & 153.9 & 113.1 & 87.22 & 66.89 & 53.02 & 44.02 & 36.54 & 30.38 & 25.89 & 22.58 & 19.97 \\
  9.00 & 411.1 & 301.4 & 219.4 & 171.6 & 124.3 & 95.52 & 73.07 & 57.43 & 47.44 & 39.14 & 32.31 & 27.46 & 23.81 & 20.94 \\
  9.25 & 457.6 & 339.0 & 246.2 & 191.7 & 138.9 & 104.5 & 80.58 & 62.78 & 51.34 & 42.10 & 34.43 & 29.22 & 25.19 & 22.04 \\
  9.50 & 508.2 & 377.0 & 276.1 & 214.1 & 155.7 & 117.0 & 89.01 & 69.27 & 56.25 & 45.37 & 37.10 & 31.32 & 26.89 & 23.32 \\
  9.75 & 547.2 & 412.0 & 307.1 & 239.8 & 173.7 & 131.4 & 100.0 & 76.87 & 62.24 & 49.59 & 40.33 & 33.85 & 28.93 & 24.79 \\
  10.00 & 589.5 & 446.5 & 335.6 & 267.7 & 193.2 & 146.7 & 112.5 & 86.46 & 69.49 & 54.71 & 44.21 & 37.08 & 31.56 & 26.67 \\
  10.25 & 635.0 & 479.3 & 362.2 & 292.3 & 215.6 & 163.5 & 124.7 & 97.02 & 78.11 & 60.81 & 48.97 & 40.76 & 34.69 & 29.01 \\
  10.50 & 673.6 & 512.3 & 386.2 & 313.5 & 234.6 & 180.4 & 138.3 & 107.1 & 86.93 & 68.04 & 54.70 & 45.39 & 38.36 & 31.74 \\
  10.75 & 715.8 & 541.6 & 412.6 & 335.6 & 253.3 & 195.2 & 153.4 & 118.5 & 95.91 & 75.64 & 60.98 & 50.50 & 42.51 & 34.85 \\
  11.00 & 762.2 & 571.6 & 440.2 & 355.9 & 269.9 & 209.3 & 165.7 & 130.3 & 105.7 & 83.14 & 66.78 & 55.83 & 46.53 & 38.24 \\
  11.25 & 813.6 & 603.7 & 464.6 & 377.3 & 284.6 & 223.5 & 177.7 & 140.4 & 114.3 & 91.54 & 73.13 & 60.75 & 50.34 & 41.48 \\
  11.50 & 869.2 & 637.0 & 489.4 & 396.0 & 300.8 & 237.0 & 187.3 & 149.9 & 121.1 & 99.24 & 79.95 & 65.55 & 54.40 & 44.25 \\
  11.75 & 920.8 & 673.7 & 516.6 & 415.2 & 317.0 & 249.3 & 197.7 & 158.9 & 128.5 & 105.0 & 84.82 & 70.35 & 58.73 & 47.19 \\
  12.00 & 975.4 & 713.8 & 546.4 & 436.2 & 332.6 & 261.8 & 209.0 & 167.0 & 136.5 & 111.3 & 89.90 & 74.95 & 62.18 & 50.27 \\
  12.25 & 1034. & 756.3 & 578.9 & 459.2 & 349.7 & 274.1 & 219.3 & 175.8 & 145.2 & 118.1 & 95.36 & 79.25 & 65.64 & 53.55 \\
  12.50 & 1095. & 801.4 & 613.5 & 484.3 & 368.4 & 287.4 & 229.3 & 184.0 & 154.3 & 125.4 & 101.1 & 83.80 & 69.32 & 56.94 \\
  12.75 & 1160. & 849.1 & 650.2 & 513.2 & 388.7 & 302.1 & 240.2 & 192.7 & 161.7 & 133.1 & 107.3 & 88.40 & 73.16 & 60.21 \\
  13.00 & 1229. & 899.8 & 689.0 & 544.1 & 411.5 & 317.9 & 252.1 & 202.1 & 169.7 & 139.7 & 113.6 & 93.20 & 77.25 & 63.71 \\
  13.25 & 1302. & 953.3 & 730.1 & 576.7 & 436.1 & 336.6 & 265.4 & 212.3 & 178.3 & 146.4 & 120.7 & 98.30 & 81.99 & 67.40 \\
  13.50 & 1379. & 1011. & 773.7 & 611.3 & 462.2 & 356.7 & 281.6 & 224.3 & 187.5 & 153.8 & 128.1 & 103.8 & 87.01 & 71.28 \\
  13.75 & 1461. & 1071. & 819.8 & 647.9 & 489.8 & 377.9 & 298.9 & 238.4 & 199.1 & 161.6 & 136.0 & 109.6 & 92.33 & 75.31 \\
  14.00 & 1547. & 1134. & 868.7 & 686.7 & 519.0 & 400.5 & 317.1 & 253.2 & 211.4 & 173.1 & 144.4 & 116.8 & 97.94 & 80.17 \\
  14.25 & 1639. & 1202. & 920.5 & 727.8 & 550.0 & 424.3 & 336.5 & 269.0 & 224.4 & 185.5 & 153.2 & 124.8 & 103.9 & 85.41 \\
  14.50 & 1736. & 1273. & 975.3 & 771.3 & 582.9 & 449.6 & 357.0 & 285.7 & 238.2 & 198.6 & 162.6 & 133.3 & 110.2 & 90.96 \\
  14.75 & 1839. & 1349. & 1034. & 817.4 & 617.7 & 476.4 & 378.6 & 303.4 & 252.9 & 212.4 & 172.6 & 142.3 & 116.9 & 96.87 \\
  15.00 & 1948. & 1429. & 1095. & 866.2 & 654.5 & 504.8 & 401.6 & 322.1 & 268.3 & 227.1 & 183.1 & 151.8 & 124.0 & 103.1 \\
\hline
X-ray\tablenotemark{b} & 3850  & 2700  & 2200  & 1700  & 1150  & 870   & 620   & 440   & 326   & 240   & 162   & 109   & 74.1  & 48.7 \\
\hline
$m'_{\rm w}$\tablenotemark{c} & 17.6  & 17.3  & 17.0  & 16.8  & 16.6  & 16.3   & 16.2   & 16.1   & 15.9   & 15.9   & 15.7   & 15.7   & 15.8  & 15.7 \\
\hline
$\log f_{\rm s}$\tablenotemark{d} & 0.81  & 0.72  & 0.62  & 0.52  & 0.42  & 0.31   & 0.21   & 0.10   & 0.01   & $-0.07$   & $-0.16$   & $-0.27$   & $-0.36$  & $-0.46$ \\
\hline
$M_{\rm w}$\tablenotemark{e} & 5.3  & 4.8  & 4.3  & 3.8  & 3.3  & 2.8   & 2.4   & 2.0   & 1.6   & 1.4   & 1.0   & 0.8   & 0.6  & 0.2 \\
\enddata
\tablenotetext{a}{chemical composition of the envelope is assumed
to be that of ``CO nova 2'' in Table \ref{chemical_composition}}
\tablenotetext{b}{duration of supersoft X-ray phase in units of days}
\tablenotetext{c}{converted magnitudes at the bottom point in Figure
\ref{v1668_cyg_vy_jhk_mag_x35z02c10o20_model_scaling_law}}
\tablenotetext{d}{stretching factors against V1668 Cyg observational
data in Figure
\ref{v1668_cyg_vy_jhk_mag_x35z02c10o20_model_scaling_law}}
\tablenotetext{e}{absolute magnitudes at the bottom point in Figure
\ref{v1668_cyg_vy_jhk_mag_x35z02c10o20_model_scaling_law} by assuming
$(m-M)_V = 14.3$  (V1668 Cyg)}
\end{deluxetable*}

% Table 3
%%%\placetable{light_curves_of_novae_ne}

\begin{deluxetable*}{llllllllllllll}
\tabletypesize{\scriptsize}
%%%\rotate
\tablecaption{Light Curves of Ne Novae\tablenotemark{a}
\label{light_curves_of_novae_ne}}
\tablewidth{0pt}
\tablehead{
\colhead{$m_{\rm ff}$} &
\colhead{0.7$M_\sun$} &
\colhead{0.75$M_\sun$} &
\colhead{0.8$M_\sun$} &
\colhead{0.85$M_\sun$} &
\colhead{0.9$M_\sun$} &
\colhead{0.95$M_\sun$} &
\colhead{1.0$M_\sun$} &
\colhead{1.05$M_\sun$} &
\colhead{1.1$M_\sun$} &
\colhead{1.15$M_\sun$} &
\colhead{1.2$M_\sun$} &
\colhead{1.25$M_\sun$} &
\colhead{1.3$M_\sun$} \\
\colhead{(mag)} &
\colhead{(day)} &
\colhead{(day)} &
\colhead{(day)} &
\colhead{(day)} &
\colhead{(day)} &
\colhead{(day)} &
\colhead{(day)} &
\colhead{(day)} &
\colhead{(day)} &
\colhead{(day)} &
\colhead{(day)} &
\colhead{(day)} &
\colhead{(day)} 
}
\startdata
  3.000     & 0.0 & 0.0 & 0.0 & 0.0 & 0.0 & 0.0 & 0.0 & 0.0 & 0.0 & 0.0 & 0.0 & 0.0 & 0.0 \\
  3.250     &  2.830     &  2.270     &  1.610     &  1.292     &  1.120     & 0.960     & 0.870     & 0.758     & 0.667     & 0.608     & 0.565     & 0.515     & 0.471     \\
  3.500     &  6.860     &  4.590     &  3.230     &  2.742     &  2.270     &  1.940     &  1.740     &  1.499     &  1.352     &  1.215     &  1.117     &  1.026     & 0.939     \\
  3.750     &  10.98     &  6.970     &  5.070     &  4.292     &  3.450     &  2.960     &  2.630     &  2.259     &  2.005     &  1.828     &  1.669     &  1.535     &  1.408     \\
  4.000     &  15.27     &  9.400     &  7.630     &  5.882     &  4.770     &  4.020     &  3.560     &  3.035     &  2.683     &  2.429     &  2.227     &  2.039     &  1.872     \\
  4.250     &  19.66     &  12.35     &  10.34     &  7.572     &  6.260     &  5.150     &  4.500     &  3.874     &  3.372     &  3.052     &  2.784     &  2.543     &  2.336     \\
  4.500     &  24.59     &  16.54     &  13.12     &  9.872     &  7.810     &  6.460     &  5.480     &  4.736     &  4.154     &  3.716     &  3.429     &  3.114     &  2.863     \\
  4.750     &  30.87     &  20.97     &  16.18     &  12.51     &  9.470     &  8.050     &  6.870     &  5.736     &  4.981     &  4.466     &  4.097     &  3.717     &  3.480     \\
  5.000     &  38.11     &  25.58     &  19.56     &  15.25     &  11.51     &  9.680     &  8.330     &  6.980     &  5.865     &  5.276     &  4.817     &  4.346     &  4.068     \\
  5.250     &  45.78     &  31.11     &  23.10     &  17.98     &  13.69     &  11.34     &  9.740     &  8.270     &  6.945     &  6.126     &  5.557     &  4.996     &  4.600     \\
  5.500     &  54.00     &  37.61     &  26.89     &  20.83     &  15.97     &  13.10     &  11.05     &  9.340     &  7.935     &  7.046     &  6.297     &  5.616     &  5.119     \\
  5.750     &  65.14     &  44.69     &  32.20     &  23.82     &  18.41     &  15.02     &  12.46     &  10.38     &  8.855     &  7.846     &  6.997     &  6.186     &  5.588     \\
  6.000     &  78.08     &  52.44     &  38.16     &  27.74     &  21.03     &  17.16     &  14.13     &  11.57     &  9.745     &  8.566     &  7.647     &  6.756     &  6.043     \\
  6.250     &  92.06     &  62.15     &  44.65     &  32.40     &  24.08     &  19.48     &  16.08     &  13.08     &  10.83     &  9.416     &  8.327     &  7.326     &  6.502     \\
  6.500     &  106.9     &  72.79     &  51.93     &  37.53     &  27.75     &  22.27     &  18.22     &  14.77     &  12.09     &  10.37     &  9.087     &  7.886     &  6.973     \\
  6.750     &  121.6     &  84.36     &  59.93     &  43.19     &  31.76     &  25.48     &  20.73     &  16.64     &  13.50     &  11.46     &  9.937     &  8.546     &  7.467     \\
  7.000     &  137.5     &  95.87     &  68.63     &  49.46     &  36.14     &  28.98     &  23.51     &  18.75     &  15.04     &  12.65     &  10.86     &  9.206     &  7.997     \\
  7.250     &  153.6     &  107.6     &  77.53     &  56.30     &  40.95     &  32.66     &  26.49     &  21.09     &  16.78     &  13.95     &  11.87     &  9.956     &  8.537     \\
  7.500     &  171.0     &  120.3     &  86.47     &  63.47     &  46.23     &  36.67     &  29.55     &  23.51     &  18.58     &  15.37     &  12.98     &  10.72     &  9.087     \\
  7.750     &  190.3     &  134.1     &  96.17     &  70.30     &  51.94     &  41.04     &  32.87     &  25.93     &  20.38     &  16.77     &  14.09     &  11.53     &  9.627     \\
  8.000     &  214.2     &  149.1     &  107.7     &  77.72     &  57.59     &  45.66     &  36.49     &  28.49     &  22.32     &  18.22     &  15.23     &  12.35     &  10.19     \\
  8.250     &  241.0     &  167.0     &  120.4     &  87.02     &  63.69     &  50.42     &  40.36     &  31.26     &  24.43     &  19.81     &  16.44     &  13.18     &  10.77     \\
  8.500     &  270.1     &  187.9     &  134.9     &  97.51     &  71.21     &  55.63     &  44.52     &  34.40     &  26.85     &  21.55     &  17.74     &  14.12     &  11.41     \\
  8.750     &  301.9     &  211.4     &  151.8     &  109.3     &  79.80     &  62.42     &  49.45     &  38.30     &  29.64     &  23.69     &  19.26     &  15.25     &  12.18     \\
  9.000     &  338.4     &  235.7     &  170.6     &  122.9     &  89.49     &  69.91     &  55.54     &  42.96     &  32.86     &  26.07     &  21.07     &  16.60     &  13.15     \\
  9.250     &  376.5     &  263.7     &  191.0     &  138.2     &  100.7     &  78.89     &  62.33     &  48.14     &  36.75     &  29.04     &  23.32     &  18.17     &  14.32     \\
  9.500     &  409.3     &  295.3     &  214.2     &  155.4     &  113.3     &  89.13     &  70.38     &  54.26     &  41.14     &  32.40     &  25.95     &  20.02     &  15.61     \\
  9.750     &  447.0     &  320.0     &  240.5     &  175.0     &  127.4     &  100.4     &  79.33     &  61.20     &  46.37     &  36.42     &  28.95     &  22.15     &  17.15     \\
  10.00     &  490.2     &  347.6     &  259.3     &  196.8     &  143.3     &  113.2     &  89.49     &  69.13     &  52.20     &  40.94     &  32.36     &  24.74     &  19.00     \\
  10.25     &  539.6     &  379.0     &  280.5     &  211.7     &  161.0     &  127.6     &  100.9     &  78.49     &  59.10     &  46.07     &  36.50     &  27.61     &  21.04     \\
  10.50     &  577.7     &  414.4     &  304.3     &  228.4     &  173.3     &  140.5     &  113.9     &  87.86     &  66.88     &  51.98     &  41.46     &  30.91     &  23.52     \\
  10.75     &  612.7     &  448.9     &  331.2     &  247.1     &  186.9     &  151.2     &  123.1     &  97.18     &  74.73     &  58.50     &  46.47     &  34.56     &  26.20     \\
  11.00     &  651.7     &  474.6     &  358.2     &  267.9     &  202.1     &  162.9     &  132.6     &  106.4     &  82.36     &  64.22     &  51.36     &  38.45     &  28.83     \\
  11.25     &  694.6     &  502.9     &  378.2     &  289.6     &  218.9     &  175.9     &  143.1     &  115.0     &  89.13     &  69.31     &  56.29     &  41.87     &  31.37     \\
  11.50     &  741.9     &  534.3     &  400.3     &  305.4     &  236.6     &  190.1     &  154.6     &  122.7     &  95.41     &  74.79     &  60.80     &  45.47     &  33.74     \\
  11.75     &  787.7     &  568.8     &  424.7     &  322.8     &  249.5     &  203.3     &  167.1     &  131.0     &  102.1     &  80.68     &  65.21     &  49.08     &  36.17     \\
  12.00     &  835.3     &  605.2     &  451.6     &  342.0     &  263.7     &  214.6     &  177.0     &  139.3     &  109.3     &  86.96     &  69.69     &  52.43     &  38.69     \\
  12.25     &  885.6     &  641.6     &  479.8     &  363.0     &  279.2     &  227.0     &  186.9     &  147.6     &  116.1     &  93.09     &  74.47     &  55.96     &  41.36     \\
  12.50     &  939.0     &  680.3     &  509.2     &  385.4     &  296.3     &  240.7     &  197.7     &  156.7     &  123.4     &  98.90     &  79.37     &  59.84     &  44.29     \\
  12.75     &  995.4     &  721.2     &  540.4     &  408.7     &  314.7     &  255.4     &  209.5     &  166.6     &  131.2     &  105.2     &  84.67     &  63.97     &  47.35     \\
  13.00     &  1055.     &  764.5     &  573.4     &  433.5     &  334.1     &  271.2     &  222.4     &  177.1     &  139.6     &  112.1     &  90.29     &  68.34     &  50.60     \\
  13.25     &  1118.     &  810.4     &  608.4     &  459.7     &  354.7     &  287.9     &  236.2     &  187.8     &  148.3     &  119.2     &  96.26     &  72.90     &  53.86     \\
  13.50     &  1185.     &  859.0     &  645.4     &  487.5     &  376.5     &  305.7     &  250.9     &  199.2     &  157.4     &  126.6     &  102.3     &  77.56     &  57.21     \\
  13.75     &  1257.     &  910.6     &  684.6     &  517.0     &  399.6     &  324.4     &  266.4     &  211.2     &  167.0     &  134.5     &  108.8     &  82.50     &  60.77     \\
  14.00     &  1332.     &  965.1     &  726.2     &  548.2     &  424.0     &  344.4     &  282.9     &  223.9     &  177.2     &  142.9     &  115.6     &  87.73     &  64.53     \\
  14.25     &  1412.     &  1023.     &  770.2     &  581.2     &  450.0     &  365.4     &  300.4     &  237.4     &  188.0     &  151.7     &  122.9     &  93.27     &  68.52     \\
  14.50     &  1496.     &  1084.     &  816.8     &  616.2     &  477.4     &  387.7     &  318.8     &  251.7     &  199.4     &  161.1     &  130.5     &  99.16     &  72.75     \\
  14.75     &  1586.     &  1149.     &  866.2     &  653.2     &  506.5     &  411.4     &  338.4     &  266.8     &  211.5     &  171.1     &  138.6     &  105.4     &  77.22     \\
  15.00     &  1681.     &  1218.     &  918.5     &  692.5     &  537.2     &  436.4     &  359.1     &  282.9     &  224.3     &  181.6     &  147.2     &  112.0     &  81.96     \\
\hline
X-ray\tablenotemark{b} & 4300  & 3100 & 2240  & 1620 & 1170   & 864   & 640   & 440   & 300   & 201   & 135   & 79.3  & 42.2 \\
\hline
$m'_{\rm w}$\tablenotemark{c} & 15.2  & 14.9 & 14.8  & 14.5 &  14.3   & 14.2   & 14.0   & 14.0   & 13.8   & 13.8   & 13.8   & 13.8  & 13.8 \\
\hline
$\log f_{\rm s}$\tablenotemark{d} & 0.72  & 0.60 & 0.48  & 0.38 & 0.27   & 0.18   & 0.09   & $-0.02$   & $-0.12$   & $-0.22$   & $-0.31$   & $-0.44$  & $-0.60$ \\
\hline
$M_{\rm w}$\tablenotemark{e} & 4.7  & 4.1 & 3.7  & 3.1 & 2.7   & 2.3   & 2.0   & 1.6   & 1.2   & 0.9   & 0.7   & 0.4  & 0.0 \\
\enddata
\tablenotetext{a}{chemical composition of the envelope is assumed
to be that of ``Ne nova 2'' in Table \ref{chemical_composition}}
\tablenotetext{b}{duration of supersoft X-ray phase in units of days}
\tablenotetext{c}{converted magnitudes at the bottom point in Figure
\ref{mass_v_uv_x_v1974_cyg_x55z02o10ne03_new_model}}
\tablenotetext{d}{stretching factor against V1974 Cyg observational data
in Figure \ref{mass_v_uv_x_v1974_cyg_x55z02o10ne03_new_model}}
\tablenotetext{e}{absolute magnitudes at the bottom point in Figure
\ref{mass_v_uv_x_v1974_cyg_x55z02o10ne03_new_model} by assuming
$(m-M)_V = 12.3$  (V1974 Cyg)}
\end{deluxetable*}

%Fig.1
%%%\placefigure{v1668_cyg_vy_jhk_mag_x35z02c10o20_model_test_all}

\begin{figure*}
\epsscale{0.85}
%\epsscale{1.0}
%\epsscale{1.15}
\plotone{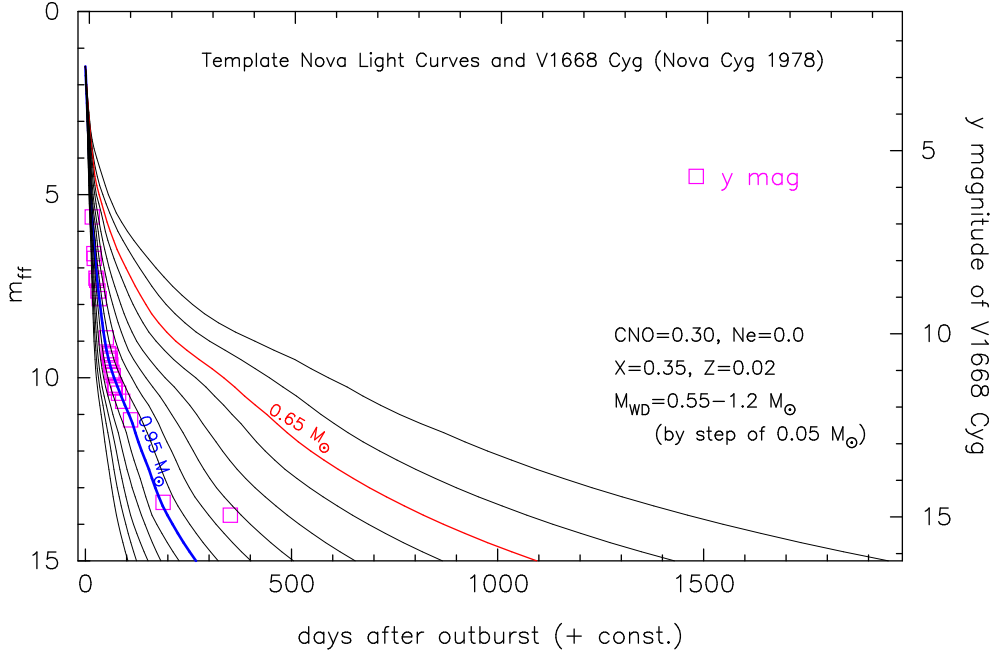}
%\plotone{f1bw.epsi}
%\plotone{v1668_cyg_vy_jhk_mag_x35z02c10o20_model_test_all_no4_color.epsi}
%\plotfiddle{evolution1.ps}{5.0cm}{270}{0.4}{0.4}{-170}{220}
\caption{
Magnitudes of our free-free emission model light curves
for various WD masses,
i.e., $0.55 - 1.2~M_\sun$ by $0.05~M_\sun$ step, numerical data 
of which are tabulated in Table \ref{light_curves_of_novae_co}.  
We adopt a chemical composition of ``CO nova 2'' in Table
\ref{chemical_composition} for these WD envelopes.
The decay timescale depends mainly on the WD mass.
We add observational $y$ magnitudes ({\it open squares})
of V1668 Cyg (Nova Cygni 1978) for comparison, which are taken 
from \citet{gal80}.  The $y$ magnitudes are shifted up by 1.2 mag
to fit them with the $0.95 ~M_\sun$ WD model (see the right axis).
Our light curve model
of $0.95 ~M_\sun$ WD (blue color line) almost perfectly
fits with the observational $y$ magnitudes except the last point.
}
\label{v1668_cyg_vy_jhk_mag_x35z02c10o20_model_test_all}
\end{figure*}

%Fig.2
%%%\placefigure{v1668_cyg_vy_jhk_mag_x35z02c10o20_model_scaling_law}

\begin{figure*}
\epsscale{0.85}
%\epsscale{1.0}
%\epsscale{1.15}
\plotone{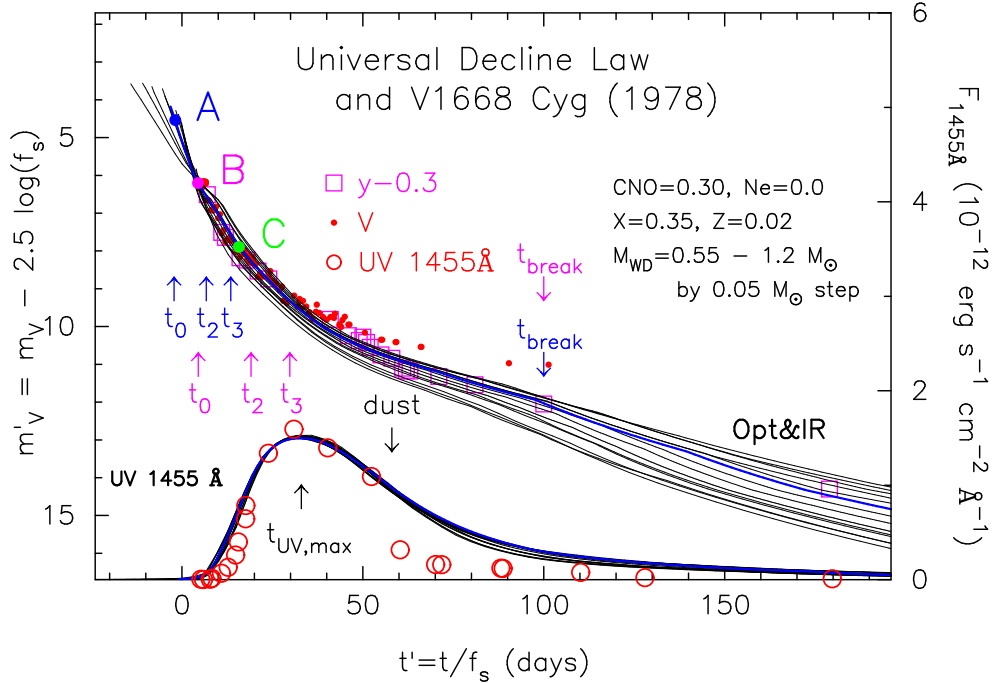}
%\plotone{f17bw.epsi}
%\plotone{v1668_cyg_vy_jhk_mag_x35z02c10o20_model_scaling_law_color.epsi}
%\plotfiddle{evolution1.ps}{5.0cm}{270}{0.4}{0.4}{-170}{220}
\caption{
Same as Figure \ref{v1668_cyg_vy_jhk_mag_x35z02c10o20_model_test_all},
but in the plane of rescaled flux, 
$-2.5 ~ \log f_{\rm s} F_\nu +{\rm const.} 
= m_V - 2.5 \log f_{\rm s} = m'_V$ and 
scaled time, $t/f_{\rm s}= t'$.
We added UV 1455\AA~ model light curves which are stretched
along time by a factor of $1/f_{\rm s}$.  This stretching factor
is determined in a way that each UV light curve is overlapped to
that of V1668 Cyg observation ({\it red big open circles}),
data of which are taken from \citet{cas02}. 
The UV model fluxes are normalized to 
the peak value of the UV observation. 
The values of $\log f_{\rm s}$ are tabulated
in Table \ref{light_curves_of_novae_co}.
The sudden drop in the UV flux of V1668 Cyg on day $\sim 60$ is
caused by formation of an optically thin dust shell (denoted
by an arrow with ``dust'').  
All the free-free emission model light curves are almost
overlapped to each other.  We call this homologous nature
of free-free light curves ``the universal decline law.''
Here $t_2$ and $t_3$ times are not uniquely determined
along our universal decline law, because these times depend
on the initial envelope
mass as indicated by two sets of ($t_0$, $t_2$, $t_3$) in the figure.
Three different initial envelope masses of 
$M_{\rm env,0} = 2.3 \times 10^{-5}$,
$1.6 \times 10^{-5}$, and 
$1.1 \times 10^{-5} M_\sun$ 
correspond to three peak brightnesses, points A, B, and C, respectively,
on the $0.95 ~M_\sun$ WD model.  We added the $V$ magnitude 
observation ({\it red filled circles}) for comparison, data of which
are taken from \citet{mal79}.  The $y$ magnitudes are shifted up by 
0.3 mag to match them with the $V$ magnitudes in the early phase.
}
\label{v1668_cyg_vy_jhk_mag_x35z02c10o20_model_scaling_law}
\end{figure*}

     The decay timescale of a light curve depends mainly on
the WD mass and weakly on the chemical composition,
whereas  the maximum brightness of a light curve depends
strongly on the initial envelope mass, $M_{\rm env, 0}$.  
In this section, we present various features, especially the
absolute magnitudes, of nova light curves
based on our universal decline law.

\subsection{Template of nova light curves}
Figure \ref{v1668_cyg_vy_jhk_mag_x35z02c10o20_model_test_all}
shows our free-free emission model light curves calculated from
Equation (\ref{wind-free-free-emission}), i.e.,
\begin{equation}
m_{\rm ff} = -2.5 ~ \log ~ \left[ {{\dot M_{\rm wind}^2}
\over {v_{\rm ph}^2 R_{\rm ph}}} \right]^{\{M_{\rm WD}\}}_{(t)}
+ ~ G^{\{M_{\rm WD}\}},
\label{template-wind-free-free-emission}
\end{equation}
 for various WD masses,
from $M_{\rm WD}= 0.55$ to $1.2~M_\sun$ by step of $0.05 ~M_\sun$ for 
a chemical composition of ``CO nova 2'' (these numerical data
are tabulated in Table \ref{light_curves_of_novae_co}).  
The subscript $(t)$ explicitly expresses that this is a function
of time while the superscript $\{M_{\rm WD}\}$ indicates a model 
parameter of WD mass.  An optically thick wind phase continues 
for a long time, from a very early phase to a late phase of
a nova outburst.  In this figure we plot free-free emission
model light curves calculated by Equation
(\ref{template-wind-free-free-emission})
from a very early phase until the end of
an optically thick wind phase.  The 15th mag point
of each light curve corresponds to the end of the wind phase.  
In other words, we define a constant of $G^{\{M_{\rm WD}\}}$ in Equation 
(\ref{template-wind-free-free-emission}) as that
the last (lowest) point of each light curve (the end
of an optically thick wind phase) is 15th mag.  
As a result, the value of $G^{\{M_{\rm WD}\}}$ is different for
different $M_{\rm WD}$.  In order to make
Table \ref{light_curves_of_novae_co} more compact,
we have defined $G^{\{M_{\rm WD}\}}$ in this way.

These light curves demonstrate that the decay timescales
of novae are sensitive to the WD mass.  Therefore we can determine
the WD mass of a nova from light curve fitting.
The $0.95 ~M_\sun$  (blue thick solid line)
or $1.0 ~M_\sun$ WD may be a typical case of
fast novae and the light curve of $0.65 ~M_\sun$ WD
(red thick solid line) may correspond to a slow nova.

These free-free emission model light curves are applicable to optical
and near infrared (IR) wavelength regions \citep[e.g.,][]{hac06kb}.
We added observational $y$ magnitudes of V1668 Cyg (Nova Cygni 1978)
to Figure \ref{v1668_cyg_vy_jhk_mag_x35z02c10o20_model_test_all},
which nicely reproduce the light curve of $0.95 ~M_\sun$ WD.
In this case, we shift the model light curve of
$0.95 ~M_\sun$ WD down by 1.2 mag
in order to fit it with the observation (see the right axis
of Figure \ref{v1668_cyg_vy_jhk_mag_x35z02c10o20_model_test_all}).
The intermediate width $y$ band is an emission line-free
optical band, so it is ideal to fit our model light curve
with $y$ magnitudes.

In this way, we can shift the template light curves to fit them
with observation for individual novae.  In other words,
one can shift the model light curve ``up and down'' against the
observational data of an individual nova.  In the case of 
V1668 Cyg, we obtain $m_y = m_{\rm ff} + 1.2$ from the 
$0.95 ~M_\sun$ WD model in Table \ref{light_curves_of_novae_co}.

%Fig.3
%%%\placefigure{v1668_cyg_vy_jhk_mag_x35z02c10o20_model_scaling_law_logt}

\begin{figure*}
\epsscale{0.85}
%\epsscale{1.0}
%\epsscale{1.15}
\plotone{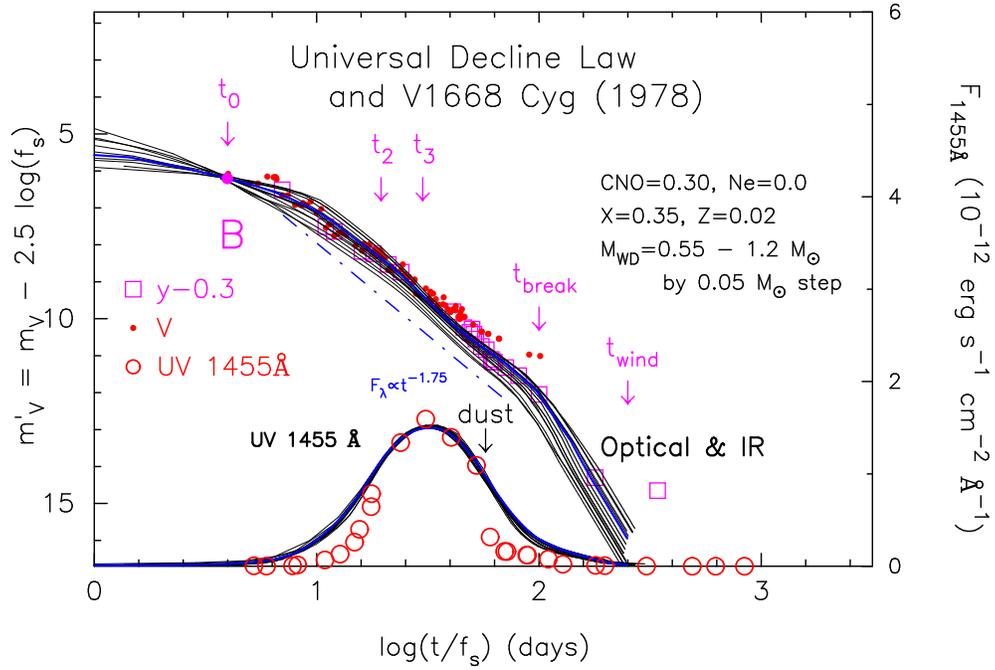}
%\plotone{f17bw.epsi}
%\plotone{v1668_cyg_vy_jhk_mag_x35z02c10o20_model_scaling_law_logt_color.epsi}
%\plotfiddle{evolution1.ps}{5.0cm}{270}{0.4}{0.4}{-170}{220}
\caption{
Same as Figure \ref{v1668_cyg_vy_jhk_mag_x35z02c10o20_model_scaling_law},
but in the logarithmic time vs. magnitude plane.
The origin of time is about 4 days before the optical maximum of
V1668 Cyg.  The right edges of free-free emission model light curves
correspond to the epoch when the optically thick winds stop (denote
by an arrow labeled $t_{\rm wind}$).
We also added a decline rate of free-free
flux, i.e., $F_\lambda \propto t^{-1.75}$ ({\it blue dash-dotted line})
in the early phase before $t_{\rm break}$.
}
\label{v1668_cyg_vy_jhk_mag_x35z02c10o20_model_scaling_law_logt}
\end{figure*}

\subsection{Brightness of free-free light curves}
\label{brightness_free-free}
The free-free emission model light curves in Figure
\ref{v1668_cyg_vy_jhk_mag_x35z02c10o20_model_test_all}
have strong similarity in their shapes.
In our previous paper \citep{hac06kb}, we showed 
that these model light curves are homologous and
their shapes are almost overlapped with each other when they
are properly squeezed or stretched along time.  In this subsection,
we reformulate this homologous properties in a more sophisticated
manner and obtain their brightnesses.

   First, we squeeze or stretch our UV 1455 \AA~  model light curves
in the direction of time and also its peak flux is normalized to
be fitted with the observational peak of V1668 Cyg.
Then we obtain all the UV light curves almost overlapped to 
each other as shown in Figure 
\ref{v1668_cyg_vy_jhk_mag_x35z02c10o20_model_scaling_law}.
We determine a stretching factor, $f_{\rm s}$, of each UV model light
curve by increasing or
decreasing it until the model light curve shape matches the observation.
Table \ref{light_curves_of_novae_co} shows logarithmic values of
$\log f_{\rm s}$, that is, positive (negative) for stretching (squeezing)
against the V1668 Cyg UV 1455 \AA~  observation.   For example,
we see that the $0.95 ~M_\sun$ WD model light curve is almost
overlapped to the V1668 Cyg observation without stretching ($f_{\rm s}
\approx 1.0$) and the $0.8 ~M_\sun$ WD model light curve evolves about
twice slower ($f_{\rm s} \approx 2.0$) than the $0.95 ~M_\sun$ WD model.
In this way, all the UV 1455 \AA~ model light curves are overlapped
to each other,  if we squeeze/stretch the time as $t' = t / f_{\rm s}$. 
Note that the stretching factor is slightly different from
that in our previous paper \citep[Table 9 of][]{hac06kb},
which was determined in the way that
each free-free model light curve is overlapped with each other
instead of UV 1455 \AA~ model light curves.

Figure \ref{v1668_cyg_vy_jhk_mag_x35z02c10o20_model_scaling_law}
also shows optical data ($y$ and $V$ magnitudes) of V1668 Cyg and
fitted free-free emission model light curve of $0.95 ~M_\sun$ WD
($f_{\rm s} \approx 1.0$).  The other free-free emission model
light curves are squeezed/stretched with the same $f_{\rm s}$
as that of UV light curves.   When we squeeze the model light
curves by a factor of $f_{\rm s}$ in the time direction
($t' = t/f_{\rm s}$ and $\nu ' = f_{\rm s} \nu$),
the flux is also changed as $F'_{\nu '} = f_{\rm s} F_\nu$
because
\begin{equation}
{{d } \over {d t'}} = f_{\rm s} {{d} \over {d {t}} }.
\label{time-derivation-flux}
\end{equation}
Therefore, we shift the model light curves in the vertical direction as
\begin{eqnarray}
m'_V =  m_V - 2.5 \log f_{\rm s},
\label{simple_final_scaling_flux}
\end{eqnarray}
and obtained Figure 
\ref{v1668_cyg_vy_jhk_mag_x35z02c10o20_model_scaling_law}.
Here $m_V$ is the apparent $V$ magnitude of an $M_{\rm WD}$ model
and $m'_V$ is its time-stretched magnitude
against a standard model of 0.95 $M_\sun$ WD ($\approx$ V1668 Cyg).
We see that all the light curves are almost overlapped to each other.
%Note that Equation (\ref{simple_final_scaling_flux}) cannot apply to
%blackbody emission.
For more details of derivation of Equation
(\ref{simple_final_scaling_flux}), see Appendix A.

The origins of magnitude in Figures
\ref{v1668_cyg_vy_jhk_mag_x35z02c10o20_model_test_all} and
\ref{v1668_cyg_vy_jhk_mag_x35z02c10o20_model_scaling_law}
are different because the magnitude in the right axis of Figure
\ref{v1668_cyg_vy_jhk_mag_x35z02c10o20_model_test_all} is calibrated
by the $y$ magnitudes whereas by the $V$ magnitudes in Figure
\ref{v1668_cyg_vy_jhk_mag_x35z02c10o20_model_scaling_law}.
We define the rescaled apparent $V$ magnitude (time-stretched 
$V$ magnitude) against the $V$ magnitudes of V1668 Cyg instead of
the $y$ magnitudes.  The apparent $V$ magnitudes are written as
\begin{equation}
m'_V = -2.5 ~ \log ~ \left[ {{\dot M_{\rm wind}^2}
\over {v_{\rm ph}^2 R_{\rm ph}}} \right]^{\{M_{\rm WD}\}}_{(t/f_{\rm s})}
+ K_V,
\label{simple_scaling_flux}
\end{equation}
where $K_V$ is a constant and is determined to match
$m'_V$ of the 0.95 $M_\sun$ WD model with
the $V$ magnitude of V1668 Cyg as shown in Figure
\ref{v1668_cyg_vy_jhk_mag_x35z02c10o20_model_scaling_law}.
Therefore, $K_V$ is also related to $G^{\{ 0.95 ~M_\sun \} }$ as 
\begin{equation}
K_V = G^{\{0.95 ~M_\sun \} } + 0.9.
\label{relation_between_constants} 
\end{equation}
This simply means
\begin{equation}
m'_V = m_V = m_{\rm ff} + 0.9,
\label{relation_between_ff_and_mv} 
\end{equation}
for 0.95 $M_\sun$ WD model with $f_{\rm s} = 1.0$.
Note that we here obtain 0.9 instead of 1.2 because observational
$m_V$ is brighter than $m_y$ by about 0.3 mag, that is,
$m_V$ and $m_y - 0.3$ are overlapped to each other in Figure
\ref{v1668_cyg_vy_jhk_mag_x35z02c10o20_model_scaling_law}.
We use the same $K_V$ for all other WD mass models
(from $M_{\rm WD}= 0.55$ to 1.2 $M_\sun$ by 0.05 $M_\sun$ step)
in Figure \ref{v1668_cyg_vy_jhk_mag_x35z02c10o20_model_scaling_law}.
All the free-free emission model light curves are almost
overlapped to each other.

It should be noticed here that the flux is condensed and
increased by a factor of $f_{\rm s}$ when the time is
stretched by a factor of $1/f_{\rm s}$ (or squeezed by a
factor of $f_{\rm s}$) {\it only under the condition that
the spectrum is not changed after time-stretching}.
This condition is satisfied in free-free emission
(optical and near IR) because the flux ($F_\nu \sim$ constant)
is independent
of the frequency $\nu$, but not satisfied in blackbody emission
because the blackbody emissivity depends on the frequency $\nu$ and,
after time-stretching, a blackbody spectrum changes with
$\nu ' = f_{\rm s} \nu$.   See Appendix A for more detailed
formulation of the scaling of free-free flux.

%Fig.4
%%%\placefigure{v1668_cyg_vy_jhk_mag_x35z02c10o20_model_real_scale_logt_no2}

\begin{figure*}
\epsscale{0.85}
%\epsscale{1.0}
%\epsscale{1.15}
\plotone{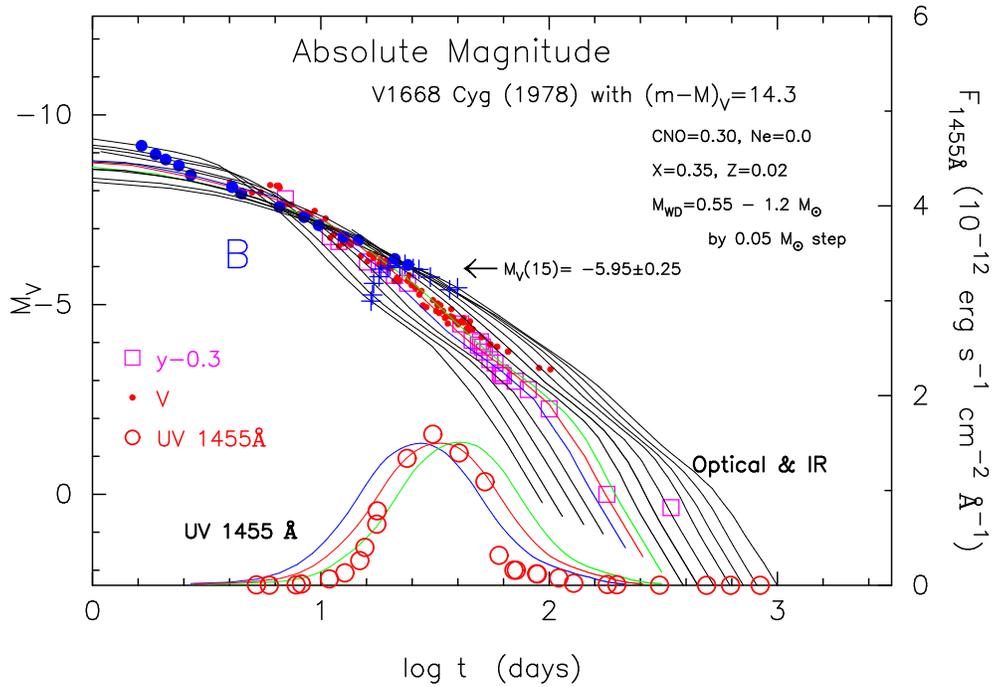}
%\plotone{f17bw.epsi}
%\plotone{v1668_cyg_vy_jhk_mag_x35z02c10o20_model_real_scale_logt_no2_color.epsi}
%\plotfiddle{evolution1.ps}{5.0cm}{270}{0.4}{0.4}{-170}{220}
\caption{
Same as Figure \ref{v1668_cyg_vy_jhk_mag_x35z02c10o20_model_scaling_law_logt},
but for absolute magnitudes and real timescales.
We have calibrated the free-free
model light curves by a distance modulus of $(m-M)_V = 14.3$ for
V1668 Cyg.  The position at point B in Figure
\ref{v1668_cyg_vy_jhk_mag_x35z02c10o20_model_scaling_law_logt}
is indicated by a blue filled circle.  We also show the magnitude,
$M_V(15)$, at 15 days after the optical maximum by a blue cross.
Their average value of $M_V(15) = -5.95 \pm 0.25$ is also indicated
in the figure for $0.7$--$1.05 ~M_\sun$ WDs.
Model UV and supersoft X-ray light curves are omitted to simplify 
the figure except three WD masses ({\it blue solid line} for 1.0,
{\it red solid line} for 0.95,
and {\it green solid line} for $0.9 ~M_\sun$).
}
\label{v1668_cyg_vy_jhk_mag_x35z02c10o20_model_real_scale_logt_no2}
\end{figure*}

\subsection{Maximum brightness vs. initial envelope mass}
Figure \ref{v1668_cyg_vy_jhk_mag_x35z02c10o20_model_scaling_law}
shows three points of A, B, and C on the same light curve of
$0.95 ~M_\sun$ WD model.  
Point B indicates the initial envelope mass of 
$M_{\rm env,0} = 1.6 \times 10^{-5} M_\sun$,
corresponding to $t_0$ at the maximum brightness 
of V1668 Cyg.  Point A indicates a much brighter maximum and
corresponds to a much larger envelope mass
($M_{\rm env,0} = 2.3 \times 10^{-5} M_\sun$) than that of point B.
On the other hand, point C corresponds to a much smaller envelope mass
($M_{\rm env,0} = 1.1 \times 10^{-5} M_\sun$).

The initial envelope mass depends on the mass accretion rate
to the WD before ignition \citep[see, e.g., Figure 9
 of][]{nom82, pri95}.   Point A is corresponding to a case
of much lower mass accretion rates to the WD.  On the other hand,
point C is corresponding to a case of much higher mass accretion
rates to the WD than that of point B, for the case of V1668 Cyg.  
This simply suggests that novae are brighter for lower mass
accretion rates.

For the other WD masses, the trend between the maximum brightness
and the initial envelope mass is the same as that of 
$0.95 ~M_\sun$.  For more (less) massive WDs, the envelope mass
at the same point B is smaller (larger) than that of $0.95 ~M_\sun$.
We will see again these relations between the maximum brightness 
and the initial envelope mass in more details in Section
\ref{mmrd_relation}.

We regard that V1668 Cyg is a typical classical nova and point B
in Figure \ref{v1668_cyg_vy_jhk_mag_x35z02c10o20_model_scaling_law}
corresponds to typical brightnesses for other WD mass models.
Therefore, we replot the same nova model light curves
in a logarithmic time in    
Figure \ref{v1668_cyg_vy_jhk_mag_x35z02c10o20_model_scaling_law_logt}.
In the overlapping procedure of UV 1455\AA\  model light curves,
we adopt the origin of time which is
different from the origin of time in $m_{\rm ff}$ of
Table \ref{light_curves_of_novae_co}.
This procedure is necessary for the horizontal axis of $\log t'$ 
because we must start all the free-free emission 
light curves from the same stage like point B (or C) in Figure
\ref{v1668_cyg_vy_jhk_mag_x35z02c10o20_model_scaling_law}.
Here we define the origin of time as about 4 days before point B
in Figure \ref{v1668_cyg_vy_jhk_mag_x35z02c10o20_model_scaling_law},
that is, the outburst day of V1668 Cyg.
In this case, the initial brightnesses of novae are almost the same
(brightness at point B) and overlapped with each other
in the $\log t'$--$m'_V$ diagram.

\subsection{Absolute magnitude of free-free light curves}
\label{absolute_magnitude}
We have given an apparent magnitude to each free-free emission 
model light curve against V1668 Cyg data.
Next step is to determine the absolute magnitudes of these
model light curves.  This can be done by using
the distance modulus of V1668 Cyg.
Figure \ref{v1668_cyg_vy_jhk_mag_x35z02c10o20_model_scaling_law}
shows observed $y$ and $V$ magnitudes of V1668 Cyg, which can be
converted to the absolute magnitude from the distance modulus
to V1668 Cyg, i.e., 
\begin{equation}
(m-M)_V = \left[5 \log(d/10) + A_V \right]_{\rm V1668~Cyg}  = 14.3.
\label{distance_modulus_v1668_cyg}
\end{equation}
Here, we adopt the distance of $d = 4.1$ kpc \citep{kat07h}
and the absorption of
$A_V = 3.1 \times E(B-V) = 3.1 \times 0.4 = 1.24$
\citep{sti81} for V1668 Cyg.
Since our $0.95 ~M_\sun$ WD model shows a good agreement
with the $y$ and $V$ magnitudes of V1668 Cyg, we determine
the absolute $V$ magnitude of our $0.95 ~M_\sun$ WD model 
light curve as follows:  
\begin{equation}
M_V = m_V - 14.3 = (m_{\rm ff} + 0.9) -14.3,
\label{absolute_mag_v1668_cyg}
\end{equation}
where $m_V = m'_V = m_{\rm ff} + 0.9$ for $0.95 ~M_\sun$ WD model 
($f_{\rm s}=1$) from Equation (\ref{relation_between_ff_and_mv}).
Then, the absolute $V$ magnitude at the last point of free-free
template light curve (the end point of Table 
\ref{light_curves_of_novae_co}) corresponds to
\begin{equation}
M_{\rm w} = m_{\rm w} - 14.3 = (m_{\rm ff} + 0.9) -14.3 = 1.6,
\label{last_point_v1668_cyg}
\end{equation}
where the subscript of ``w'' means ``wind'' which corresponds to
$m_{\rm ff} = 15$ at the end point (Table \ref{light_curves_of_novae_co}).
Using $M_{\rm w}$, we restore the entire absolute $V$ magnitudes of 
the $0.95 ~M_\sun$ WD model by
\begin{equation}
M_V = (m_{\rm ff} - 15) + M_{\rm w},
\label{resotre_abs_mag_v1668_cyg}
\end{equation}
and it is shown by a red solid line in Figure
\ref{v1668_cyg_vy_jhk_mag_x35z02c10o20_model_real_scale_logt_no2}.

\begin{figure}
%\epsscale{0.6}
%\epsscale{0.85}
\epsscale{1.0}
%\epsscale{1.15}
\plotone{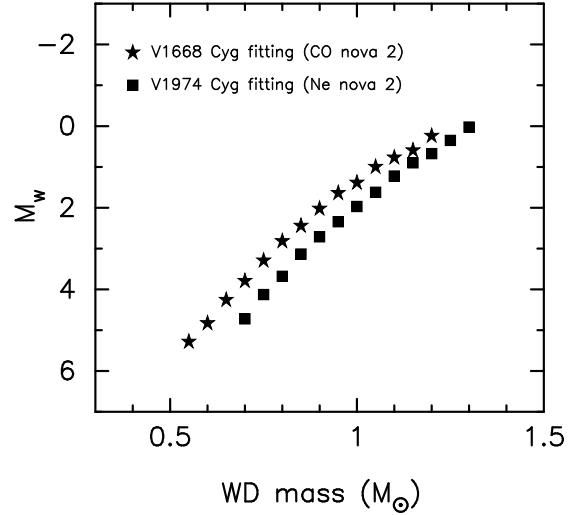}
%\plotone{f5bw.epsi}
%\plotone{absolute_mag_v1974cyg_v1668_cyg_scale_color.epsi}
%\plotfiddle{evolution1.ps}{5.0cm}{270}{0.4}{0.4}{-170}{220}
\caption{
The absolute magnitude at the end point of each free-free emission
model light curve, $M_{\rm w}$, is plotted against the WD mass for two
different chemical compositions, i.e., ``CO nova 2'' 
({\it star marks}) and ``Ne nova 2'' ({\it squares}).
See text for details. 
}
\label{absolute_mag_v1974cyg_v1668_cyg_scale}
\end{figure}

For the other WD mass models,
from Equation (\ref{simple_final_scaling_flux})
we simply derive the following relations, i.e., 
\begin{equation}
M_V = m'_V + 2.5 \log f_{\rm s} - 14.3,
\label{absolute_mag_common_v1668_cyg}
\end{equation}
\begin{equation}
M_{\rm w} = m'_{\rm w} + 2.5 \log f_{\rm s} - 14.3,
\label{absolute_bottom_mag_v1668_cyg}
\end{equation}
based on the V1668 Cyg data.  
We plot $M_{\rm w}$ against various WD masses in Figure
\ref{absolute_mag_v1974cyg_v1668_cyg_scale}.
Using $M_{\rm w}$, we have
\begin{equation}
M_V = (m_{\rm ff} - 15) + M_{\rm w},
\label{resotre_abs_bottom_mag_v1668_cyg}
\end{equation}
where $\log f_{\rm s}$, $m'_{\rm w}$, and $M_{\rm w}$
are tabulated in Table \ref{light_curves_of_novae_co}
for all other WD mass models
from $M_{\rm WD}= 0.55$ to 1.2 $M_\sun$ by 0.05 $M_\sun$ step.
Here prime symbol means the values for squeezed/stretched
light curves.  We have restored the entire absolute $V$ magnitudes
for $0.55$--$1.2 ~M_\sun$ WDs in Figure
\ref{v1668_cyg_vy_jhk_mag_x35z02c10o20_model_real_scale_logt_no2}.

It is interesting to examine whether or not the absolute magnitudes
of our free-free emission model light curves are consistent with
various empirical relations of classical novae that have ever been
proposed.  Very popular relations between the maximum magnitude
and the rate of decline (MMRD) will be discussed in Section
\ref{mmrd_relation}.  Here we check the empirical formula
that the absolute magnitude at 15 days after
optical maximum, $M_V(15)$, is almost common among various novae
\citep[e.g.,][]{bus55, cap89, coh85, dow00, van87}.  This relation
was first proposed by \citet{bus55} with $M_V(15)= -5.2 \pm 0.1$,
then followed by \citet{coh85} with $M_V(15)= -5.60 \pm 0.43$,
by \citet{van87} with $M_V(15)= -5.23 \pm 0.39$,
by \citet{cap89} with $M_V(15)= -5.69 \pm 0.42$,
and by \citet{dow00} with $M_V(15)= -6.05 \pm 0.44$.
We have obtained $M_V(15)= -5.7 \pm 0.3$ for $0.55$--$1.2 ~M_\sun$
WDs (14 light curves) with equal weight for all WD mass models,
being roughly consistent with the above empirical estimates
as shown by blue crosses in Figure 
\ref{v1668_cyg_vy_jhk_mag_x35z02c10o20_model_real_scale_logt_no2}.
For the statistical point of view, carbon-oxygen (CO) novae have
typical masses of $0.7$--$1.05 ~M_\sun$.  If we take this rage of
WD masses (equal weight), the value becomes $M_V(15)= -5.95 \pm 0.25$
as indicated in Figure
\ref{v1668_cyg_vy_jhk_mag_x35z02c10o20_model_real_scale_logt_no2},
being close to the value proposed by \citet{dow00}.

%Fig.5
%%%\placefigure{absolute_mag_v1974cyg_v1668_cyg_scale}

%Fig.6
%%%\placefigure{mass_v_uv_x_v1974_cyg_x55z02o10ne03_new_model}

\begin{figure*}
\epsscale{0.85}
%\epsscale{1.0}
%\epsscale{1.15}
\plotone{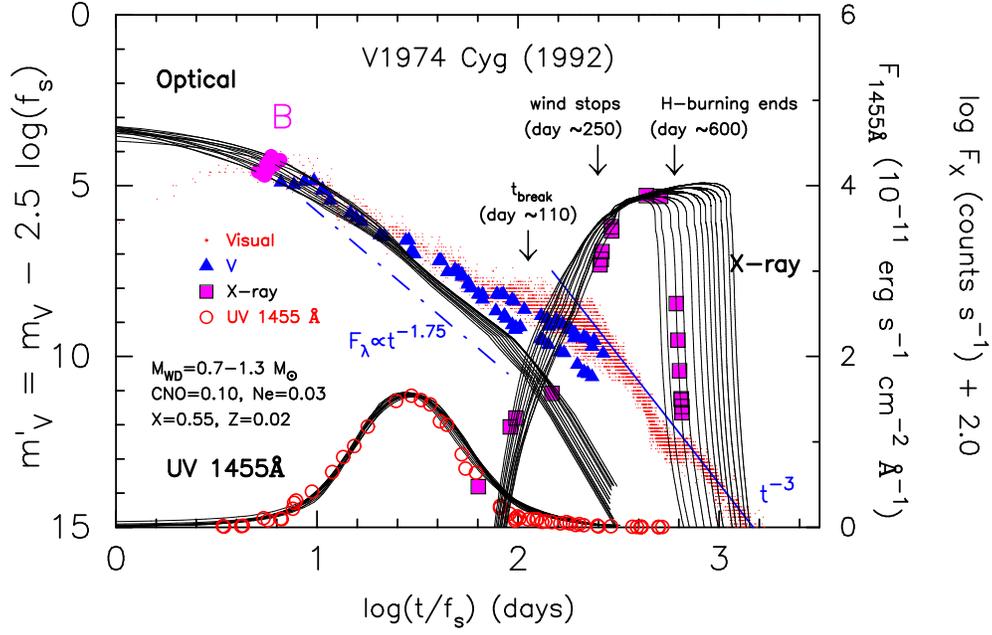}
%\plotone{f17bw.epsi}
%\plotone{mass_v_uv_x_v1974_cyg_x55z02o10ne03_new_model_color.epsi}
%\plotfiddle{evolution1.ps}{5.0cm}{270}{0.4}{0.4}{-170}{220}
\caption{
Same as Figure \ref{v1668_cyg_vy_jhk_mag_x35z02c10o20_model_scaling_law_logt},
but models for ``Ne nova 2.'' Observational data of the neon nova
V1974 Cyg is added for comparison.  Thirteen different WD mass models
($0.7-1.3 ~M_\sun$ by $0.05 ~M_\sun$ step) are plotted
in the logarithmic time--magnitude plane, data of which
are tabulated in Table \ref{light_curves_of_novae_ne}.  
We squeeze or stretch the model light curve along time to fit with
the observational data of V1974 Cyg, i.e.,
{\it ROSAT} X-ray observation,
optical, and {\it IUE} UV 1455\AA~ light curves.
UV 1455\AA~ model fluxes are normalized to have the same peak value
as that of {\it IUE} data, peak of which is 
$1.54 \times 10^{-11}$ erg s$^{-1}$ cm$^{-2}$ \AA$^{-1}$.
Fluxes of supersoft X-ray are normalized to fit it with 
the peak of {\it ROSAT} X-ray data.  The observational $V$ magnitudes
are taken from \citet{cho93}, visual data from
the American Association of Variable Star Observers (AAVSO), {\it IUE}
UV 1455 \AA~ data from \citet{cas02}, and {\it ROSAT} X-ray data 
from \citet{kra96}.  We also added two decline rates of free-free
flux, i.e., $F_\lambda \propto t^{-1.75}$ ({\it blue dash-dotted line})
in the early phase before $t_{\rm break}$, 
and $F_\lambda \propto t^{-3}$({\it blue solid straight line})
of Equation (\ref{expansion-free-free-emission})
in the expanding nebular phase.
Here we set point B on each model light curve near $t_0$.
}
\label{mass_v_uv_x_v1974_cyg_x55z02o10ne03_new_model}
\end{figure*}

%Fig.7
%%%\placefigure{mass_v_uv_x_v1974_cyg_x55z02o10ne03_real_scale_model}

\begin{figure*}
\epsscale{0.85}
%\epsscale{1.0}
%\epsscale{1.15}
\plotone{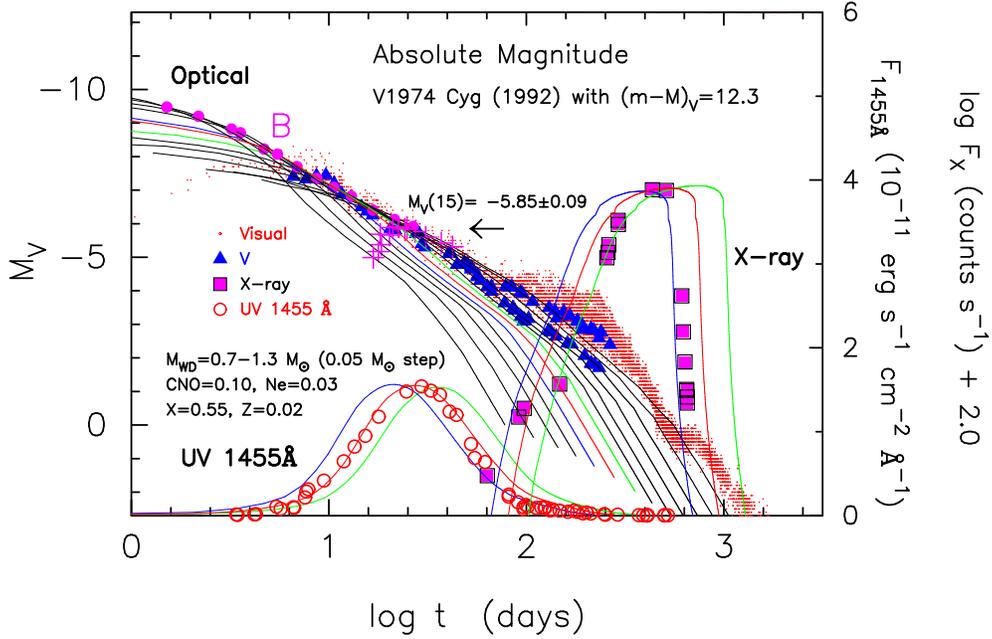}
%\plotone{f17bw.epsi}
%\plotone{mass_v_uv_x_v1974_cyg_x55z02o10ne03_real_scale_model_color.epsi}
%\plotfiddle{evolution1.ps}{5.0cm}{270}{0.4}{0.4}{-170}{220}
\caption{
Same as Figure \ref{mass_v_uv_x_v1974_cyg_x55z02o10ne03_new_model},
but for absolute magnitudes and real timescales.
We have calibrated the free-free
model light curves by a distance modulus of $(m-M)_V = 12.3$ for
V1974 Cyg.  The positions at point B in Figure
\ref{mass_v_uv_x_v1974_cyg_x55z02o10ne03_new_model}
are indicated by a magenta filled circle.  We also show the magnitude,
$M_V(15)$, at 15 days after the optical maximum by a magenta cross.
Their average value of $M_V(15) = -5.85 \pm 0.09$ is also indicated
in the figure for $0.9$--$1.15 ~M_\sun$ WDs.
Model UV and supersoft X-ray light curves are omitted to simplify 
the figure except
three WD masses ({\it blue solid line} for 1.1,
{\it red solid line} for 1.05,
and {\it green solid line} for $1.0 ~M_\sun$).
}
\label{mass_v_uv_x_v1974_cyg_x55z02o10ne03_real_scale_model}
\end{figure*}

\subsection{Absolute magnitudes for neon novae}
   In this subsection, we study light curves of neon (Ne) novae and show
that the essential results on the free-free emission model light curves
are the same as those for CO novae studied in the previous subsections.
Table \ref{light_curves_of_novae_ne} lists free-free model light curves
in the form of $m_{\rm ff}$, definition of which is the same
(see Equation (\ref{template-wind-free-free-emission})) but for
the chemical composition of ``Ne nova 2'' in Table 
\ref{chemical_composition}.

Figure \ref{mass_v_uv_x_v1974_cyg_x55z02o10ne03_new_model} shows
free-free (optical), UV 1455 \AA, and supersoft X-ray model
light curves as well as observational data of V1974 Cyg (Nova Cygni 1992).
We plot these light curves
in $\log t'$--$m'_V$ plane and show nova evolutions until more
later phases.
In the overlapping procedure of UV 1455\AA\  model light curves,
we adopt the origin of time which is
different from the origin of time in $m_{\rm ff}$ of
Table \ref{light_curves_of_novae_ne}.
This procedure is necessary for the same reason as in 
Figure \ref{v1668_cyg_vy_jhk_mag_x35z02c10o20_model_scaling_law_logt}.
As a result, the peak brightness
of each light curve is almost the same in the $\log t'$--$m'_V$ plane
as shown in Figure \ref{mass_v_uv_x_v1974_cyg_x55z02o10ne03_new_model}.
In this figure, we adopt $M_{\rm env, 0}= 1.5 \times 10^{-5}
M_\sun$ for $1.05 ~M_\sun$ WD model ($f_{\rm s} \approx 1.0$) as
a starting point of the model light curves, corresponding to
$m_{\rm ff} = 4.1$ of $1.05 ~M_\sun$ WD model
in Table \ref{light_curves_of_novae_ne}.

We also obtain the absolute magnitudes of free-free model light curves
in the same way as the CO novae mentioned in the previous subsections,
but for the neon nova V1974 Cyg as a comparison.
Adopting the distance of $d = 1.8$ kpc \citep{cho97}
and the absorption of $A_V = 1.0$ \citep{cho93} for V1974 Cyg,
we obtain the distance modulus of
\begin{equation}
(m-M)_V = \left[5 \log(d/10) + A_V \right]_{\rm V1974~Cyg}  = 12.3,
\label{distance_modulus_v1974_cyg}
\end{equation}
for V1974 Cyg.  Then, we specify the absolute
magnitude at the bottom of our free-free model light curves as
\begin{equation}
M_{\rm w} = m'_{\rm w} + 2.5 \log f_{\rm s} - 12.3,
\label{absolute_bottom_mag_v1974_cyg}
\end{equation}
for V1974 Cyg
and tabulate $f_{\rm s}$, $m'_{\rm w}$, and $M_{\rm w}$ in
Table \ref{light_curves_of_novae_ne}.  We plot 
$M_{\rm w}$ for both CO and Ne novae in
Figure \ref{absolute_mag_v1974cyg_v1668_cyg_scale}.
This clearly shows that the brightness $M_{\rm w}$
is smoothly increasing from $\sim 5$ mag to $\sim 0$ mag with
$M_{\rm WD}$ from 0.55 to 1.2 $M_\sun$ for ``CO nova 2''
(from 0.7 to 1.3 $M_\sun$ for ``Ne nova 2'').

Now, we can determine the absolute magnitude of free-free model
light curves, $M_V = (m_{\rm ff} -15.0) + M_{\rm w}$
from Table \ref{light_curves_of_novae_ne},
for various WD masses as shown
in Figure \ref{mass_v_uv_x_v1974_cyg_x55z02o10ne03_real_scale_model}.

We also check whether or not $M_V(15)$ is almost common among 
various neon novae. 
We have obtained $M_V(15)= -5.6 \pm 0.3$ for $0.70$--$1.3 ~M_\sun$
WDs (13 light curves with equal weight for all WD mass models), being
roughly consistent with the empirical estimates (see Section 
\ref{absolute_magnitude}).
For the statistical point of view, neon novae have
typical masses of $0.9$--$1.15 ~M_\sun$.  If we take this rage of
WD masses (equal weight for 6 light curves),
the value becomes $M_V(15)= -5.85 \pm 0.09$ as indicated in Figure
\ref{mass_v_uv_x_v1974_cyg_x55z02o10ne03_real_scale_model}.

If we take all the CO and Ne nova light curves with equal weight
($14 + 13$ WD mass models), the average value becomes
$M_V(15)= -5.7 \pm 0.3$.
Thus, the absolute magnitudes of our free-free emission model
light curves are consistent with the $M_V(15)$ empirical relations.

%Fig.8
%%%\placefigure{v1668_cyg_vy_jhk_mag_x35z02c10o20}

\begin{figure*}
\epsscale{0.85}
%\epsscale{1.0}
%\epsscale{1.15}
\plotone{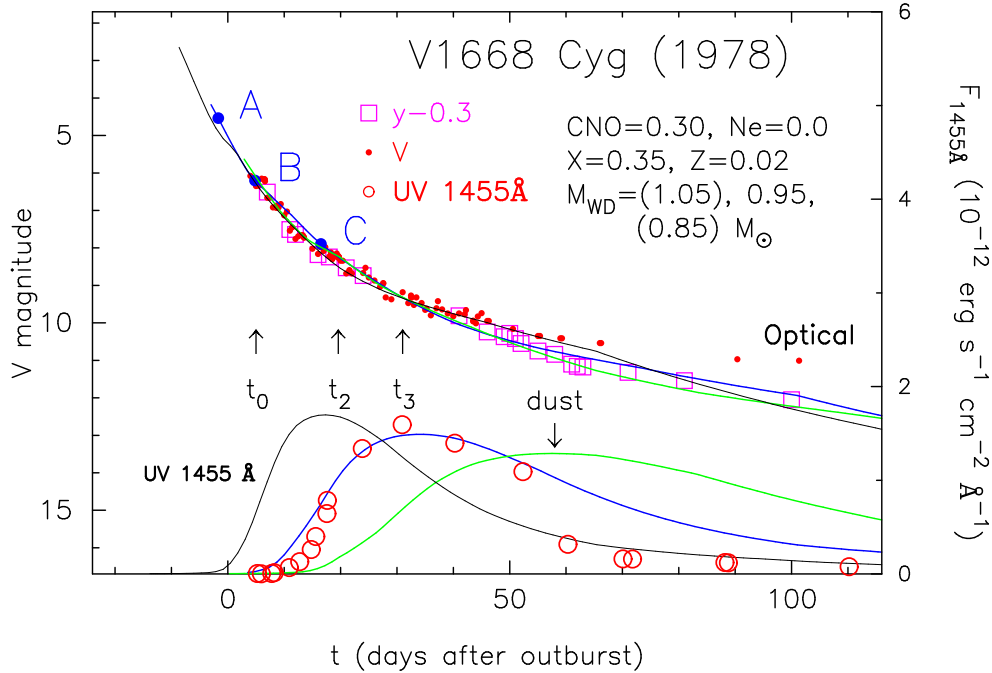}
%\plotone{f14bw.epsi}
%\plotone{v1668_cyg_vy_jhk_mag_x35z02c10o20_model_new_color.epsi}
%\plotfiddle{evolution1.ps}{5.0cm}{270}{0.4}{0.4}{-170}{220}
\caption{
Close-up view of three light curves in
Figures \ref{v1668_cyg_vy_jhk_mag_x35z02c10o20_model_test_all}
and \ref{v1668_cyg_vy_jhk_mag_x35z02c10o20_model_scaling_law}.
Three models are for 1.15 ({\it black thin line}), 
0.95 ({\it blue thick line}), and $0.85 ~M_\sun$
({\it Green thick line}) WDs.  None of them are squeezed
or stretched.   The model of $0.95 ~M_\sun$ WD reasonably reproduces
both the optical $y$ (and $V$) and UV 1455 \AA~ light curves
of V1668 Cyg.
In order to fit free-free model light curves of 1.15 and $0.85 ~M_\sun$
WDs with the observation, we shift the $1.05 ~M_\sun$ up and leftward,
and the $0.85 ~M_\sun$ down and rightward.
As a result, two free-free model light curves reasonably follow
the $y$ magnitudes but the two UV light curves
do not fit with the observation of V1668 Cyg.
Three arrows indicate $t_0$, $t_2$, and $t_3$ times of V1668 Cyg.
The times of $t_2$ and $t_3$ are determined along our model light curve
of the $0.95 ~M_\sun$ WD, which we call ``intrinsic'' times.
Our estimated ``intrinsic'' values of V1668 Cyg 
are $t_2 = 14.4$ and $t_3 = 26$ days.  We also add observational
$V$ magnitudes ({\it red small filled circles}), which
deviate from $y$ magnitudes in the later phase due mainly to
contribution from strong emission lines such as [\ion{O}{3}].
Fortunately this deviation has no effects on the estimate
of $t_2$ and $t_3$ times because
it starts about 40 days after the outburst in this case.
Points A, B, and C correspond to the same initial envelope masses
as in Figure \ref{v1668_cyg_vy_jhk_mag_x35z02c10o20_model_scaling_law}
along the $0.95 ~M_\sun$ WD model.
The magnitude at the end of a wind phase (outside the figure)
is $m_{\rm w}= 14.9$, 15.9, 
and 17.1 in this plot,
for $M_{\rm WD}= 1.05$, 0.95, and $0.85 ~M_\sun$,
respectively.  So the distance moduli of these three light curves are
$(m-M)_V = m_{\rm w} - M_{\rm w}
= 14.9 - 1.0= 13.9$,  $15.8 - 1.6= 14.3$,  $17.1 - 2.4= 14.7$,
respectively.  
}
\label{v1668_cyg_vy_jhk_mag_x35z02c10o20}
\end{figure*}

% Fig.9
%\placefigure{light_curve_v598_pup_v1500_cyg}

\begin{figure}
%\epsscale{1.0}
\epsscale{1.15}
\plotone{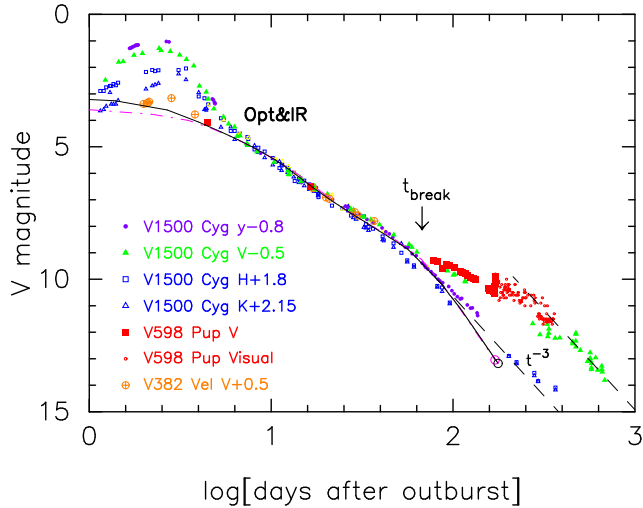}
%\plotone{f1bw.epsi}
%\plotone{light_curve_v598_pup_v1500_cyg_color.epsi}
%\plotfiddle{evolution1.ps}{5.0cm}{270}{0.4}{0.4}{-170}{220}
\caption{
Light curves for V1500 Cyg as an example of $t_{\rm break}$.
Optical $V$ ({\it filled triangles}), $y$ ({\it filled circles}),
near infrared (IR) $H$ ({\it open squares}), and $K$ ({\it open
triangles}) magnitudes of the fast nova V1500 Cyg.
Light curves of free-free emission are essentially
independent of wavelength, so that $V$, $y$, $H$, and $K$ light curves
of V1500 Cyg are almost overlapped with each other from several days
to $\sim 70$ days after the outburst.
Then, the $V$ light curve deviates from the others
($y$, $H$, and $K$) mainly because the wide $V$ band is heavily
contaminated by strong emission lines such as [\ion{O}{3}].
Two other nova light curves of V598 Pup and V382 Vel show an almost
identical decline from several days to several tens of days
after the outburst, that is, the free-free part of light curves.
Optical $V$ ({\it filled squares}) and visual ({\it small open
circles}) of V598 Pup (see Fig. 
\ref{all_mass_v598_pup_v1500_cyg_x55z02o10ne03} for more details of
V598 Pup) and $V$ ({\it open circles with plus})
of V382 Vel (see Fig. \ref{all_mass_v382_vel_v1500_cyg_x55z02o10ne03}
for more details of V382 Vel).
{\it Black solid line}: free-free model light curve
of $1.15 ~M_\sun$ WD with a composition of 
``Ne nova 2.'' 
{\it Magenta dash-dotted line}: free-free model light curve
of $1.05 ~M_\sun$ WD with a composition of 
``CO nova 2.''
{\it Large open circle}: the epoch
when the wind stops.  {\it Dashed line}:  model
light curve based on free-free emission following
a law of $F_{\lambda} \propto t^{-3}$
(Equation [\ref{expansion-free-free-emission}] after winds stop).
Near IR data of V1500 Cyg are taken from \citet{enn77},
\citet{gal76}, and \citet{kaw76};
$y$ and $V$ data of V1500 Cyg are from \citet{loc76} and 
\citet{tem79}.
}
\label{light_curve_v598_pup_v1500_cyg}
\end{figure}

\subsection{Consistency check among nova light curves}
\label{checking_absolute_magnitude}
We have proposed new methods for obtaining the absolute magnitude of
novae in the previous subsections.  Using these absolute magnitudes, 
we can estimate the distance to novae.  In this subsection,
we will check the accuracy of our new methods by comparing three
nova light curves.

First of all, we explain how to select a model that reproduces well
observational light curves.  Figure \ref{v1668_cyg_vy_jhk_mag_x35z02c10o20}
shows our method in the case of V1668 Cyg.   We shift the model light
curve ``back and forth'' and ``up and down'' to fit with
the observational $y$ magnitudes.   Together with the UV
1455\AA\  light curve, we finally determine by eyes
the $0.95 ~M_\sun$ WD as the best-reproducing one among
the light curves tabulated in Table \ref{light_curves_of_novae_co}
($0.55-1.2 ~M_\sun$ by $0.05 ~M_\sun$ step).  In this figure,
we added two other light curves of different WD masses, i.e.,
$0.85$ and $1.05 ~M_\sun$ WDs.  Here, we use different phases
of the model light curves
in Figure \ref{v1668_cyg_vy_jhk_mag_x35z02c10o20_model_test_all}
to fit with the $y$ observation; the $0.85 ~M_\sun$ WD model is shifted
rightward (back) but the $1.05 ~M_\sun$ WD model is
shifted leftward (forth).  As a result, the three model light
curves of $0.85$, $0.95$, and $1.05 ~M_\sun$ WDs give 
similar decline rates from maximum until day $\sim 40$.
In other words, we can find similar shapes in different phases
of different WD mass light curves
in Figure \ref{v1668_cyg_vy_jhk_mag_x35z02c10o20_model_test_all}.
In this sense, it is hard to choose the best-reproducing model 
only from early phase optical and near IR data
if neither UV 1455 \AA~ nor X-ray data are available.

It should be also noted here that the ambiguity of WD mass 
estimate of $\mp ~0.1 ~M_\sun$ introduces an error of about
$\pm ~0.4$ mag in the distance modulus of $(m-M)_V$, 
when only early optical (or IR) light curves are available.
In Figure \ref{v1668_cyg_vy_jhk_mag_x35z02c10o20},
we obtain $(m-M)_V = m_{\rm w} - M_{\rm w} =
15.8 - 1.6 =14.3$ for $M_{\rm WD}= 0.95 ~M_\sun$, but
$(m-M)_V = 14.9 -1.0 = 13.9$ for $M_{\rm WD}= 1.05 ~M_\sun$
or $(m-M)_V = 17.1 - 2.4 = 14.7$ for $M_{\rm WD}= 0.85 ~M_\sun$.

Now let us go to the first example of
a very fast nova V1500 Cyg (Nova Cygni 1975).
Figure \ref{light_curve_v598_pup_v1500_cyg} demonstrates that
optical $V$ (filled triangles), $y$ (filled circles),
near infrared $H$ (open squares), and $K$ (open triangles)
magnitudes are almost overlapped with each other
from $\sim 5$ days until $\sim 100$ days after the outburst.
This wavelength-free decline is a characteristic property
of free-free emission.  This impressive part is reproduced well
by Equation (\ref{wind-free-free-emission})
with the models of $M_{\rm WD}= 1.05 ~M_\sun$ (``CO nova 2'')
and $M_{\rm WD}= 1.15 ~M_\sun$ (``Ne nova 2'').  V1500 Cyg is a
superbright nova, in which the peak luminosity was
about 4 mag brighter than the Eddington limit of a 1.0 $M_\sun$ WD
\citep[e.g.,][]{fer86}, and the spectra of the superbright part
(first 5 days) are approximated by blackbody emission
\citep[e.g.,][]{gal76}.  Therefore, our free-free
emission model cannot be applicable to the superbright part. 

After day $\sim 70$, the $V$ light curve gradually deviates from
the other light curves ($y$, $H$, and $K$)
because emission lines dominate the optical fluxes.
The $V$ magnitudes are 2 mag or more brighter than
the $y$, $H$, and $K$ magnitudes in the late phase.
As extensively discussed in \citet{hac06kb, hac07k} and
\citet{hac08kc}, $V$ magnitudes are often heavily contaminated
by strong emission lines such as [\ion{O}{3}] $\lambda\lambda$
4959, 5007 \citep[see][for V598 Pup spectra]{rea08},
which eventually dominate over the continuum, causing
an increasing deviation from the free-free emission model.

Comparing the apparent magnitude of V1500 Cyg and the absolute 
magnitude of the model light curve of $1.05 ~M_\sun$ 
(Table \ref{light_curves_of_novae_co}), we can estimate 
the distance modulus.  Figure \ref{light_curve_v598_pup_v1500_cyg}
shows that the $1.05 ~M_\sun$ WD model is a best-reproducing model 
(magenta dash-dotted line) among
other WD mass models in Table \ref{light_curves_of_novae_co} for
chemical composition of ``CO nova 2.''  Using $M_{\rm w}= 1.0$
in Table \ref{light_curves_of_novae_co}
and $m_{\rm w}= 13.0 + 0.5 = 13.5$ from Figure
\ref{light_curve_v598_pup_v1500_cyg},
we obtain the distance modulus to V1500 Cyg, i.e.,
$(m-M)_V = m_{\rm w} - M_{\rm w}= 13.5 - 1.0 = 12.5$.
If we adopt another chemical composition of ``Ne nova 2,''
we obtain a best-reproducing model of $1.15 ~M_\sun$ WD (black solid line).
Then we obtain the distance modulus of
$(m-M)_V = m_{\rm w} - M_{\rm w}= 13.6 - 0.9 = 12.7$
with $m_{\rm w}= 13.1 + 0.5 = 13.6$ 
(Figure \ref{light_curve_v598_pup_v1500_cyg})
and $M_{\rm w}= 0.9$ (Table \ref{light_curves_of_novae_ne}).
The difference between CO and neon nova models is only 0.2 mag.
Therefore, this method for obtaining the distance modulus is 
rather robust against the ambiguity of chemical composition
of a nova envelope.

The distance to V1500 Cyg was estimated to be $d = 1.5 \pm 0.2$ kpc
from the nebular (ejecta) expansion parallax \citep[e.g.,][]{sla95}.
The absorption was obtained to be $A_V = 1.60 \pm 0.16$
by \citet{lan88}, so that the distance modulus to V1500 Cyg is
$(m-M)_V = 12.5 \pm 0.4$, being consistent with the value of 12.5 
(``CO nova 2'') and 12.7 (``Ne nova 2'')
estimated above based on our new method.

Thus, the absolute magnitude estimates of V1500 Cyg based on
our free-free emission model light curves are consistent
with the distance derived from the nebular expansion
parallax method \citep{sla95} within an accuracy of 0.2 mag in the distance
modulus of $(m-M)_V$.
This in turn supports our assumptions and
simplifications on free-free emission of novae.

\begin{figure}
%\epsscale{1.0}
\epsscale{1.15}
\plotone{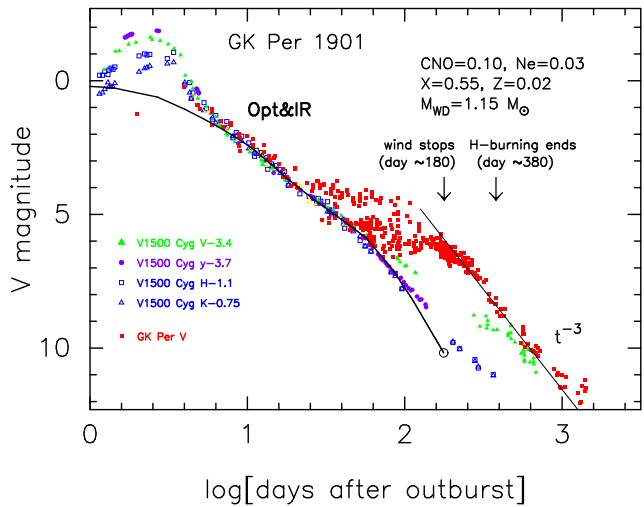}
%\plotone{f23bw.epsi}
%\plotone{all_mass_gk_per_v1500_cyg_x55z02o10ne03_color.epsi}
%\plotfiddle{evolution1.ps}{5.0cm}{270}{0.4}{0.4}{-170}{220}
\caption{
Same as Figure \ref{light_curve_v598_pup_v1500_cyg}, but 
we add the $V$ light curve of GK Per (Nova Persei 1901) for comparison.
We fit the smooth light curve of V1500 Cyg with the bottom line
of the GK Per light curve during the transition oscillations.
Optical $V$ data of GK Per are the same as those in Figure 2
of \citet{hac07k}.  
}
\label{all_mass_gk_per_v1500_cyg_x55z02o10ne03}
\end{figure}

The second example is GK Per (Nova Persei 1901).  We assume
that, when two novae have a similar decline rate (or a similar
timescale) of free-free
light curve, their brightnesses are the same
during the period in which the two free-free light curves
are overlapped.   Figure \ref{all_mass_gk_per_v1500_cyg_x55z02o10ne03}
shows such an example, a comparison between the light curves
of GK Per and V1500 Cyg.  GK Per has a decline timescale very 
similar to that of V1500 Cyg.  Even in a typical transition phase
(from day $\sim 25$ until day $\sim 150$), 
we are able to nicely fit our universal decline law with
the observation, especially the bottom line of
each oscillation until day $\sim 100$ as shown in the figure.
The top line connecting maxima of each oscillation deviates
largely from our free-free model light curve.  This deviation
is caused mainly by the increase in continuum flux itself.
The effect of strong emission lines starts from day $\sim 100$,
where the bottom line of each oscillation starts to deviate
from our model light curve.  

Our best-reproducing model is a $1.15 ~M_\sun$ WD among other WD masses
in Table \ref{light_curves_of_novae_ne}.  Then we obtain the
distance modulus of
\begin{equation}
\left[ (m-M)_V \right]_{\rm FF} 
= m_{\rm w} - M_{\rm w}= 10.2 - 0.9 = 9.3
\end{equation}
from $m_{\rm w}= 10.2$ in
Figure \ref{all_mass_gk_per_v1500_cyg_x55z02o10ne03}
and $M_{\rm w}= 0.9$ in Table \ref{light_curves_of_novae_ne},
where FF means the method of our free-free (FF) light curve fitting.

     Direct comparison of the brightness between GK Per and
V1500 Cyg gives the difference of their distance moduli.
From Figure \ref{all_mass_gk_per_v1500_cyg_x55z02o10ne03}, we have
\begin{equation}
\left[5 \log(d/10) + A_V \right]_{\rm V1500~Cyg}
 - \left[5 \log(d/10) + A_V \right]_{\rm GK~Per}  = +3.4,
\label{distance_modulus_v1500cyg_gkper}
\end{equation}
where $d$ is the distance in units of pc and $A_V$ is 
the absorption in the $V$ band.
With $d = 1.5$ kpc and $A_V = 1.6$ for V1500 Cyg
and $d = 0.455$ kpc \citep{sla95}
and $A_V = 3.1 E(B-V)= 3.1 \times 0.3 = 0.9$  \citep{wu89}
for GK Per,
we obtain
\begin{equation}
(m-M)_V =\left[5 \log(d/10) + A_V \right]_{\rm V1500~Cyg}= 12.5,
\label{distance_modulus_v1500_cyg}
\end{equation}
and
\begin{equation}
(m-M)_V = \left[5 \log(d/10) + A_V \right]_{\rm GK~Per}  = 9.2.
\label{distance_modulus_gk_per}
\end{equation}
Therefore, the difference of $12.5 - 9.2 = 3.3$ is consistent with
the difference in Equation (\ref{distance_modulus_v1500cyg_gkper})
within an accuracy of 0.1 mag and the distance modulus of
Equation (\ref{distance_modulus_gk_per}) is also consistent with
the result of the above FF method (9.3 mag)
within 0.1 mag.  Thus, we confirm that the brightnesses of
the two novae are the same within an uncertainty of $\pm 0.2$ mag.
This good agreement supports our assumption 
that two novae have the same brightness during their
overlapping period if the two novae have a similar timescale as shown
in Figure \ref{mass_v_uv_x_v1974_cyg_x55z02o10ne03_real_scale_model}.

% Fig.10
%\placefigure{all_mass_gk_per_v1500_cyg_x55z02o10ne03}

\begin{figure}
%\epsscale{1.0}
\epsscale{1.15}
\plotone{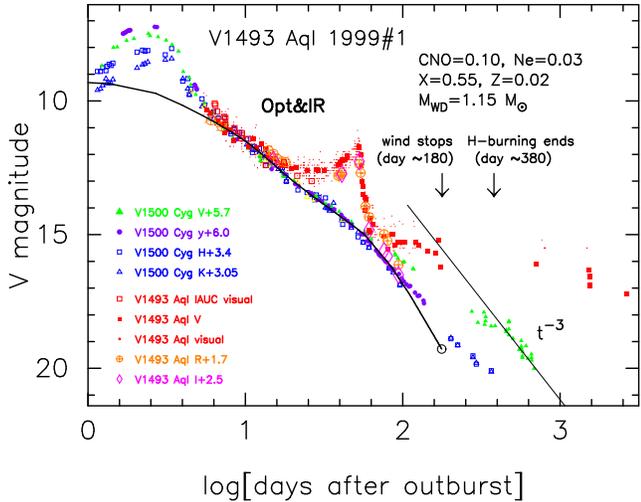}
%\plotone{f23bw.epsi}
%\plotone{all_mass_v1493_aql_v1500_cyg_x55z02o10ne03_color.epsi}
%\plotfiddle{evolution1.ps}{5.0cm}{270}{0.4}{0.4}{-170}{220}
\caption{
Same as Figure \ref{light_curve_v598_pup_v1500_cyg}, but 
we add light curves of V1493 Aql 1999\#1 for comparison.
We fit the smooth light curve of V1500 Cyg with the bottom line
of the V1493 Aql light curve.
Optical $V$ ({\it filled squares}), $R$ ({\it open circles with
a plus}), $I$ ({\it open diamonds}), and visual ({\it very small
open circles}) are taken from the American Association
of Variable Star Observers (AAVSO) and the
Variable Star Observers League in Japan (VSOLJ).
Some visual magnitude estimates are also taken from
IAU Circulars 7225, 7228, 7232, 7258, 7273, and 7313
({\it open squares}).
Here we assume the outburst day (day 0) of
$t_{\rm OB}=$ JD 2451369.0 (1999 July 9.5 UT) for V1493 Aql.
}
\label{all_mass_v1493_aql_v1500_cyg_x55z02o10ne03}
\end{figure}

     The third example is V1493 Aql (Nova Aquilae 1999 No.1).
This nova shows a peculiar secondary peak 
(Figure \ref{all_mass_v1493_aql_v1500_cyg_x55z02o10ne03}),
which is seen in all the bands, i.e., $V$, $R$,
and $I$ bands.  \citet{hac09k} explained this kind of
secondary peak by an additional energy source
from strong magnetic field (and rotation).  Therefore, we
regard that the nova light curves come back to a
free-free emission light curve after the additional energy
input stopped.  We see that,
just after the secondary peak, the light curve sharply drops to merge 
into the light curve of V1500 Cyg.  So we expect that 
V1493 Aql has a similar decline rate to V1500 Cyg, i.e.,
our free-free model light curve of $M_{\rm WD}= 1.15 ~M_\sun$ 
with ``Ne nova 2'' agrees well with the observation.
The effect of strong emission lines starts only from day $\sim 100$,
where the $V$ and visual magnitudes begin to depart from
the $I$ magnitude.

Using a $1.15 ~M_\sun$ WD model in Table \ref{light_curves_of_novae_ne},
we obtain the distance modulus of
\begin{equation}
\left[ (m-M)_V \right]_{\rm FF} 
= m_{\rm w} - M_{\rm w}= 19.3 - 0.9 = 18.4
\end{equation}
from $m_{\rm w}= 19.3$ in
Figure \ref{all_mass_v1493_aql_v1500_cyg_x55z02o10ne03}
and $M_{\rm w}= 0.9$ in Table \ref{light_curves_of_novae_ne}.

     Direct comparison of the brightness between V1493 Aql and
V1500 Cyg gives the difference of their distance moduli as shown in
Figure \ref{all_mass_v1493_aql_v1500_cyg_x55z02o10ne03}, i.e.,
\begin{equation}
\left[5 \log(d/10) + A_V \right]_{\rm V1500~Cyg}
 - \left[5 \log(d/10) + A_V \right]_{\rm V1493~Aql}  = -5.7.
\end{equation}
With $(m-M)_V = 12.5$ from Equation (\ref{distance_modulus_v1500_cyg})
for V1500 Cyg, we obtain the distance modulus of
\begin{equation}
\left[ (m-M)_V \right]_{\rm LC}  = 
\left[ 5 \log(d/10) + A_V \right]_{\rm V1493 Aql} = 18.2,
\label{distance_absorption_v1493aql_V}
\end{equation}
where LC means the method of our direct light curve (LC) fitting.
The distance modulus of this nova was estimated to be
$(m-M)_V = 17.8$ \citep{bon00}, $(m-M)_V = 18.8$ \citep{ven04}, 
$(m-M)_V = 18.4$ \citep{ark02}, and $(m-M)_V = 18.6$ \citep{bur08},
based on the MMRD relations.  The simple average values of these
four observational estimates becomes $(m-M)_V = 18.4 \pm 0.4$,
being consistent with our estimate of the FF and LC methods.
If we adopt $A_V = 3.1 E(B-V)= 3.1 \times 1.5 = 4.7$ \citep{ark02},
the distance is $d = 5.5$ and $5.0$ kpc for FF and LC method,
respectively.  These values are consistent with $d= 5.4$ kpc
obtained by \citet{ark02} based on the MMRD relations.

In the present paper, 
we proposed two new methods for obtaining distance modulus
of a nova: one is fitting our free-free (FF) emission model
light curve with the observation (we call this FF method).
The other is fitting the light curve (LC) of a distance-unknown
nova with the other distance-well-known nova (we call this LC method).  
From the above confirmations, we may conclude that our new methods 
give a reasonable distance.  It should be addressed that
our new methods for obtaining distance moduli of novae are
applicable even when we missed the optical maxima of
nova light curves.  In Section \ref{light_curve_of_ten_novae},
we apply these two methods to ten novae.

% Fig.11
%\placefigure{all_mass_v1493_aql_v1500_cyg_x55z02o10ne03}

\section{Characteristic Timescales of Nova Light Curves}
\label{various_timescales}

%     Figures \ref{t2t3_t_break}--\ref{t2t3_t_h_burning}
There are several timescales that characterize nova light curves;
$t_0$ is the time in units of day at optical maximum
(usually $V$ magnitude); $t_2$ and $t_3$ are the times in units of day,
in which $V$ magnitude decays by 2 and 3 mag, respectively,
from its maximum (from $t_0$);
$t_{\rm break}$ is the epoch when our free-free model light curves
(or optical $y$, near infrared $I$, $J$, $H$, $K$ light curves)
bend sharply as shown in Figures 
\ref{v1668_cyg_vy_jhk_mag_x35z02c10o20_model_scaling_law_logt},
\ref{mass_v_uv_x_v1974_cyg_x55z02o10ne03_new_model},
and \ref{light_curve_v598_pup_v1500_cyg}; 
$t_{\rm wind}$ is the epoch when optically thick winds stop; 
and $t_{\rm H-burning}$ is the epoch when hydrogen shell-burning
ends.  \citet{hac06kb, hac07k} and \citet{hac08kc}
calculated free-free model light curves for various WD masses with
many sets of chemical composition.
They fitted their model light curves with observation
and derived $t_{\rm break}$, $t_{\rm wind}$, and $t_{\rm H-burning}$
for many novae.  Table \ref{light_curve_parameters_novae} summarizes
such values of $t_{\rm break}$, $t_{\rm wind}$, and $t_{\rm H-burning}$ 
for 15 classical novae, taken from our published papers as well as
$t_2$ and $t_3$ for the same 15 novae, which are taken from literature.
Figures \ref{t2t3_t_break}--\ref{t2t3_t_h_burning} show
these five timescales for the 15 novae.

  When the optically thick wind stops, the photospheric radius
shrinks drastically and the photospheric temperature increases
up to $20-30$ eV \citep[e.g.,][]{hac03kb} so that the supersoft X-ray 
flux emerges (see Appendix B for the photospheric radius which is
shrinking with the temperature of the WD envelope during
a supersoft X-ray phase).  We regard that $t_{\rm wind}$ is almost
equal to the emerging time of supersoft X-ray, $t_{\rm X-on}$.
After the hydrogen shell-burning stops, the WD envelope quickly cools
down and the supersoft X-ray phase ends. So we regard that
$t_{\rm H-burning}$ is almost equal to the decay time of
supersoft X-ray, $t_{\rm X-off}$.
Among these five timescales of $t_2$, $t_3$, 
$t_{\rm break}$, $t_{\rm wind}$, and $t_{\rm H-burning}$, 
the first three timescales, $t_2$, $t_3$, and $t_{\rm break}$,
are much shorter than the other two, $t_{\rm wind}$
and $t_{\rm H-burning}$.

% Table 4
%\placetable{light_curve_parameters_novae}

\begin{deluxetable*}{llrrrrrr}
\tabletypesize{\scriptsize}
\tablecaption{Light Curve Parameters of Novae in Literature
\label{light_curve_parameters_novae}}
\tablewidth{0pt}
\tablehead{
\colhead{object} &
\colhead{$t_2$} &
\colhead{$t_3$} &
\colhead{ref.\tablenotemark{a}} &
\colhead{$t_{\rm break}$} &
\colhead{$t_{\rm wind}$}  &
\colhead{$t_{\rm H-burning}$} &
\colhead{ref.\tablenotemark{b}} \\
\colhead{} &
\colhead{(days)} &
\colhead{(days)} &
\colhead{} &
\colhead{(days)} &
\colhead{(days)} &
\colhead{(days)} &
\colhead{} 
}
\startdata
GK Per 1901 & 7 & 13 & 1 & 73 & 182 & 382 & 5 \\
V1500 Cyg 1975 & 2.9 & 3.6 & 2 & 70 & 180 & 380 & 8  \\
V1668 Cyg 1978 & 12.2 & 24.3 & 3 & 110 & 280 & 720 & 8 \\
V1974 Cyg 1992 & 16 & 42 & 1 & 96 & 250 & 600 & 8 \\
V382 Vel 1999 & 6 & 12 & 4 & 49 & 120 & 220 & 9 \\
V2361 Cyg 2005 & 6 & 8 & 5 & 69 & 169 & 340 & 5\\
V382 Nor 2005 & 12 & 18 & 5 & 73 & 182 & 382 & 5\\
V5115 Sgr 2005 & 7 & 14 & 5 & 60 & 145 & 280 & 5\\
V378 Ser 2005 & 44 & 90 & 5 & 257 & 858 & 2560 & 5\\
V5116 Sgr 2005\#2 & 20 & 33 & 5 & 114 & 319 & 757 & 5 \\
V1188 Sco 2005 & 7 & 21 & 5 & 44 & 110 & 190 & 5 \\
V1047 Cen 2005 & 6 & 26 & 5 & 257 & 858 & 2560 & 5 \\
V476 Sct 2005 & 15 & 28 & 6 & 100 & 260 & 590 & 5\\
V1663 Aql 2005 & 13 & 26 & 5 & 100 & 260 & 590 & 5 \\
V477 Sct 2005\#2 & 3 & 6 & 7 & 32 & 80 & 121 & 5
\enddata
\tablenotetext{a}{$t_2$ and $t_3$ times at observational face value
taken from literature of
1-\citet{dow00}, 2-\citet{war95}, 
3-\citet{mal79}, 4-\citet{van01}, 5-\citet{hac07k}, 6-\citet{mun06b},
7-\citet{mun06a}}
\tablenotetext{b}{$t_{\rm break}$, $t_{\rm wind}$, and
$t_{\rm H-burning}$ taken from 5-\citet{hac07k}, 8-\citet{hac06kb}, 
9-present work}
\end{deluxetable*}

\subsection{Relations between various timescales}
\label{relation_among_five_timescales}
\begin{figure}
%\epsscale{0.6}
%\epsscale{1.0}
\epsscale{1.15}
\plotone{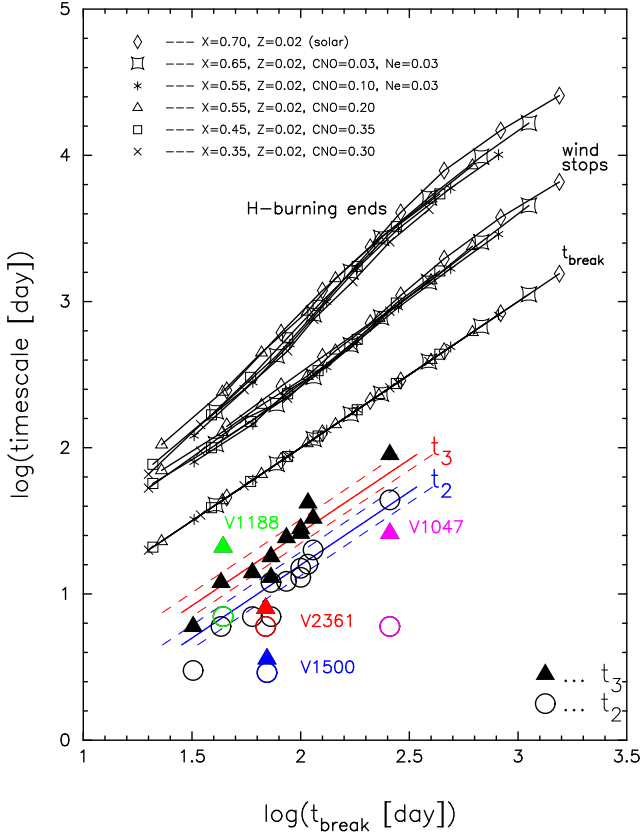}
%\plotone{f4bw.epsi}
%\plotone{t2t3_t_break_color.epsi}
%\plotfiddle{evolution1.ps}{5.0cm}{270}{0.4}{0.4}{-170}{220}
\caption{
Five characteristic nova timescales  against $t_{\rm break}$.
From top to bottom, hydrogen shell-burning
($t_{\rm H-burning}$), optically thick wind ($t_{\rm wind}$),
break point of nova light curve ($t_{\rm break}$), 
$t_3$, and $t_2$.
Supersoft X-ray on/off correspond to $t_{\rm wind}$ and
$t_{\rm H-burning}$, respectively.
Each symbol in the upper left corner represents our various WD
models with different chemical compositions 
taken from \citet{hac06kb}.
We also plot observational face values of $t_3$ ({\it filled triangles})
and $t_2$ ({\it open circles}) for 15 novae in Table
\ref{light_curve_parameters_novae}. 
Thick red solid line: central values of Equation (\ref{t_3-t_break}). 
Upper/lower red dashed lines: upper/lower limit values
of Equation (\ref{t_3-t_break}). 
Thick blue solid line: central values of $t_2$-$t_{\rm break}$ 
relation.  Upper/lower blue dashed lines: upper/lower values
of $t_2$-$t_{\rm break}$ relation. 
Colored marks indicate $t_2$ and $t_3$ times that deviate largely
from the statistical relations.  {\it Green}: visual
magnitudes of V1188 Sco are so contaminated by emission lines that
$t_3$ is largely deviated from the central line.
{\it Red}: visual and $V$ light curves of V2361 Cyg are
affected by dust formation.
{\it Blue}: V1500 Cyg is a superbright nova and its face
values of $t_2$ and $t_3$ times do not follow the universal 
decline law.
{\it Magenta}: visual and $V$ light curves of V1047 Cen is also
a superbright nova.
}
\label{t2t3_t_break}
\end{figure}

\begin{figure}
%\epsscale{0.6}
%\epsscale{1.0}
\epsscale{1.15}
\plotone{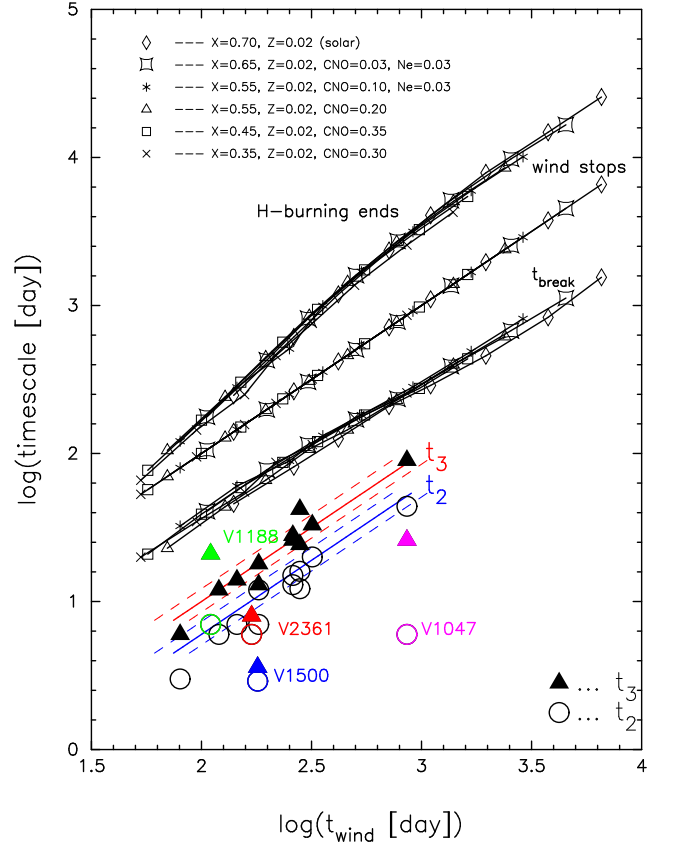}
%\plotone{f4bw.epsi}
%\plotone{t2t3_t_wind_color.epsi}
%\plotfiddle{evolution1.ps}{5.0cm}{270}{0.4}{0.4}{-170}{220}
\caption{
Same as Figure \ref{t2t3_t_break}, but against $t_{\rm wind}$.
Thick red solid line: central values of Equation (\ref{t_3-t_wind}). 
Upper/lower red dashed lines indicate upper/lower limit values
of Equation (\ref{t_3-t_wind}).  
Thick blue solid line: central values of $t_2$-$t_{\rm wind}$ 
relation.  Upper/lower blue dashed lines: upper/lower values
of $t_2$-$t_{\rm wind}$ relation. 
See text for more details. 
}
\label{t2t3_t_wind}
\end{figure}

\begin{figure}
%\epsscale{0.6}
%\epsscale{1.0}
\epsscale{1.15}
\plotone{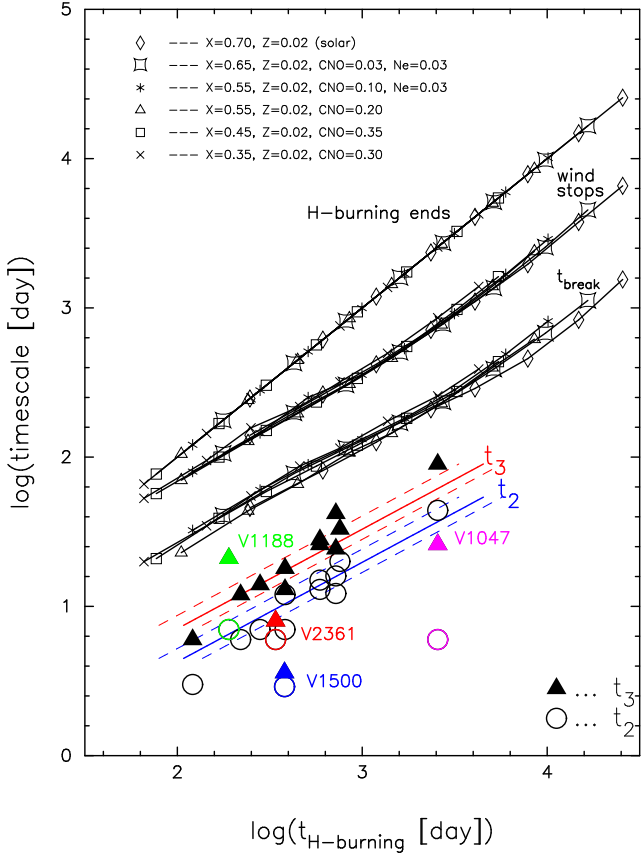}
%\plotone{f5bw.epsi}
%\plotone{t2t3_t_h_burning_color.epsi}
%\plotfiddle{evolution1.ps}{5.0cm}{270}{0.4}{0.4}{-170}{220}
\caption{
Same as Figure \ref{t2t3_t_break}, but against $t_{\rm H-burning}$.   
Thick red solid line: central values of Equation (\ref{t_3-t_h_burning}). 
Upper/lower red dashed lines indicate upper/lower limit values
of Equation (\ref{t_3-t_h_burning}).
Thick blue solid line: central values of $t_2$-$t_{\rm H-burning}$ 
relation.  Upper/lower blue dashed lines: upper/lower values
of $t_2$-$t_{\rm H-burning}$ relation. 
See text for more details. 
}
\label{t2t3_t_h_burning}
\end{figure}

     In what follows we introduce several relations between
these five timescales, $t_2$, $t_3$, $t_{\rm break}$, $t_{\rm wind}$,
and $t_{\rm H-burning}$.
     In Figure \ref{t2t3_t_break}, we plot $t_{\rm H-burning}$,
$t_{\rm wind}$, and $t_{\rm break}$ against $t_{\rm break}$, which 
are theoretical ones calculated from our free-free model light curves
\citep{hac06kb, hac07k}.
These values depend on $M_{\rm WD}$ and the chemical composition, but are
actually independent of $M_{\rm env,0}$.
This is because the difference in $M_{\rm env,0}$ makes
a large difference in the optical maximum but affects only the very
early phase of nova outbursts as shown in Figures 
\ref{v1668_cyg_vy_jhk_mag_x35z02c10o20_model_scaling_law} and 
\ref{v1668_cyg_vy_jhk_mag_x35z02c10o20} and the difference in
the epoch of optical maximum is so short (a few to several days)
that it can be neglected compared with the timescales of
$t_{\rm H-burning}$, $t_{\rm wind}$, and $t_{\rm break}$
(several tens of days or a few hundred days).  So we can derive
relations among $t_{\rm H-burning}$, $t_{\rm wind}$,
and $t_{\rm break}$ from our models (independently of observations)
as shown in Figures \ref{t2t3_t_break}--\ref{t2t3_t_h_burning}.
Once we determine $t_{\rm break}$ of individual novae 
from observations, we can theoretically predict the duration of
a luminous supersoft X-ray phase from Figure \ref{t2t3_t_break}.

From Figure \ref{t2t3_t_break}, we obtain approximate relations
between $t_{\rm break}$, turn-on ($t_{\rm X-on} \approx t_{\rm wind}$),
and turnoff ($t_{\rm X-off} \approx t_{\rm H-burning}$)
times as
\begin{equation}
t_{\rm X-on} \approx t_{\rm wind}
= \left( 2.6 \pm 0.4 \right) t_{\rm break},
\label{t_break-t_wind}
\end{equation}
\begin{equation}
t_{\rm X-off} \approx t_{\rm H-burning}
  = \left( 0.71 \pm 0.1 \right) (t_{\rm break})^{1.5},
\label{t_break-t_h_burning}
\end{equation}
and
\begin{equation}
t_{\rm X-off} \approx t_{\rm H-burning}
  = \left( 0.17 \pm 0.023 \right) (t_{\rm wind})^{1.5},
\label{t_wind-t_h_burning}
\end{equation}
for $25 \lesssim t_{\rm break} \lesssim 250$ days.

It should be noted that the proportionality of $t_{\rm wind}$ 
to $t_{\rm break}$, i.e., $t_{\rm wind}/t_{\rm break} \approx 2.6$,
is a direct result of our scaling law (the universal decline law),
because both $t_{\rm break}$ and $t_{\rm wind}$ are during 
the optically thick wind phase which is the subject of
universal decline law.  Then the 
$t_{\rm wind}$ is directly proportional to $t_{\rm break}$.
On the other hand, the turnoff time of hydrogen shell burning,
$t_{\rm H-burning}$, is not proportional to $t_{\rm break}$ or
$t_{\rm wind}$, because the duration of static hydrogen shell
burning phase depends on hydrogen content $X$ (amount of
fuel) of the envelope and WD mass (temperature)
in a different way from the optically thick wind phase,
in which the decay rate is determined mainly by wind mass-loss
not by hydrogen burning rate
(see Figure \ref{mass_v_uv_x_v1974_cyg_x55z02o10ne03_new_model}).
Therefore, Equations (\ref{t_break-t_h_burning}) 
and (\ref{t_wind-t_h_burning}) are empirical laws obtained
numerically, the power of which is not 1.0 but approximately 1.5.

It is, however, not easy to obtain $t_{\rm break}$
because nova light curves such as $V$ magnitudes are often
heavily contaminated by strong emission lines in the later phase
and this contamination easily clouds the break point of light curves
as can be seen in Figure \ref{light_curve_v598_pup_v1500_cyg}.
In the next subsection, we derive relations between $t_3$ time
and two of $t_{\rm wind}$ and $t_{\rm H-burning}$ times, 
because $t_3$ time is determined relatively easily 
from early optical observation.

% Fig.12
%\placefigure{t2t3_t_break}

% Fig.13
%\placefigure{t2t3_t_wind}

% Fig.14
%\placefigure{t2t3_t_h_burning}

% Table 5
%\placetable{t2_t3_txonoff}

\begin{deluxetable*}{llllllll}
\tabletypesize{\scriptsize}
\tablecaption{ Prediction Formulae vs. Nova Decay Timescales
in Literature
\label{t2_t3_txonoff}}
\tablewidth{0pt}
\tablehead{
\colhead{object} &
\colhead{...} &
\colhead{$t_2$} &
\colhead{$t_3$} &
\colhead{$t_2-t_3$} &
\colhead{$t_3-t_{\rm Xon}$} &
\colhead{$t_3-t_{\rm Xoff}$} &
\colhead{reference\tablenotemark{a}} \\
\colhead{} &
\colhead{...} &
\colhead{(day)} &
\colhead{(day)} &
\colhead{(eq.[\ref{t_2-t_3}])} &
\colhead{(eq.[\ref{t_3-t_wind}])} &
\colhead{(eq.[\ref{t_3-t_h_burning}])} &
\colhead{} 
}
\startdata
V598 Pup 2007\#2 & ... & -- & -- & -- & -- & -- & -- \\
V382 Vel 1999 & ... & 7.5 & 15  & $<$yes$>$\tablenotemark{b}
 & -- & no & \citet{bur08} \\
V382 Vel 1999 & ... & 4.5 & 9  & $<$yes$>$ & -- & $<$yes$>$
 & \citet{del02} \\
V382 Vel 1999 & ... & 6 & 12  & yes & -- & yes & \citet{van01} \\
V382 Vel 1999 & ... & -- & 12.5  & -- & -- & yes & \citet{lil00} \\
V382 Vel 1999 & ... & 6 & 10  & yes & -- & yes & \citet{del99} \\
V4743 Sgr 2002\#3 & ... & 12 & 22  & yes & -- & $<$yes$>$
 & \citet{bur08} \\
V4743 Sgr 2002\#3 & ... & -- & 15  & -- & -- & $<$yes$>$
 & \citet{nie03} \\
V4743 Sgr 2002\#3 & ... & 9 & 16  & yes & -- & yes & \citet{mor03} \\
V1281 Sco 2007\#2 & ... & -- & -- & -- & -- & -- & -- \\
V597 Pup 2007\#1 & ... & 2.5 & -- & -- & -- & -- & \citet{nai09} \\
V1494 Aql 1999\#2 & ... & 8 & 16  & $<$yes$>$ & no & no & \citet{bur08} \\
V1494 Aql 1999\#2 & ... & 7 & 23  & no & yes & $<$yes$>$ & \citet{ven04} \\
V1494 Aql 1999\#2 & ... & 6.1 & 15.8  & no & no & no & \citet{bar03} \\
V1494 Aql 1999\#2 & ... & 6 & 15  & no & no & no & \citet{ark02} \\
V1494 Aql 1999\#2 & ... & 6.6 & 16  & no & no & no & \citet{kis00} \\
V2467 Cyg 2007 & ... & 7.3 & 15.1 & $<$yes$>$
 & -- & no & \citet{lyn09} \\
V2467 Cyg 2007 & ... & 7.6 & 14.6 & yes & -- & no & \citet{pog09} \\
V2467 Cyg 2007 & ... & 8 & -- & -- & -- & -- & \citet{sen08} \\
V5116 Sgr 2005\#2 & ... & 6.5 & 20.2 & no & -- & no & \citet{dob08} \\
V5116 Sgr 2005\#2 & ... & 18 & 45  & no & -- & no & \citet{bur08} \\
V5116 Sgr 2005\#2 & ... & 20 & 33  & yes & -- & yes & \citet{hac07k} \\
V574 Pup 2004 & ... & 37 & 85  & no & no & no & \citet{bur08} \\
V574 Pup 2004 & ... & 13 & 58  & no & no & no & \citet{siv05} \\
V458 Vul 2007 & ... & 8 & 31 & no & no & -- & \citet{wes08} \\
V458 Vul 2007 & ... & 7 & 15 & $<$yes$>$ & no & -- & \citet{pog08} \\
V458 Vul 2007 & ... & 7 & 18 & no & no & -- & \citet{tar08} \\
\\
V1974 Cyg 1992 & ... & 22 & 48  & no & no & no & \citet{bur08} \\
V1974 Cyg 1992 & ... & 16 & 42  & no & no & no & \citet{war95} \\
%%%GQ Mus 1983 & ... & 18 & 48  & no & -- & no & \citet{whi84} \\
V1668 Cyg 1978 & ... & 12.2 & 24.3  & yes & -- & -- & \citet{mal79} \\
V1500 Cyg 1975 & ... & 2.9 & 3.6  & -- & -- & -- & \citet{war95}
\enddata
\tablenotetext{a}{reference for quoted values of $t_2$ and $t_3$
at observational face value}
\tablenotetext{b}{``yes'' satisfies eqs.(\ref{t_2-t_3}),
(\ref{t_3-t_wind}), and (\ref{t_3-t_h_burning}) but ``no'' does not
and parenthesis means slightly outside the border}
\end{deluxetable*}

\subsection{Prediction formulae by intrinsic $t_3$ time}
\label{prediction_by_t3}
It is very useful if we can predict $t_{\rm X-on}$
($\approx t_{\rm wind}$) and $t_{\rm X-off}$
($\approx t_{\rm H-burning}$) from $t_3$ time, 
because $t_3$ is popular and widely used to define the nova
speed class.  In this subsection, we derive relations between
$t_3$ and $t_{\rm X-on}$, and between $t_3$ and $t_{\rm X-off}$.

As already explained in the previous subsection,
$t_{\rm break}$, $t_{\rm wind}$, and $t_{\rm H-burning}$
are practically independent of the initial envelope mass,
$M_{\rm env,0}$, but $t_3$ time depends on the peak
brightness of an individual nova.  As shown in Figures 
\ref{v1668_cyg_vy_jhk_mag_x35z02c10o20_model_scaling_law}
and \ref{v1668_cyg_vy_jhk_mag_x35z02c10o20},
the initial envelope mass $M_{\rm env,0}$ is larger,
the optical maximum is brighter.
Therefore, $t_2$ and $t_3$ depend on $M_{\rm env,0}$.
This is the reason why we cannot theoretically derive $t_2$
and $t_3$.  We need information of maximum brightness
to specify $t_3$ time for individual novae.

Now we go back to the well-observed V1668 Cyg which we regard as
a typical classical nova.  Thus, we adopt point B in Figures
\ref{v1668_cyg_vy_jhk_mag_x35z02c10o20_model_scaling_law} and 
\ref{v1668_cyg_vy_jhk_mag_x35z02c10o20} as a typical (an average)
case of maximum brightness.
In this case, $t_2 = 14.4$ and $t_3 = 26$ days are derived from
Figures \ref{v1668_cyg_vy_jhk_mag_x35z02c10o20_model_scaling_law} and 
\ref{v1668_cyg_vy_jhk_mag_x35z02c10o20}.
Then, we have $t_3/t_{\rm break} = 26/100$ (see Table 
\ref{fitting_t2_t3_tb_txonoff}
for V1668 Cyg timescales).  Using the universal
decline law,  we obtain
a relation between $t_3$ and $t_{\rm break}$ as
\begin{equation}
{{t_3} \over {t_{\rm break}}} = {{t_3 / f_{\rm s}}
\over {t_{\rm break} / f_{\rm s}}} = {{26 \pm 4} \over {100}}
= 0.26 \pm 0.04,
\label{t_3-t_break}
\end{equation}
for $8 \lesssim t_3 \lesssim 80$ days,
where the upper and lower bounds are corresponding to $\pm 0.5$ mag
brighter/darker (more massive/less massive initial envelope mass)
points than that of point B.  This relation is indicated by a red
thick solid line in Figure \ref{t2t3_t_break} together with
the upper and lower red dashed lines that denote upper/lower values
($\pm 0.04$) of Equation (\ref{t_3-t_break}).

There is a relation between $t_2$ and $t_3$ derived from the scaling
law of our universal decline law, which is also well used in
literature.   \citet{hac07k} have already
derived
\begin{equation}
t_2  = \left( 0.6 \pm 0.08 \right) t_3,
\label{t_2-t_3}
\end{equation}
for the optical light curves that follow the free-free model
light curve with $8 \lesssim t_3 \lesssim 80$ days.
This relation has been empirically known \citep{cap90}
and theoretically confirmed by \citet{hac06kb} 
based on the universal decline law with a slope of
$F_\lambda \propto t^{-1.75}$ as shown in Figures
\ref{v1668_cyg_vy_jhk_mag_x35z02c10o20_model_scaling_law_logt} and
\ref{mass_v_uv_x_v1974_cyg_x55z02o10ne03_new_model}.
Inserting the central value of Equation (\ref{t_2-t_3}) into
Equation (\ref{t_3-t_break}), we obtain relations between 
$t_{\rm break}$ and $t_2$ relations  by blue solid and
dashed lines as shown in Figure \ref{t2t3_t_break}.

Figure \ref{t2t3_t_break} also shows $t_2$ and $t_3$ times
for individual novae in Table \ref{light_curve_parameters_novae}.
Four novae indicated by colors are far from the relation of 
Equation (\ref{t_3-t_break}),
but there is a special
reason in each nova as described in the figure caption.
The other novae reasonably follow the above relation with the
reasonable scatter corresponding to $\pm 0.5$ mag brighter/darker
of the peak brightness.  This supports that V1668 Cyg has
an average $M_{\rm env,0}$.

     In the same way, we derive a relation between $t_3$ and
$t_{\rm wind}$ from Equations (\ref{t_break-t_wind}) and
(\ref{t_3-t_break}), i.e.,
\begin{equation}
%%%{{t_3} \over {t_{\rm wind}}} = {{t_3 ~ f_{\rm s}}
%%%\over {t_{\rm wind} ~ f_{\rm s}}} = {{26 \pm 4} \over {250}}
%%%= 0.10 \pm 0.013,
t_{\rm X-on} \approx t_{\rm wind}
= \left( 10.0 \pm 1.8 \right) t_3,
\label{t_3-t_wind}
\end{equation}
for $8 \lesssim t_3 \lesssim 80$ days.
Figure \ref{t2t3_t_wind} shows the five timescales
against $t_{\rm wind}$ and we also see a reasonable fit
of our relation (Equation (\ref{t_3-t_wind})) with the observation.

     From Equations (\ref{t_break-t_h_burning}) and
(\ref{t_3-t_break}), we derive a relation 
between $t_3$ and $t_{\rm H-burning}$, i.e.,
\begin{equation}
t_{\rm X-off} \approx t_{\rm H-burning}
  = \left( 5.3 \pm 1.4 \right) (t_3)^{1.5},
\label{t_3-t_h_burning}
\end{equation}
for $8 \lesssim t_3 \lesssim 80$ days.
Figure \ref{t2t3_t_h_burning} shows the five timescales
against $t_{\rm H-burning}$ and we can also see a reasonable fit
of Equation (\ref{t_3-t_h_burning}) with the observation.

However, it is not always easy to determine $t_3$ (or $t_2$) time
from observation.  In fact, Table \ref{t2_t3_txonoff} shows
nova timescales $t_2$ and $t_3$ of 11 novae taken from literature,
which are sometimes very different from author to author, and
sometimes very different from our estimates in Table
\ref{fitting_t2_t3_tb_txonoff}.  Therefore, we propose a 
more robust way.  Instead of observational face values
of $t_2$ and $t_3$, we will use ``intrinsic''  $t_2$ and $t_3$,
which are obtained from our model light curves fitted with observation.
For example, we use $t_2 = 14.4$ and $t_3 = 26$ days for 
V1668 Cyg obtained from Figures
\ref{v1668_cyg_vy_jhk_mag_x35z02c10o20_model_scaling_law} and 
\ref{v1668_cyg_vy_jhk_mag_x35z02c10o20} instead of observational
face values.  This ``intrinsic'' $t_2$ and $t_3$ times are
especially useful when the observational data deviate from our model
light curve near the peak.
In what follows, we will see how to determine ``intrinsic''
$t_3$.

Figure \ref{v1668_cyg_vy_jhk_mag_x35z02c10o20} explained
how to select our best model light curve from
early phase observations.  It also shows that
three different WD mass (0.85, 0.95, and $1.05 ~M_\sun$) 
models can reproduce the early phase
light curve and we can hardly select one only from the early
phase optical and IR light curves.  However, 
this uncertainty of mass determination is not a problem
in the estimation of ``intrinsic'' $t_3$ time,
because these three model light curves,
of course, give similar values of $t_3$ time
($t_3 \sim 26 \pm 2$ days) along our model light curves.

Using relations of
Equations (\ref{t_break-t_wind})--(\ref{t_3-t_h_burning}),
we can predict the period of a supersoft X-ray phase
of individual novae from $t_3$ (or $t_2$, $t_{\rm break}$)
time.  
In Section \ref{light_curve_of_ten_novae}, we will show that
our formulae are much better fitted with observation if we
use intrinsic $t_3$ time than if we use face values of $t_3$.
Therefore, in what follows in this paper,
we use intrinsic $t_3$ time in order to predict
$t_{\rm X-on} \approx t_{\rm wind}$
and $t_{\rm X-off} \approx t_{\rm H-burning}$
from Equations (\ref{t_3-t_wind}) and (\ref{t_3-t_h_burning}). 

We assumed here that a nova has a typical $M_{\rm env,0}$,
i.e., has a peak brightness similar to that of V1668 Cyg
(i.e., point B in the rescaled $t/f_{\rm s}$--$m'_V$ plane, Figure
\ref{v1668_cyg_vy_jhk_mag_x35z02c10o20_model_scaling_law}).
If the rescaled peak magnitude is much brighter/darker than
that of V1668 Cyg (for example, points A/C in
Figure \ref{v1668_cyg_vy_jhk_mag_x35z02c10o20_model_scaling_law}),
X-ray on and off times, $t_{\rm X-on} \approx t_{\rm wind}$
and $t_{\rm X-off} \approx t_{\rm H-burning}$,
can deviate from the prediction of
Equations (\ref{t_3-t_wind}) and (\ref{t_3-t_h_burning}). 
In this sense, our prediction formulae (\ref{t_3-t_wind}) and
(\ref{t_3-t_h_burning}) are statistical ones.

% Fig.15
%\placefigure{max_t3_scale}

\begin{figure*}
\epsscale{0.85}
%\epsscale{1.0}
%\epsscale{1.15}
\plotone{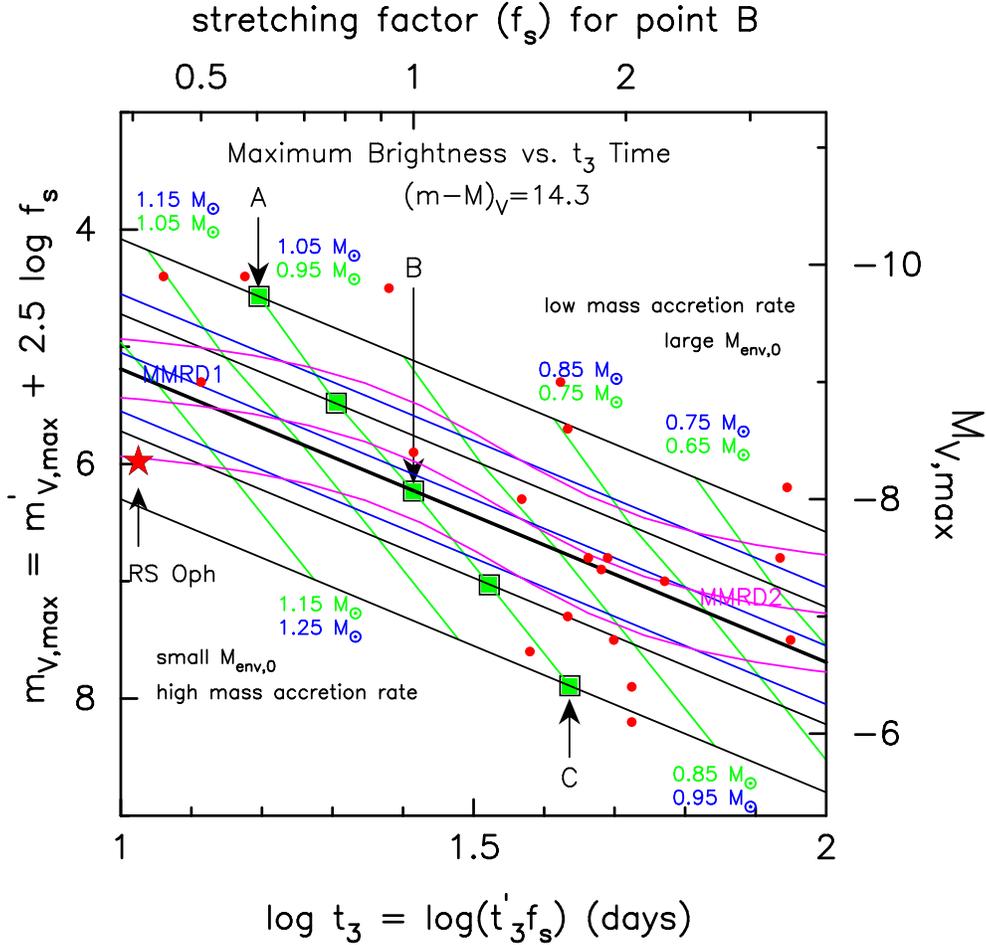}
%\plotone{f5bw.epsi}
%\plotone{max_t3_scale_color.epsi}
%\plotfiddle{evolution1.ps}{5.0cm}{270}{0.4}{0.4}{-170}{220}
\caption{
The maximum magnitude--rate of decline (MMRD) relation against
$t_3$ time based on the universal decline law.
Point B represents a nova on a $0.95 M_\sun$ WD with a
typical initial envelope mass.
The black solid line passing through point B 
is corresponding to an MMRD relation of novae having a typical initial
envelope mass but different WD mass. 
The black solid line passing through point A is an MMRD relation 
for a much larger initial
envelope mass than that of point B (much lower mass accretion rates).
The black solid line passing through point C is an MMRD relation
for a much smaller initial envelope mass than that of
point B (much higher mass accretion rates).  Therefore,
the region between upper and lower black solid lines represents
our theoretical MMRD relation. 
Green solid lines connect the same WD mass, from 1.15 to
0.65 $M_\sun$ by 0.1 $M_\sun$ step ({\it green attached number})
for ``CO nova 2,'' 
but from 1.25 to 0.75 $M_\sun$ by 0.1 $M_\sun$ step ({\it blue attached
number}) for 
``Ne nova 2.''   The absolute magnitude $M_{V, {\rm max}}$ at
maximum is obtained from calibration with V1668 Cyg of
the distance modulus $(m-M)_V = 14.3$.  
{\it Red filled star marks}: the recurrent nova RS Oph as
an example for a very high mass accretion rate
and very short timescale $t_3$ (very small $f_{\rm s}$).
Two well-known MMRD relations are added:  
Kaler-Schmidt's law \citep{sch57} of
Equation (\ref{kaler-schmidt-law}) labeled ``MMRD1''
({\it blue solid lines}),
and Della Valle \& Livio's (1995) law of
Equation (\ref{della-valle-livio-law}) labeled ``MMRD2''
({\it magenta solid lines}).
We add upper/lower bounds of $\pm 0.5$ mag for each of these
two MMRD relations.  
{\it Red filled circles}: We also add MMRD points
for individual novae taken from Table 5 of \citet{dow00}.
}
\label{max_t3_scale}
\end{figure*}

\section{MMRD relation derived from a universal decline law}
\label{mmrd_relation}

The Maximum Magnitude--Rate of Decline (MMRD) relation is often used
to estimate the distance to novae
\citep[e.g.,][]{coh88, del95, dev78, dow00, sch57}.
In this section, we derive
theoretical MMRD relations based on our universal decline law.

\subsection{Average MMRD relation}
Figures \ref{v1668_cyg_vy_jhk_mag_x35z02c10o20_model_real_scale_logt_no2}
and \ref{mass_v_uv_x_v1974_cyg_x55z02o10ne03_real_scale_model}
clearly shows a trend that a more massive WD has a brighter maximum
magnitude (smaller $M_{V, {\rm max}}$) and a faster decline rate
(smaller $t_3$ time).  The relation between $t_3$ and 
$M_{V, {\rm max}}$ for novae is called
``the Maximum Magnitude--Rate of Decline'' (MMRD) relation.
In order to derive theoretical MMRD relations, we go back to Figure
\ref{v1668_cyg_vy_jhk_mag_x35z02c10o20_model_scaling_law} and
utilize the universal decline law starting from point B (a typical
case of nova envelope mass).

We have already shown that all the free-free emission model light 
curves for different WD masses overlap with each other
(they converge to $0.95 ~M_\sun$ WD in Figure
\ref{v1668_cyg_vy_jhk_mag_x35z02c10o20_model_scaling_law} against
V1668 Cyg data) when they are squeezed in the direction
of time by a factor of $f_{\rm s}$, data of which are tabulated
in Table \ref{light_curves_of_novae_co}.
Now we assume that all the free-free emission model light curves
reaches the same maximum point, like point B in Figure
\ref{v1668_cyg_vy_jhk_mag_x35z02c10o20_model_scaling_law}.
In such a case, the apparent maximum brightness $m_{V, {\rm max}}$
of each light curve with different WD masses is expressed as 
$m_{V,{\rm max}} = m'_{V,{\rm max}} + 2.5 \log f_{\rm s}$.
The $t_3$ time of each model light curve with different WD masses
is squeezed to be $t_3 = f_{\rm s} t'_3$.
Eliminating $f_{\rm s}$ from these two relations, we have
\begin{equation}
m_{V,{\rm max}} = 2.5 \log t_3 + m'_{V,{\rm max}} - 2.5 \log t'_3.
\label{theoretical_apparent_MMRD_relation}
\end{equation}
Since point B corresponds to the optical peak of V1668 Cyg
($f_{\rm s}=1$ in Figure
\ref{v1668_cyg_vy_jhk_mag_x35z02c10o20_model_scaling_law}),
we have already known  $t'_3 = 26$ days and $m'_{V,{\rm max}} = 6.2$.
Substituting these values of $(t'_3 = 26, m'_{V,{\rm max}}=6.2)$
into Equation (\ref{theoretical_apparent_MMRD_relation}), we have
\begin{equation}
m_{V,{\rm max}} = 2.5 \log t_3 + 6.2 - 2.5 \log 26
= 2.5 \log t_3 + 2.7.
\end{equation}
This is our theoretical apparent MMRD ration for a starting point B
in Figure \ref{v1668_cyg_vy_jhk_mag_x35z02c10o20_model_scaling_law}.
Next, using the distance modulus of $(m-M)_V$, we obtain our theoretical
MMRD relation as
\begin{eqnarray}
M_{V,{\rm max}} & = & m_{V,{\rm max}}  - (m-M)_V \cr
 & = & 2.5 \log t_3 + m'_{V,{\rm max}} - 2.5 \log t'_3 - (m-M)_V.
\label{theoretical_MMRD_relation}
\end{eqnarray}
Substituting $(m-M)_V = 14.3$, the distance modulus of V1668 Cyg,
into Equation (\ref{theoretical_MMRD_relation}), we have
\begin{equation}
M_{V,{\rm max}} = 2.5 \log t_3 + 6.2 - 2.5 \log 26 - 14.3
= 2.5 \log t_3 -11.6.
\end{equation}
Figure \ref{max_t3_scale} shows this theoretical MMRD relation
for novae that has a typical (average) $M_{\rm env,0}$
like V1668 Cyg.  In the right axis we convert the apparent
magnitude to the absolute magnitude by using the distance modulus
of V1668 Cyg.  Point B in Figure \ref{max_t3_scale} corresponds to 
V1668 Cyg ($f_{\rm s}=1$: $0.95 ~M_\sun$, in which
nova light curves start from point B in Figure
\ref{v1668_cyg_vy_jhk_mag_x35z02c10o20_model_scaling_law}).
The other WD mass cases starting from the same point B in
Figure \ref{v1668_cyg_vy_jhk_mag_x35z02c10o20_model_scaling_law}
have different $f_{\rm s} \ne 1$.  
Changing $f_{\rm s}$ (or $M_{\rm WD}$), we have
a line passing through point B in Figure \ref{max_t3_scale}.
This is our theoretical MMRD relation, the parameter of which
is $M_{\rm WD}$ (or $f_{\rm s}$).  We also show the equivalent
parameter, stretching factor $f_{\rm s}$ (against point B),
in the upper axis of the same figure.

The black thick solid line passing through point B shows quite
a good agreement with the central blue and
magenta lines, which are empirical MMRD relations 
obtained from observation as will be explained bellow.

\subsection{Scatter of MMRD relations from average}
If the initial envelope mass $M_{\rm env,0}$ is larger
or smaller than that of point B (which is an average value,
see Figure \ref{v1668_cyg_vy_jhk_mag_x35z02c10o20_model_scaling_law}),
we obtain different MMRD relations.  In Figure \ref{max_t3_scale},
we show five points of green filled squares corresponding to
different initial envelope mass of $M_{\rm env, 0}$.
Point A $(t'_3 = 16$ days, $m'_{V,{\rm max}}=4.6)$
corresponds to a larger envelope mass at ignition (point A in
Figure \ref{v1668_cyg_vy_jhk_mag_x35z02c10o20_model_scaling_law}).
The black solid line passing through point A is other
$m_{V,{\rm max}}$--$\log t_3$ relation for a much larger initial
envelope mass of $M_{\rm env,0}$ (at point A in Figure
\ref{v1668_cyg_vy_jhk_mag_x35z02c10o20_model_scaling_law}), i.e.,
\begin{equation}
M_{V,{\rm max}} = 2.5 \log t_3 + 4.6 - 2.5 \log 16 - 14.3
= 2.5 \log t_3 -12.7.
\end{equation}
On the other hand,
the black solid line passing through point C 
$(t'_3 = 43$ days, $m'_{V,{\rm max}}=7.9)$ is also different
$m_{V,{\rm max}}$--$\log t_3$ relation
for a much smaller initial envelope mass (at point C in Figure
\ref{v1668_cyg_vy_jhk_mag_x35z02c10o20_model_scaling_law}), i.e.,
\begin{equation}
M_{V,{\rm max}} = 2.5 \log t_3 + 7.9 - 2.5 \log 43 - 14.3
= 2.5 \log t_3 -10.5.
\end{equation}
For a given WD mass, $M_{\rm env,0}$ is larger (smaller)
for a smaller (larger) accretion rate.  Points A, B, and C correspond
to different ignition masses for a given WD mass.
When the accretion rate is much smaller than that for point B,
the brightness reaches, for example, point A
at maximum and then we have a shorter $t_3$ time
(see Figure \ref{v1668_cyg_vy_jhk_mag_x35z02c10o20_model_scaling_law}).
In the case of a larger accretion rate than that for point B,
the brightness goes up only
to point C at maximum and we have a longer $t_3$ time.
So, even for a given $M_{\rm WD}$ and chemical composition
of the envelope, we have different $t_3$ time, depending
on $M_{\rm env,0}$.
In other words, a variation of mass accretion rates from
an average value makes a divergence (or width) in the MMRD 
relation for individual novae.

The five points including points A, B, and C are calculated 
for V1668 Cyg model light curves, all of which correspond to
the 0.95 $M_\sun$ WD of ``CO nova 2'' ($f_{\rm s} = 1$). 
The green solid line connecting these five points
represents the $M_{V, \rm{max}}$--$\log t_3$ relation
for 0.95 $M_\sun$ WD of ``CO nova 2.''
We then obtain other $M_{V, \rm{max}}$--$\log t_3$ lines
for other WD masses as shown by green lines
(attached numbers represent WD masses)
in Figure \ref{max_t3_scale}. 
In the case of neon novae, we also obtain similar lines,
which are also shown in Figure \ref{max_t3_scale}. 
These lines are almost overlapped to those of ``CO nova 2''
but different WD masses
(blue attached numbers for WD masses).

\subsection{Comparison with other MMRD relations}
Our MMRD ($M_{V, \rm{max}}$--$\log t_3$) relation passing through
point B is in a good agreement with two well-known MMRD relations
empirically obtained:
one is Kaler-Schmidt's law \citep{sch57}, i.e.,
\begin{equation}
M_{V, {\rm max}} = -11.75 + 2.5 \log t_3,
\label{kaler-schmidt-law}
\end{equation}
and the other is Della Valle \& Livio's (1995) law, i.e.,
\begin{equation}
M_{V, {\rm max}} = -7.92 -0.81 \arctan \left(
{{1.32-\log t_2} \over {0.23}} \right),
\label{della-valle-livio-law}
\end{equation}
where $M_{V, {\rm max}}$ is the absolute $V$ magnitude at maximum
and we use Equation (\ref{t_2-t_3}) to calculate $t_2$ from
$t_3$ in Equation (\ref{della-valle-livio-law}) 
in Figure \ref{max_t3_scale}.  Kaler-Schmidt's law is
denoted by a blue solid line with two attendant blue solid lines
corresponding to $\pm 0.5$ mag brighter/darker cases.  
Della Valle \& Livio's law is indicated by
magenta solid lines.  These two well-known empirical MMRD
relations are very close to our theoretical MMRD relation
at average.

We also add observational MMRD points ($M_{V, \rm{max}}$ and $t_3$)
for individual novae, data of which are taken from
Table 5 of \citet{dow00}, although their $t_3$ are
not intrinsic but observational face values.
It is clearly shown that a large scatter of individual points
(red filled circles)
from the two empirical MMRD relations falls into between
upper/lower cases of our MMRD relations for the
largest/smallest initial envelope mass of $M_{\rm env,0}$.
This simply means that there is a second parameter to
specify the MMRD relation for individual novae.
The main parameter is the WD mass represented by 
stretching factor $f_{\rm s}$.  The second parameter
is the initial envelope mass (or the mass accretion rate to the WD).
This second parameter can reasonably explain the scatter
of individual novae from the empirical MMRD relations ever proposed.

It is interesting to place the position of the recurrent nova RS Oph
in this diagram.  RS Oph locates in the smallest corner of $t_3$ time
and a relatively darker level of $M_{V, {\rm max}}$,
near the $M_{V, {\rm max}}$--$\log t_3$ relation
passing through point C.  This indicates a much smaller envelope mass at
the optical maximum, suggesting that the mass accretion rate to the WD
is very high.  This situation is very consistent with
the total picture of recurrent novae; a very massive WD close to the
Chandrasekhar mass and a high mass accretion rate
to the WD \citep[e.g.,][]{hac01kb}.  In this figure, we adopt
$M_{V, {\rm max}} = -8.3$ and $t_3 = 10.5$ days with
the distance of $d= 1.6$ kpc \citep{hje86}, 
absorption of $A_V = 3.1 E(B-V) = 3.1 \times 0.73 = 2.3$
\citep{sni87a, sni87b}, $m_{V, {\rm max}}= 5.0$ \citep{ros87},
and $t_3 = 10.5$ days from optical light curve fitting with
our free-free model light curves \citep{hac01kb, hac06kk, hac07kl}.

\section{Light Curve Analysis of Ten Novae with Supersoft X-ray Emission}
\label{light_curve_of_ten_novae}

In this section, we examine ten classical novae which we have not
yet analyzed in our series of papers,
i.e., V598 Pup, V382 Vel, V4743 Sgr, V1281 Sco, V597 Pup,
V1494 Aql, V2467 Cyg, V5116 Sgr, V574 Pup, and V458 Vul,
in the order of increasing timescale.  All are supersoft X-ray
detected novae, so we can check whether or not our prediction
formulae of a supersoft X-ray phase work on these novae.

% Table 6
%\placetable{fitting_t2_t3_tb_txonoff}

\begin{deluxetable*}{llllllllll}
\tabletypesize{\scriptsize}
\tablecaption{Timescales of Novae Estimated from Universal 
Decline Law
\label{fitting_t2_t3_tb_txonoff}}
\tablewidth{0pt}
\tablehead{
\colhead{object} &
\colhead{...} &
\colhead{$t_2$\tablenotemark{a}} &
\colhead{$t_3$\tablenotemark{a}} &
\colhead{$t_{\rm break}$} &
\colhead{$t_{\rm wind}$} &
\colhead{$t_{\rm H-burning}$} &
\colhead{$t_{\rm emerge}$} &
\colhead{$M_{\rm WD}$} &
\colhead{chem.comp.\tablenotemark{b}} \\
\colhead{} &
\colhead{} &
\colhead{(day)} &
\colhead{(day)} &
\colhead{(day)} &
\colhead{(day)} &
\colhead{(day)} &
\colhead{(day)} &
\colhead{($M_\sun$)} &
\colhead{} 
}
\startdata
V598 Pup 2007\#2 & ... & -- & -- & 40 & 90 & 150 & -- & 1.28 & Ne nova 2 \\
V382 Vel 1999 & ... & 6.6 & 12.4 & 52 & 120 & 220 & 32 & 1.23 & Ne nova 2 \\
V4743 Sgr 2002\#3 & ... & 9.4 & 17 & 70 & 180 & 380 & 36 & 1.15 & Ne nova 2 \\
V1281 Sco 2007\#2 & ... & 9.5 & 16.7 & 78 & 190 & 425 & -- & 1.13 & Ne nova 2 \\
V597 Pup 2007\#1 & ... & 8.7 & 16.5 & 74 & 200 & 430 & 52 & 1.2 & Ne nova 3 \\
V1494 Aql 1999\#2 & ... & 11.1 & 21.2 & 96 & 260 & 670 & 66 & 1.13 & Ne nova 3 \\
V1494 Aql 1999\#2 & ... & 11.7 & 22.1 & 102 & 260 & 670 & 72 & 1.06 & Ne nova 2 \\
V1494 Aql 1999\#2 & ... & 14.7 & 26.6 & 120 & 285 & 675 & 95 & 0.92 & CO nova 2 \\
V2467 Cyg 2007 & ... & 14.3 & 25.3 & 110 & 290 & 765 & 68 & 1.11 & Ne nova 3 \\
V2467 Cyg 2007 & ... & 14.8 & 27.9 & 113 & 290 & 765 & 75 & 1.04 & Ne nova 2 \\
%%%V2467 Cyg 2007 & ... & 15.0 & 28.0 & 108 & 290 & 760 & 84 & 0.94 & CO nova 3 \\
V2467 Cyg 2007 & ... & 17.6 & 31.7 & 135 & 310 & 750 & 96 & 0.90 & CO nova 2 \\
V5116 Sgr 2005\#2 & ... & 16.7 & 30.4 & 127 & 355 & 1000 & 91 & 1.07 & Ne nova 3 \\
V5116 Sgr 2005\#2 & ... & 16.2 & 30.5 & 134 & 350 & 995 & 100 & 1.0 & Ne nova 2 \\
V5116 Sgr 2005\#2 & ... & 20.0 & 35.8 & 169 & 390 & 1010 & 130 & 0.85 & CO nova 2 \\
V574 Pup 2004 & ... & 18.6 & 35.0 & 144 & 390 & 1140 & -- & 1.05 & Ne nova 3 \\
V458 Vul 2007 & ... & 20.4 & 37.2 & 145 & 415 & 1250 & 62 & 0.93 & CO nova 4 \\
\\
V1974 Cyg 1992 & ... & 16 & 26 & 108 & 250 & 600 & 90 & 1.05 & Ne nova 2 \\
%%%GQ Mus 1983 & ... & 67 & 122 & 260 & 1000 & 3300 & 330 & 0.7 & CO nova 3 \\
V1668 Cyg 1978 & ... & 14.4 & 26 & 100 & 280 & 720 & 87 & 0.95 & CO nova 3 \\
V1500 Cyg 1975 & ... & 7.2 & 13 & 70 & 180 & 380 & 50 & 1.15 & Ne nova 2 
\enddata
\tablenotetext{a}{$t_2$ and $t_3$ times are all intrinsic values
estimated along our theoretical light curve}
\tablenotetext{b}{see Table \ref{chemical_composition}}
\end{deluxetable*}

% Table 7
%\placetable{physical_properties_of_novae}

\begin{deluxetable*}{llllllllll}
\tabletypesize{\scriptsize}
\tablecaption{Summary of Physical Properties of Novae 
\label{physical_properties_of_novae}}
\tablewidth{0pt}
\tablehead{
\colhead{object} &
\colhead{...} &
\colhead{$M_{\rm WD}$} &
\colhead{$M_2$} &
\colhead{$P_{\rm orb}$} &
\colhead{$a$} &
\colhead{$R_1^*$} &
\colhead{$M_{\rm wind}$}\tablenotemark{a} &
\colhead{$M_{\rm env,0}$}\tablenotemark{b} &
\colhead{chem.comp.\tablenotemark{c}} \\
\colhead{} &
\colhead{} &
\colhead{($M_\sun$)} &
\colhead{($M_\sun$)} &
\colhead{(day)} &
\colhead{($R_\sun$)} &
\colhead{($R_\sun$)} &
\colhead{($10^{-5}M_\sun$)} &
\colhead{($10^{-5}M_\sun$)} &
\colhead{} 
}
\startdata
V598 Pup 2007\#2 & ... & 1.28 & -- & -- & -- & -- & 0.31 & 0.34 & Ne nova 2 \\
V382 Vel 1999 & ... & 1.23 & 0.31 & 0.1461 & 1.35 & 0.68 & 0.43 & 0.48 & Ne nova 2 \\
V4743 Sgr 2002\#3 & ... & 1.15 & 0.70 & 0.2799 & 2.2 & 0.93 & 0.69 & 0.78 & Ne nova 2 \\
V1281 Sco 2007\#2 & ... & 1.13 & -- & -- & -- & -- & 0.92 & 1.02 & Ne nova 2 \\
V597 Pup 2007\#1 & ... & 1.2 & 0.22 & 0.1112 & 1.1 & 0.58 & 1.06 & 1.14 & Ne nova 3 \\
V1494 Aql 1999\#2 & ... & 1.13 & 0.28 & 0.1346 & 1.24 & 0.62 & 1.04 & 1.12 & Ne nova 3 \\
V1494 Aql 1999\#2 & ... & 1.06 & 0.28 & 0.1346 & 1.22 & 0.60 & 1.11 & 1.26 & Ne nova 2 \\
V1494 Aql 1999\#2 & ... & 0.92 & 0.28 & 0.1346 & 1.17 & 0.57 & 1.75 & 2.00 & CO nova 2 \\
V2467 Cyg 2007 & ... & 1.11 & 0.35 & 0.1596 & 1.40 & 0.67 & 1.08 & 1.23 & Ne nova 3 \\
V2467 Cyg 2007 & ... & 1.04 & 0.35 & 0.1596 & 1.38 & 0.65 & 1.16 & 1.34 & Ne nova 2 \\
%%%V2467 Cyg 2007 & ... & 0.94 & 0.35 & 0.1596 & 1.35 & 0.63 & 1.38 & 1.60 & CO nova 3 \\
V2467 Cyg 2007 & ... & 0.90 & 0.35 & 0.1596 & 1.33 & 0.62 & 1.67 & 1.90 & CO nova 2 \\
V5116 Sgr 2005\#2 & ... & 1.07 & 0.25 & 0.1238 & 1.15 & 0.58 & 1.25 & 1.43 & Ne nova 3 \\
V5116 Sgr 2005\#2 & ... & 1.00 & 0.25 & 0.1238 & 1.13 & 0.56 & 1.30 & 1.53 & Ne nova 2 \\
V5116 Sgr 2005\#2 & ... & 0.85 & 0.25 & 0.1238 & 1.08 & 0.53 & 2.17 & 2.52 & CO nova 2 \\
V574 Pup 2004 & ... & 1.05 & -- & -- & -- & -- & 1.25 & 1.46 & Ne nova 3 \\
V458 Vul 2007 & ... & 0.93 & 0.74\tablenotemark{d} & 0.5895 & 3.51 & 1.40 & 1.26 & 1.53 & CO nova 4 \\
\\
V1974 Cyg 1992 & ... & 1.05 & 0.15 & 0.0813 & 0.85 & 0.44 & 1.0 & 1.17 & Ne nova 2 \\
%%%GQ Mus 1983 & ... & 0.7 & 0.10 & 0.0594 & 0.59 & 0.33 & 1.8 & 2.5 & CO nova 3 \\
V1668 Cyg 1978 & ... & 0.95 & 0.29 & 0.1384 & 1.21 & 0.58 & 1.1 & 1.3 & CO nova 3 \\
V1500 Cyg 1975 & ... & 1.15 & 0.29 & 0.1396 & 1.28 & 0.64 & 0.71 & 0.8 & Ne nova 2 
\enddata
\tablenotetext{a}{ejected mass by winds}
\tablenotetext{b}{envelope mass at the optical maximum}
\tablenotetext{c}{see Table \ref{chemical_composition}
 for chemical composition}
\tablenotetext{d}{upper limit from $q = M_2/M_{\rm WD} < q_{\rm crit}= 0.8$}
\end{deluxetable*}

% Fig.16
%\placefigure{all_mass_v598_pup_v1500_cyg_x55z02o10ne03}

\begin{figure}
%\epsscale{1.0}
\epsscale{1.15}
\plotone{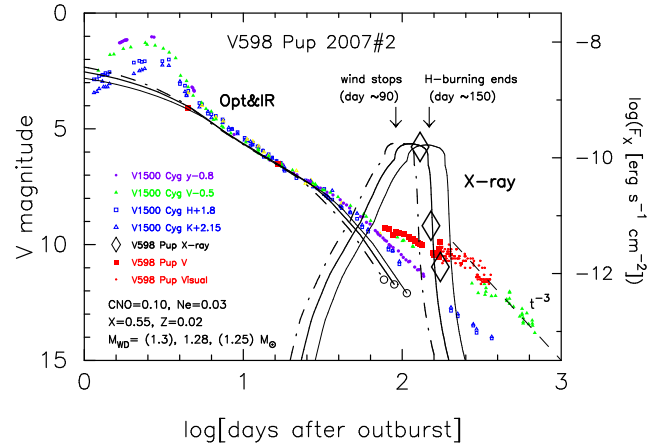}
%\plotone{f7bw.epsi}
%\plotone{all_mass_v598_pup_v1500_cyg_x55z02o10ne03_color.epsi}
%\plotfiddle{evolution1.ps}{5.0cm}{270}{0.4}{0.4}{-170}{220}
\caption{
Optical and supersoft X-ray light curves for V598 Pup.
We plot free-free model light curves (labeled ``Opt\&IR'')
and $0.2-0.6$~keV supersoft X-ray fluxes (labeled ``X-ray'')
of three WD mass models of $M_{\rm WD}= 1.3 ~M_\sun$
 ({\it dash-dotted line}),
$1.28 ~M_\sun$ ({\it thick solid line}), and $1.25 ~M_\sun$ 
({\it thin solid line}),
for the envelope chemical composition of ``Ne nova 2.''
We select a best model of $1.28 ~M_\sun$ WD among three WD
mass models, in order to reproduce the supersoft X-ray data.  
The supersoft X-ray data is sandwiched between
the 1.3 and $1.25 ~M_\sun$ WDs, so we safely regard that
$M_{\rm WD}= 1.28 \pm 0.03 ~M_\sun$.
The X-ray flux data
({\it large diamonds}) without error bars are taken from 
\citet{rea07, rea08}.  Optical $V$ data of V598 Pup are taken from
\citet{poj07} and visual data are from the American Association
of Variable Star Observers (AAVSO).
Arrows indicate two epochs when the optically thick wind stops
about 90 days after the outburst and when the steady hydrogen shell
burning atop the WD ends about 150 days after the outburst for our
$1.28 ~M_\sun$ WD model.  We added optical and near IR data of
V1500 Cyg for comparison
(see Figure \ref{light_curve_v598_pup_v1500_cyg}).
}
\label{all_mass_v598_pup_v1500_cyg_x55z02o10ne03}
\end{figure}

\subsection{V598 Pup 2007\#2}
\label{v598_pup_model}

V598 Puppis (Nova Puppis 2007 No.2) was serendipitously
discovered by the {\it XMM-Newton} slew survey \citep{rea08}.
\citet{tor07} subsequently found an optically bright counter part
of this X-ray object.  The spectra obtained on November
16.34 UT (day $\sim 170$) reveal numerous features
suggestive of a nova that already entered the auroral phase
\citep[see also][]{rea08}.  Analysis of
All Sky Automated Survey (ASAS) archive showed that
the $V$ magnitude rose to $m_V= 4.1$ from $m_V \gtrsim 14$ 
between June 1.966 UT and June 5.968 UT \citep{poj07}.
Since we cannot specify the exact date of the outburst,
we assume here that the outburst day is $t_{\rm OB}=$ JD~2454253.0
(2007 June 1.5 UT).

The X-ray satellite {\it Swift} detected X-ray flux from V598 Pup,
which declined rapidly from $F_{\rm X}{\rm (0.2-2~keV)}
 = 1.5 \times 10^{-10}$ (day $\sim 130$) to 
$6.7 \times 10^{-12}$ (day $\sim 150$)
and then to $1.3 \times 10^{-12}$ erg~s$^{-1}$~cm$^{-2}$
(at day $\sim 170$ after the outburst) \citep{rea07}
as shown in Figure \ref{all_mass_v598_pup_v1500_cyg_x55z02o10ne03}.
This quick drop in the X-ray flux clearly showed that hydrogen
burning on the WD ended around 150 days after the outburst.

Figure \ref{all_mass_v598_pup_v1500_cyg_x55z02o10ne03} shows
observed optical light curves of V598 Pup as well as the X-ray data.
Unfortunately there are only
two observational points available before 77 days after the outburst.
To understand the characteristic properties of V598 Pup light curves,
we add observational points of V1500 Cyg that show a 
very similar light curve in the later phase.
We can see that two points of $V$ mag in the early phase are just
on our free-free model light curve but 
the $V$ data after day $\sim 70$ are those in the nebular phase
and already deviated from the expected free-free light curve.
This deviation often occurs in novae when $V$ magnitude is contaminated
by strong emission lines such as [\ion{O}{3}] in the nebular phase
\citep[see][for emission lines of V598 Pup]{rea08},
as already discussed in \citet{hac06kb, hac07k}
and \citet{hac08kc}. 
Thus we regard that the light curve of V598 Pup decays very similarly
to that of V1500 Cyg, which follows our universal decline law 
from day $\sim 5$ to day $\sim 70$.

No abundance determinations are available for V598 Pup so far.
We here assume two typical cases, one is for oxygen-neon novae of
``Ne nova 2''and the other is for carbon-oxygen novae of
``CO nova 2'' as listed in Table \ref{chemical_composition}.

The emergence/decay times of supersoft X-ray are a good indicator
of the WD mass \citep{hac06kb, hac08kc}.
We have calculated several models for different WD masses
with these two chemical compositions.  Among the models of
$M_{\rm WD}= 1.3$, $1.28$, and $1.25 ~M_\sun$, we have
selected $M_{\rm WD}= 1.28~M_\sun$ as a best-reproducing model
for ``Ne nova 2,'' but, $M_{\rm WD}= 1.2 ~M_\sun$ 
for ``CO nova 2'' as a best-reproducing one among $M_{\rm WD}= 1.25$,
$1.2$, and $1.15 ~M_\sun$, which are not shown in the figure. 
The best models are selected by careful eye fitting
with the supersoft X-ray data because there are only three points.
The optical fitting cannot be used for the selection
because all of three model light curves go through the two
observational $V$ points as seen in Figure 
\ref{all_mass_v598_pup_v1500_cyg_x55z02o10ne03}.
We define the ``error'' in the WD mass determination 
($\pm 0.03 ~M_\sun$) from larger and smaller WD masses,
$M_{\rm WD}= 1.3$ and $1.25 ~M_\sun$, safely cover the decay 
phase of supersoft X-ray flux for the fixed chemical composition.

The estimated WD mass of $1.2 ~M_\sun$ for ``CO nova 2''
exceeds the upper limit mass for CO WDs born in a binary, i.e.,
$M_{\rm CO} \lesssim 1.07 ~M_\sun$ \citep{ume99},
so we regard that the WD of V598 Pup is very likely
an ONeMg WD.  Therefore, we adopt the $1.28 ~M_\sun$ WD of
``Ne nova 2'' as a best model, which has
$t_{\rm break} =40$, $t_{\rm wind} =90$, and
$t_{\rm H-burning} =150$ days after the outburst (see Figure 
\ref{all_mass_v598_pup_v1500_cyg_x55z02o10ne03}). 

We have also calculated the ejecta mass of 
$M_{\rm wind} \sim 3 \times 10^{-6} M_\sun$,
which is lost by the optically thick wind \citep[see,
e.g.,][for more detail of ejecta mass calculation]{kat94h, hac06kb}.
However, it should be noted that 
$M_{\rm wind}$ could be the subject to significant
uncertainty because we missed the real optical maximum.
Various physical parameters of novae thus obtained are 
summarized in Tables \ref{fitting_t2_t3_tb_txonoff} and
\ref{physical_properties_of_novae}.

We are not able to determine $t_2$ and $t_3$ from the light curve
because of lack of visual magnitudes around the optical peak.

     The mass determination of a WD depends weakly on the chemical
composition of the envelope, especially, on the hydrogen content
$X$ \citep{hac06kb, hac07k, hac08kc, hac09k}.  We have estimated
this dependency from two other chemical composition models, i.e.,
``Ne nova 1'' and ``Ne nova 3'' in Table \ref{chemical_composition}.
We similarly obtain a best-reproducing model of $M_{\rm WD}= 1.2 ~M_\sun$
for $X=0.35$ (``Ne nova 1'') and $M_{\rm WD}= 1.32 ~M_\sun$
for $X=0.65$ (``Ne nova 3'').   The resultant dependency
can be approximated as
\begin{equation}
M_{\rm WD}(X) \approx \left\{ 1.28 + 0.4 (X-0.55) \right\} ~M_\sun,
\end{equation}
for V598 Pup.  Here, we have already obtained
$M_{\rm WD}(X=0.55) = 1.28 ~M_\sun$.  Therefore, we may conclude
that the WD mass is $M_{\rm WD}= 1.28 \pm 0.04 ~M_\sun$ for
a typical hydrogen content between $X=0.45 - 0.65$ for V598 Pup.
The WD mass estimate depends also on the CNO abundance, but
the difference among various $X_{\rm CNO}$ models
in Table \ref{chemical_composition} is smaller than the difference
among various hydrogen content ($X$) models
\citep[see][for more detail of $X$ dependency]{hac06kb}.

The distance to V598 Pup can be estimated by two methods even if
the optical maximum is not available.
The first one is to use the absolute magnitude of our free-free
model light curves in Table \ref{light_curves_of_novae_ne}.
We use our best-reproducing model
in Figure \ref{all_mass_v598_pup_v1500_cyg_x55z02o10ne03}
and measure the brightness at the end of the free-free (FF) light
curve, $m_{\rm w}$, and obtain the distance modulus to V598 Pup as
\begin{equation}
\left[ (m-M)_V \right]_{\rm FF} = m_{\rm w} - M_{\rm w} 
= 11.8 - 0.1 = 11.7.
\end{equation}
Here we adopt $m_{\rm w}= 11.8$ from Figure
\ref{all_mass_v598_pup_v1500_cyg_x55z02o10ne03} and
use the absolute magnitude of $M_{\rm w} = 0.1$ at
the end point of $1.28 ~M_\sun$ WD model light curve,
which is calculated by interpolation
between the $1.25 ~M_\sun$ and $1.3 ~M_\sun$ WD models
in Table \ref{light_curves_of_novae_ne}.
We call this ``FF method'' as introduced in Section
\ref{checking_absolute_magnitude}.
It should be addressed that the FF method is rather robust
and gives $(m-M)_V = 11.6 - 0.0 =11.6$ for $1.3 ~M_\sun$ WD
or $(m-M)_V = 12.2 - 0.4 =11.8$ for $1.25 ~M_\sun$ WD
even if we miss a best model.  We also emphasize that 
this FF method does not require the optical maximum like the
MMRD relations, so it can be used even if the optical maximum  
is not available like V598 Pup.

The next method is to use direct comparison with the other nova
light curve the distance modulus of which is already known.
We have already done this method in Section 
\ref{checking_absolute_magnitude}
(see Figures
\ref{mass_v_uv_x_v1974_cyg_x55z02o10ne03_real_scale_model},
\ref{all_mass_gk_per_v1500_cyg_x55z02o10ne03}, and
\ref{all_mass_v1493_aql_v1500_cyg_x55z02o10ne03}).
The direct comparison between V1500 Cyg and V598 Pup 
in Figure \ref{all_mass_v598_pup_v1500_cyg_x55z02o10ne03} gives
the difference of distance moduli of these two novae, i.e.,
\begin{equation}
\left[5 \log(d/10) + A_V \right]_{\rm V1500~Cyg}
 - \left[5 \log(d/10) + A_V \right]_{\rm V598~Pup}  = 0.5.
\end{equation}
When we adopt $(m-M)_V = 12.5$ from Equation 
(\ref{distance_modulus_v1500_cyg})
for V1500 Cyg, we obtain 
\begin{equation}
\left[ (m-M)_V \right]_{\rm LC}  = 
\left[ 5 \log(d/10) + A_V \right]_{\rm V598~Pup} = 12.0,
\label{distance_absorption_v598pup_lc}
\end{equation}
where LC means the direct light curve (LC) fitting.
It should be again addressed that the LC method does not
need the optical maximum like the MMRD relations, so it can be used
even if we missed the optical maximum like V598 Pup.

The difference of 0.3 mag in the distance modulus estimation
can be understood as follows.  These two nova light curves look
very similar in the later phase but the fitting mass is different
between $1.28 ~M_\sun$ (V598 Pup) and $1.15 ~M_\sun$ (V1500 Cyg).
The absolute magnitude at 20 days, for example, is about 0.4 mag
brighter in the $1.15 ~M_\sun$ WD model than in the $1.25 ~M_\sun$ as
shown in Figure \ref{mass_v_uv_x_v1974_cyg_x55z02o10ne03_real_scale_model}.
We have experienced that the same thing happened when we use fitting
only of optical light curves as shown in
Figure \ref{v1668_cyg_vy_jhk_mag_x35z02c10o20}.
In this figure, we obtain $\pm ~0.4$ mag larger $(m-M)_V$ when
the fitting mass is missed by $\mp ~0.1 ~M_\sun$, respectively.
In this sense, the LC method may have an error of 0.4 mag or so
even if two nova light curves look very similar in the later phase.

The above two estimates are listed in Table
\ref{distance_moduli_of_novae}, being roughly consistent with
each other, i.e., $(m-M)_V = 11.9 \pm 0.2$ (averaged with equal weight).
These two distance moduli give the distance of
$d = 2.1 \pm 0.2$ kpc for $A_V = 0.27$ \citep{rea08}.
%%% d =2.2, 2.2, 2.3, 1.8 kpc
This distance is consistent with various estimates discussed
by \citet{rea08}.

% Fig.17
%\placefigure{all_mass_v382_vel_v1500_cyg_x55z02o10ne03}

\begin{figure}
%\epsscale{1.0}
\epsscale{1.15}
\plotone{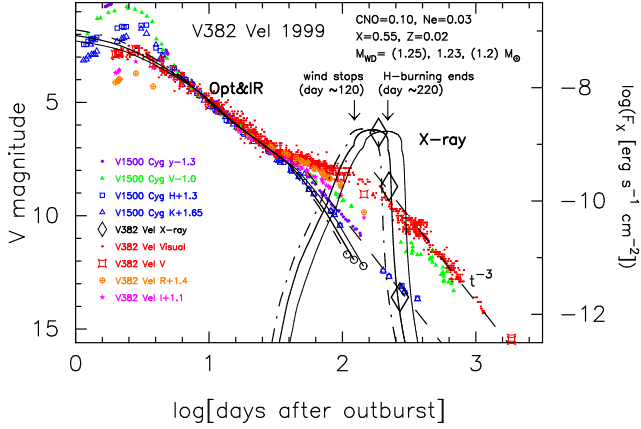}
%\plotone{f8bw.epsi}
%\plotone{all_mass_v382_vel_v1500_cyg_x55z02o10ne03_color.epsi}
%\plotfiddle{evolution1.ps}{5.0cm}{270}{0.4}{0.4}{-170}{220}
\caption{
Same as Figure \ref{all_mass_v598_pup_v1500_cyg_x55z02o10ne03} but
for V382 Vel.  Observational data of
optical $V$ ({\it boxes with sharp corners})
and $R$ ({\it circles with a plus}), near infrared
$I$ ({\it star marks}),
and supersoft X-ray ({\it large diamonds}).
We plot three model light curves of $M_{\rm WD}= 1.25$ ({\it thick
dash-dotted lines}), $1.23$ ({\it thick solid lines}), and
$1.2 ~M_\sun$ ({\it thin solid lines}) WDs 
for the envelope chemical composition of ``Ne nova 2.''
We select the $1.23 ~M_\sun$ WD as a best model among three
of 1.25, 1.23, and $1.2 ~M_\sun$ WDs 
and can reproduce the supersoft X-ray data (three points).
The supersoft X-ray data is sandwiched between
the 1.25 and $1.2 ~M_\sun$ WDs, so we safely regard that
$M_{\rm WD}= 1.23 \pm 0.03 ~M_\sun$.
Observational V382 Vel data of $V$, $R$, and $I$ magnitudes are
taken from IAU Circulars 7176, 7179, 7196, 7209,
7216, 7226, 7232, 7238, and 7277.  The darkest observational
$V$ magnitudes at day $\sim 1850$ are taken from \citet{wou05}.
Visual magnitudes ({\it small circles}) are from AAVSO.
The X-ray absorbed flux data ({\it large open diamonds})
of V382 Vel are taken from \citet{ori02} and \citet{bur02}
without error bars.
}
\label{all_mass_v382_vel_v1500_cyg_x55z02o10ne03}
\end{figure}

\subsection{V382 Vel 1999}

V382 Velorum (Nova Velorum 1999) was discovered independently
by Williams and Gilmore \citep{wil99}.  The nova reached
$m_V = 2.7$ at maximum on 1999 May 23 and then decayed to
$m_V = 16.6$ \citep{wou05} as shown in Figure 
\ref{all_mass_v382_vel_v1500_cyg_x55z02o10ne03}.
V382 Vel is a fast nova and identified as a neon nova \citep{woo99}.
We here adopt a set of chemical composition
of a typical neon nova, ``Ne nova 2'' in Table 
\ref{chemical_composition}, based on the composition analyses
by \citet{sho03} and by \citet{aug03}
\citep[see Table 1 of][]{hac06kb}.
We also assume that the outburst day is $t_{\rm OB}=$JD~2451319.0
(1999 May 20.5 UT) from information of IAU Circulars 7176 and 7184,
i.e., $m_{\rm vis} > 12$ on May 20.57 UT,
$m_{\rm vis} = 4.3$ on May 21.62 UT,
$m_{\rm vis} = 3.6$ on May 21.725 UT,
and $m_V = 3.1$ on May 22.396 UT.

Supersoft X-rays were detected by {\it BeppoSAX} about six months
after the outburst \citep{ori02}.  After that, the flux decayed rapidly
as shown in Figure \ref{all_mass_v382_vel_v1500_cyg_x55z02o10ne03}.
The {\it Chandra} observations by \citet{bur02} and by \citet{nes05}
suggested that the hydrogen burning ended around 220 days after the
outburst.  The theoretical supersoft X-ray phase is very sensitive
to the WD mass whereas it depends weakly on the chemical composition
of the envelope.
Our X-ray light curve fitting suggests that 1) the WD mass
is as massive as $M_{\rm WD}= 1.23 \pm 0.03 ~M_\sun$
and 2) the supersoft X-ray phase started from day $\sim 120$.
We have also estimated the ejecta mass of
$M_{\rm wind} \sim 4.3 \times 10^{-6} M_\sun$ (Table
\ref{physical_properties_of_novae}) for the $1.23 ~M_\sun$ WD model.
\citet{del02} derived the ejecta mass of 
$M_{\rm ej}\sim 6 \times 10^{-6} M_\sun$ from the flux of
H$\alpha$ emission, which is consistent with our estimate.

  Our model of $1.23 ~M_\sun$ WD shows
$t_{\rm break} = 52$, $t_{\rm wind} = 120$, and
$t_{\rm H-burning} = 220$ days after the outburst
(see Figure \ref{all_mass_v382_vel_v1500_cyg_x55z02o10ne03}). 
Since the $V$ magnitude reached its maximum $t_0 = 2.6$ days
after the outburst, we obtain ``intrinsic'' $t_2 = 6.6$ and
$t_3 = 12.4$ days along our model light curve.
For this nova, since the observed data decays almost along our model
light curves, the ``face'' values of $t_2$ and $t_3$ are
practically the same as those of our ``intrinsic'' $t_2$ and $t_3$.
It should be noted here that the assumed epoch of
$t_{\rm OB}=$ JD~2451319.0 (1999 May 20.5 UT) hardly affect
our estimation of $t_2$ and $t_3$ times because we do not use
the value at $t_{\rm OB}$. 
Usually we move the model light curve ``back and forth''
and ``up and down'' against the observed value
as shown in Figure \ref{v1668_cyg_vy_jhk_mag_x35z02c10o20},
and this process simply means that we fit our model light curve
with the decay phase after maximum.

    In Figure \ref{all_mass_v382_vel_v1500_cyg_x55z02o10ne03},
we add $V$ and $y$ light curves of V1500 Cyg for comparison.
From the difference between $V$ and $y$ bands, we can
roughly evaluate the influence of strong emission lines that
contribute extra flux to $V$ magnitude above free-free
continuum flux.  Comparing the light curves of V1500 Cyg,
we regard that V382 Vel light curve is contaminated by strong
emission lines from $\sim 40$ days after the outburst;
all the $V$, $R$, and $I$ magnitudes begins to diverge from
each other and start to deviate from
our universal decline law (thick solid line).

     The WD mass estimate depends weakly
on the chemical composition.  Changing hydrogen content $X$,
we obtain a best-reproducing model of $M_{\rm WD}= 1.13 ~M_\sun$
for $X=0.35$ (``Ne nova 1'') and $M_{\rm WD}= 1.28 ~M_\sun$
for $X=0.65$ (``Ne nova 3'').  Thus, we obtain the dependency
of the WD mass on $X$, i.e.,
\begin{equation}
M_{\rm WD}(X) \approx \left\{ 1.23 + 0.5 (X-0.55) \right\} ~M_\sun,
\end{equation}
for V382 Vel.  This relation on the WD mass is a bit steeper
than that for V598 Pup.   We may also conclude that
the WD mass is $M_{\rm WD}= 1.23 \pm 0.05 ~M_\sun$ for
a typical hydrogen content between $X=0.45 - 0.65$ for V382 Vel,
which corresponds to the diversity in the composition 
determination between $X= 0.66$ \citep{sho03} and 
$X= 0.47$ \citep{aug03}.

The distance modulus to V382 Vel can be calculated in the same
two ways as those in the previous subsection (V598 Pup).
First, the FF method gives
\begin{equation}
\left[ (m-M)_V \right]_{\rm FF} = m_{\rm w} - M_{\rm w} 
= 11.9 - 0.5 = 11.4,
\end{equation}
for the $1.23 ~M_\sun$ WD model.
Even if we use two other models of $1.25 ~M_\sun$ and
$1.2 ~M_\sun$ WDs in Figure 
\ref{all_mass_v382_vel_v1500_cyg_x55z02o10ne03},
this method gives a similar result of
$(m-M)_V = 11.7 - 0.4 = 11.3$ for $1.25 ~M_\sun$,  or
$(m-M)_V = 12.2 - 0.7 = 11.5$ for $1.2 ~M_\sun$ WD.

  Second, the LC method also gives us the absolute brightness, i.e.,
\begin{equation}
\left[5 \log(d/10) + A_V \right]_{\rm V1500~Cyg}
 - \left[5 \log(d/10) + A_V \right]_{\rm V382~Vel}  = 1.0.
\label{distance_v1500cyg_v382vel}
\end{equation}
With  $(m-M)_V = 12.5$ from Equation
(\ref{distance_modulus_v1500_cyg}) for V1500 Cyg,
the above Equation (\ref{distance_v1500cyg_v382vel}) yields
\begin{equation}
\left[ ( m - M )_V \right]_{\rm LC} = 
\left[ 5 \log(d/10) + A_V \right]_{\rm V382~Vel} = 11.5.
\label{distance_absorption_v382vel_lc}
\end{equation}

Kaler-Schmidt's law 
% (Equation [\ref{kaler-schmidt-law}])
with $t_3 = 12.4$ days and Della Valle \& Livio's law
% (Equation [\ref{della-valle-livio-law}])
with $t_2 = 6.6$ days
and $m_{V, {\rm max}} = 2.6$ \citep{gil99} give
\begin{equation}
\left[  \left( m - M \right)_{V, {\rm max}} \right]_{\rm MMRD}= 
\left[ 5 \log(d/10) + A_V \right]_{\rm MMRD} = 11.6 ~(11.4),
\end{equation}
where the value outside parenthesis is calculated from
Kaler-Schmidt's law of Equation (\ref{kaler-schmidt-law}) and   
the value in parenthesis is calculated from
Della Valle \& Livio's law of Equation (\ref{della-valle-livio-law}).
These two values are consistent with the FF and LC results.

These four values give a distance modulus of
$(m-M)_V = 11.5 \pm 0.1$ for these four different estimations
with equal weight and a distance of $d = 1.5 \pm 0.1$ kpc 
%%% d= 1.6, 1.5, 1.6, 1.4 kpc
for  $A_V = 3.1 \times E(B-V)=3.1 \times 0.2 = 0.62$ \citep{sho03}.

These values are consistent with the distance of
$1.66 \pm 0.11$ kpc derived from the maximum magnitude vs.
rate of decline (MMRD) relation \citep{del02}.  
On the other hand, \citet{sho03} obtained a bit larger
distance of $d = 2.5$ kpc, assuming that the UV flux of V382 Vel
is equal to that of V1974 Cyg. They took a distance of 3.1 kpc to
V1974 Cyg but this value is much larger than a reasonable one
of 1.8 kpc \citep{cho97}.
If we take the distance of 1.8 kpc instead of 3.1 kpc to V1974 Cyg,
Shore et al.'s method gives a much shorter distance of 1.5 kpc
to V382 Vel, which is consistent with our estimates.

Finally, we introduce some characteristic features concerning
binary nature, that is, the emergence of the companion
from the WD photosphere.  We define this characteristic time as
the emergence time of the companion, i.e., $t_{\rm emerge}$.
     The orbital period of $P_{\rm orb}= 0.1461$~days 
(3.51 hr) was derived by \citet{bal06} from the orbital modulations
with a full amplitude of $0.02-0.03$ mag while a bit longer orbital
period of $P_{\rm orb}= 0.1581$ days was obtained by \citet{wou05}.
So, we estimate the epoch when the companion emerges from
the nova envelope.   If the mass of the donor star can be estimated
from Warner's (1995) empirical formula, i.e.,
\begin{equation}
{{M_2} \over {M_\sun}} \approx 0.065 \left({{P_{\rm orb}}
\over {\rm hours}}\right)^{5/4},
\mbox{~for~} 1.3 < {{P_{\rm orb}} \over {\rm hours}} < 9
\label{warner_mass_formula}
\end{equation}
we have $M_2 = 0.31 ~M_\sun$, which corresponds to
the separation of $a = 1.35 ~R_\sun$ for $M_1 = 1.23 ~M_\sun$,
and the effective Roche lobe radius of the primary component
(WD) of $R_1^* =   0.68 ~R_\sun$.
%% M1  M2  P-ORB(HOUR)
%% 1.23 0.31 3.51
%% A=   1.348    RoR1=  0.6752    RoR2=  0.3616    Ro
It is about $t_{\rm emerge} \approx 32$ days
when the photospheric radius of the nova envelope shrinks
to near the orbit (when $R_{\rm ph} \approx a$).
This becomes $t_{\rm emerge} \approx 28$ days
if we take the orbit plus the companion's radius (when
$R_{\rm ph} \approx a + R_2^*$).

Hard X-ray ($2-4$ keV) flux of V382 Vel was detected 
by {\it ASCA} and {\it RXTE} \citep{muk01} about 20 days after
the outburst \citep[day 18 with {\it BeppoSAX} by ][]{ori01b}
and reached maximum at day $\sim 30$ and stayed at the same level
and then began to decrease at day $\sim 50$. 
\citet{muk01} interpreted the origin of hard X-rays 
by a model of internal shocks \citep{fri87}.
This shock may be formed by collision between two ejecta
shells as explained by \citet{muk01}, but there is
another possibility that the shock is formed by collision between
nova winds (optically thick winds) and the companion
star \citep[see, e.g.,][]{hac05k, hac06kb, hac08kc}.

If it is the second case, the hard X-ray emergence should be
coincident with the emergence of the companion from the WD
photosphere because hard X-ray is probably absorbed deep inside 
the nova photosphere.  The hard X-ray emergence time of $\sim 20$ days
is roughly consistent with $t_{\rm emerge} \approx 30$ days.
On the other hand, the hard X-ray flux should be declining 
as the wind mass-loss rate is decreasing.  This corresponds to
the break point of free-free emission light curve, because
the wind mass-loss rate is rapidly decreasing
after the break point of $t_{\rm break} = 52$ days.
Therefore, the decrease in the X-ray flux at day 50 is also
consistent with our wind model.  After $t_{\rm break}$,
the hard X-ray becomes weak but lasts about 120 days
after the outburst, in other words, until the optically
thick wind stopped at $t_{\rm wind} = 120$ days.

% Fig.18
%\placefigure{all_mass_v4743_sgr_v1500_cyg_x55z02o10ne03}

\begin{figure}
%\epsscale{1.0}
\epsscale{1.15}
\plotone{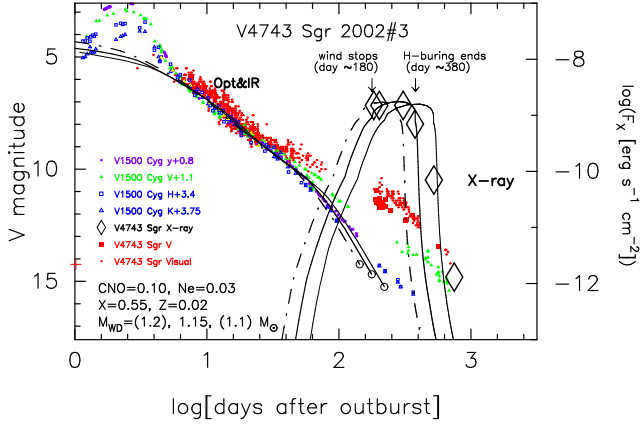}
%\plotone{f9bw.epsi}
%\plotone{all_mass_v4743_sgr_v1500_cyg_x55z02o10ne03_color.epsi}
%\plotfiddle{evolution1.ps}{5.0cm}{270}{0.4}{0.4}{-170}{220}
\caption{
Same as Figure \ref{all_mass_v598_pup_v1500_cyg_x55z02o10ne03}
but for V4743 Sgr.  
We plot three model light curves of $M_{\rm WD}= 1.2$ ({\it thick
dash-dotted lines}), $1.15$ ({\it thick solid lines}), and
$1.1 ~M_\sun$ ({\it thin solid lines}) WDs 
for the envelope chemical composition of ``Ne nova 2.''
We select the $1.15 ~M_\sun$ WD as a best model among three
of 1.2, 1.15, and $1.1 ~M_\sun$ WDs 
and can reproduce the supersoft X-ray data.
Visual ({\it small open circles})
and $V$ ({\it filled squares}) data of V4743 Sgr
are taken from IAU Circulars 7975, 7976, and 7982,
and from AAVSO.  
The X-ray flux data ({\it large open diamonds}) of V4743 Sgr 
without error bars are taken from \citet{ori03}, \citet{ori04}
and \cite{nes07a}.
}
\label{all_mass_v4743_sgr_v1500_cyg_x55z02o10ne03}
\end{figure}

\subsection{V4743 Sgr 2002\#3}
%%%\label{subsection_v4743_sgr}

V4743 Sgr (Nova Sagittarii 2002 No.3)
was discovered by Haseda on 2002 September 20.431 UT
% 2452537.5 + 0.431 = 2452537.931
% 2452537.931 - 2452533.0 = 4.931
(JD 2452537.931) at mag about 5.0 \citep{has02}.
\citet{kat02} reported a prediscovery magnitude of mag 5.5 on
September 18.465 UT.
This object was not detected by Brown \citep{bro02}
on September 9.6 UT (limiting mag about 12.0), so we here
assume that $t_{\rm OB}=$~JD 2452533.0 (2002 September 15.5 UT)
is the outburst day.

Bright supersoft X-ray phases were reported 
at day $\sim 180$ (2003 March 19.396 UT) by \citet{sta03}
\citep[see also][]{nes03}
% 2452717.896 - 2452533.0 = 184.896
and at day $\sim 200$ 
% 2452734.421 - 2452533.0 = 201.421
(2003 April 4.921 UT) by \citet{ori03}.
\citet{ori04} reported that the supersoft X-ray flux
% 2453278.5 + 0.783 = 2453279.283
% 2453279.283 - 2452533.0 = 746.283
has already declined on 2004 September 30.783 UT (day $\sim 750$)
by a factor of 1000.  
Figure \ref{all_mass_v4743_sgr_v1500_cyg_x55z02o10ne03}
shows the optical and supersoft X-ray light curves.
The supersoft X-ray phase started at least $\sim 180$ days
after the outburst and ended at day $\sim 400$.

     The model of $1.15 ~M_\sun$ WD is a best one among three
1.2, 1.15, and $1.1 ~M_\sun$ WDs, as shown in 
Figure \ref{all_mass_v4743_sgr_v1500_cyg_x55z02o10ne03}.
Here we assume a chemical composition of typical neon novae,
``Ne nova 2.''  The $1.15 ~M_\sun$ WD model has
$t_{\rm break} = 70$, $t_{\rm wind} = 180$, and
$t_{\rm H-burning} = 380$ days. 
Adopting a magnitude of 5.0 on day 3.6 along our model light
curve of $1.15 ~M_\sun$ WD, we obtain $t_2 = 9.4$ days and
$t_3 = 17$ days as listed in Table \ref{fitting_t2_t3_tb_txonoff}.
We have estimated the ejecta mass of
$M_{\rm wind} \sim 7 \times 10^{-6} M_\sun$.
The visual ($V$) magnitudes are roughly fitted with our model
light curve until about day 40,
but are gradually departing from it after that.  At this stage,
the nova probably entered the transition/nebular phase and
the deviation comes from the contribution of strong emission lines.

     We have also checked the accuracy of WD mass determination that
comes from the ambiguity of chemical composition.
We obtain a best-reproducing model of $M_{\rm WD}= 1.03 ~M_\sun$ for 
$X=0.35$ (``Ne nova 1''),  and $M_{\rm WD}= 1.21 ~M_\sun$ for 
$X=0.65$ (``Ne nova 3'').  Then, we have an approximate relation of
\begin{equation}
M_{\rm WD}(X) \approx \left\{ 1.15 + 0.6 (X-0.55) \right\}  ~M_\sun,
\end{equation}
for V4743 Sgr.  This relation of the WD mass is steeper (on $X$)
than that for V598 Pup and V382 Vel.   We may also conclude that
the WD mass is $M_{\rm WD}= 1.15 \pm 0.06 ~M_\sun$ for
a typical hydrogen content between $X=0.45 - 0.65$ of V4743 Sgr.

From the FF method, we obtain the distance modulus to V4743 Sgr, i.e.,
\begin{equation}
\left[ (m-M)_V \right]_{\rm FF} = m_{\rm w} - M_{\rm w} 
= 14.7 - 0.9 = 13.8,
\end{equation}
for the $1.15 ~M_\sun$ WD.
Two other attendant models of $1.2 ~M_\sun$ and
$1.1 ~M_\sun$ WDs in Figure 
\ref{all_mass_v4743_sgr_v1500_cyg_x55z02o10ne03}
give similar results of
$(m-M)_V = 14.3 - 0.7 = 13.6$ for $1.2 ~M_\sun$ or
$(m-M)_V = 15.2 - 1.2 = 14.0$ for $1.1 ~M_\sun$ WD.

      The LC method gives 
\begin{equation}
\left[5 \log(d/10) + A_V \right]_{\rm V1500~Cyg}
 - \left[5 \log(d/10) + A_V \right]_{\rm V4743~Sgr}  = -1.1.
\end{equation}
With  $(m-M)_V = 12.5$ from Equation
(\ref{distance_modulus_v1500_cyg}) for V1500 Cyg, we obtain
\begin{equation}
\left[ \left( m - M \right)_V \right]_{\rm LC}= 
\left[ 5 \log(d/10) + A_V \right]_{\rm V4743~Sgr} = 13.6.
\end{equation}

Kaler-Schmidt's law 
% (Equation [\ref{kaler-schmidt-law}])
with $t_3 = 17$ days and Della Valle \& Livio's law
% (Equation [\ref{della-valle-livio-law}])
with $t_2 = 9.4$ days
and $m_{V, {\rm max}} = 5.0$ yield
\begin{equation}
\left[ \left( m - M \right)_{V, {\rm max}} \right]_{\rm MMRD}=
\left[ 5 \log(d/10) + A_V \right]_{\rm MMRD} = 13.7 ~(13.7),
\end{equation}
being consistent with the FF and LC results.
Then, the distance modulus to V4743 Sgr is 
$(m-M)_V = 13.7 \pm 0.1$ and the distance is
%%% d = 4.4, 3.7, 3.8, 3.8 kpc
 $d = 3.8 \pm 0.2$ kpc for $A_V = 3.1 E(B-V) = 3.1 \times 0.25 = 0.78$
\citep{van07}.
Our estimate is also consistent with the recent results 
of $E(B-V)= 0.25$ and $d = 3.9 \pm 0.3$ kpc by \citet{van07}.

     The orbital period of $P_{\rm orb}= 0.2799$~days 
(6.72 hr) was derived by \citet{kan06} from the orbital modulations
with a full amplitude of 0.05 mag (2003) -- 0.15 mag (2005).  Then,
we estimate the epoch when the companion emerges from
the nova envelope.  From Equation (\ref{warner_mass_formula}),
we have $M_2 = 0.70 ~M_\sun$, which corresponds to
$a = 2.2 ~R_\sun$ and $R_1^* =   0.93 ~R_\sun$ for $M_1 = 1.15 ~M_\sun$.
%% M1  M2  P-ORB(HOUR)
%% 1.15 0.7 6.72
%% A=   2.210    RoR1=  0.9338    RoR2=  0.7445    Ro
It is $t_{\rm emerge} \approx 36$ days
at $R_{\rm ph} \sim a$.

% Fig.19
%\placefigure{all_mass_v1281_sco_v1500_cyg_x55z02o10ne03}

\begin{figure}
%\epsscale{1.0}
\epsscale{1.15}
\plotone{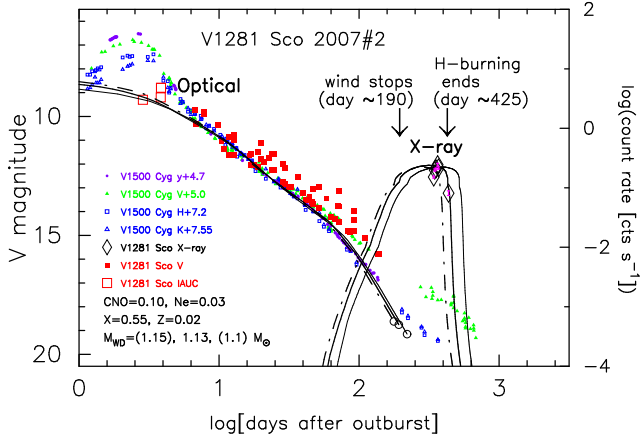}
%\plotone{f10bw.epsi}
%\plotone{all_mass_v1281_sco_v1500_cyg_x55z02o10ne03_color.epsi}
%\plotfiddle{evolution1.ps}{5.0cm}{270}{0.4}{0.4}{-170}{220}
\caption{
Same as Figure \ref{all_mass_v598_pup_v1500_cyg_x55z02o10ne03}
but for V1281 Sco.
Observational data of optical $V$ ({\it red filled squares})
are taken from AAVSO and Association Francaise des Observateurs
d'Etoiles Variables (AFOEV).
The X-ray count rate data ({\it large open diamonds})
with error bars ({\it magenta bars}) of V1281 Sco are taken from 
an automatic analyzer of the {\it Swift} web page \citep{eva09}.
We plot three model light curves of $M_{\rm WD}= 1.15$ ({\it thick
dash-dotted lines}), $1.13$ ({\it thick solid lines}), and
$1.1 ~M_\sun$ ({\it thin solid lines}) WDs 
for the envelope chemical composition of ``Ne nova 2.''
We select the $1.13 ~M_\sun$ WD as a best model among three
of 1.15, 1.13, and $1.1 ~M_\sun$ WDs and can reproduce
the supersoft X-ray data.
}
\label{all_mass_v1281_sco_v1500_cyg_x55z02o10ne03}
\end{figure}

\subsection{V1281 Sco 2007\#2}
%%%\label{subsection_v1281_sco}

V1281 Sco (Nova Scorpii 2007 No.2) was discovered by Nakamura
on 2007 February 19.86 UT
% 2454150.5 + 0.8593 = 2454151.3593
% 2454151.3593 - 2454148.5 = 2.8593
(JD 2454151.3593) at mag about 9.3 \citep{yam07}.
%\citet{yam07} reported a prediscovery magnitude of mag 5.5 on
%September 18.465 UT.
This object was not detected by Nakamura \citep{yam07}
on February 14.86 UT (limiting mag about 12.0) and also by Fujita
\citep{nai07} on February 18.85 UT (limiting mag about 11.6), so
we here assume that $t_{\rm OB}=$~JD 2454148.5
(2007 February 17.0 UT) is the outburst day.

A bright supersoft X-ray phase was reported 
at day $\sim 340$ (2008 January 24.18 UT) by \citet{nes08a}.
% 2454489.68 - 2454148.5 = 341.18
To search for the end epoch of supersoft X-ray phase, i.e.,
$t_{\rm H-burning}$, we use an automatic analyzer in the
{\it Swift} web page\footnote{http://www.swift.ac.uk/}
\citep{eva09}.  The X-ray (0.3--10 keV) count rates with the
{\it Swift} XRT are plotted
in Figure \ref{all_mass_v1281_sco_v1500_cyg_x55z02o10ne03}
together with its optical light curve.
We have checked the early data points of X-ray obtained by
automatic analyzer are very similar to the values reported
by \citet{nes08a}.  The epoch of $t_{\rm H-burning}$ is about 425 
days after the outburst, if we regard that the last observation
indicates the decay of supersoft X-ray flux.

     Our best-reproducing light curve of a $1.13 ~M_\sun$ WD is plotted in 
Figure \ref{all_mass_v1281_sco_v1500_cyg_x55z02o10ne03} for
a chemical composition of a typical neon nova, ``Ne nova 2'', which
shows $t_{\rm break} = 78$, $t_{\rm wind} = 190$, and
$t_{\rm H-burning} = 425$ days after the outburst. 
We determined the WD mass to be $1.13 \pm 0.03 ~M_\sun$ mainly from
the supersoft X-ray data.
Adopting a magnitude of 9.2 on day 3.0 along our model light
curve of $1.13 ~M_\sun$ WD, we obtain $t_2 = 9.5$ and
$t_3 = 16.7$ days as listed in Table \ref{fitting_t2_t3_tb_txonoff}.
We have estimated the ejecta mass of
$M_{\rm wind} \sim 9 \times 10^{-6} M_\sun$.
The visual ($V$) magnitudes are roughly fitted with our model
light curve until about day 80,
but are gradually departing from it after that.  
This deviation comes from the contribution of strong emission lines.

     We have also examined the accuracy of WD mass determination
that comes from the ambiguity of chemical composition.
We obtain a best-reproducing model of $M_{\rm WD}= 1.01 ~M_\sun$ for 
$X=0.35$ (``Ne nova 1''),  and $M_{\rm WD}= 1.2 ~M_\sun$ for 
$X=0.65$ (``Ne nova 3'').  Then, we have 
\begin{equation}
M_{\rm WD}(X) \approx \left\{ 1.13 + 0.6 (X-0.55) \right\} ~M_\sun,
\end{equation}
for V1281 Sco.  We may conclude that
the WD mass is $M_{\rm WD}= 1.13 \pm 0.06 ~M_\sun$ for
a typical hydrogen content between $X=0.45$ -- 0.65 of V1281 Sco.

From the FF method, we obtain the distance modulus to V1281 Sco, i.e.,
\begin{equation}
\left[ (m-M)_V \right]_{\rm FF} = m_{\rm w} - M_{\rm w} 
= 18.8 - 1.0 = 17.8,
\end{equation}
for the $1.13 ~M_\sun$ WD model.  Two other attendant models
of $1.15 ~M_\sun$ and $1.1 ~M_\sun$ WDs in Figure 
\ref{all_mass_v1281_sco_v1500_cyg_x55z02o10ne03} 
give similar results of
$(m-M)_V = 18.6 - 0.9 = 17.7$ for $1.15 ~M_\sun$ or
$(m-M)_V = 19.1 - 1.2 = 17.9$ for $1.1 ~M_\sun$ WD.

      The LC method gives 
\begin{equation}
\left[5 \log(d/10) + A_V \right]_{\rm V1500~Cyg}
 - \left[5 \log(d/10) + A_V \right]_{\rm V1281~Sco}  = -5.0.
\end{equation}
With  $(m-M)_V = 12.5$ from Equation
(\ref{distance_modulus_v1500_cyg}) for V1500 Cyg,
we obtain
\begin{equation}
\left[ ( m - M )_V \right]_{\rm LC}= 
\left[ 5 \log(d/10) + A_V \right]_{\rm V1281~Sco} = 17.5.
\end{equation}

Kaler-Schmidt's law 
% (Equation [\ref{kaler-schmidt-law}])
with $t_3 = 16.7$ days and Della Valle \& Livio's law
% (Equation [\ref{della-valle-livio-law}])
with $t_2 = 9.5$ days
and $m_{V, {\rm max}} = 9.2$ (at $t_0 = 3.0$ days) give
\begin{equation}
\left[ \left( m - M \right)_{V, {\rm max}} \right]_{\rm MMRD}=
\left[ 5 \log(d/10) + A_V \right]_{\rm MMRD} = 17.9 ~(17.9),
\end{equation}
being consistent with the FF and LC results, where the difference
between the LC method and the others is about $\pm ~0.4$ mag 
but within the ambiguity in this method (see Sections 
\ref{prediction_by_t3} and \ref{v598_pup_model}).
Then the distance modulus is $(m-M)_V = 17.8 \pm 0.2$
and the distance to V1281 Sco is $d = 13 \pm 1$ kpc 
%%% d= 13.4, 11.6, 14, 14 kpc
for $A_V = 3.1 E(B-V)= 3.1 \times 0.7 = 2.17$ \citep{rus07a}.

% Fig.20
%\placefigure{all_mass_v597_pup_v1500_cyg_x35z02c10o20}

\begin{figure}
%\epsscale{1.0}
\epsscale{1.15}
\plotone{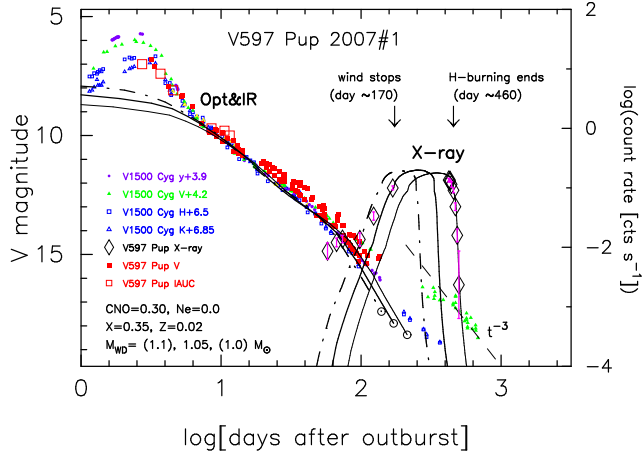}
%\plotone{f11bw.epsi}
%\plotone{all_mass_v597_pup_v1500_cyg_x35z02c10o20_color.epsi}
%\plotfiddle{evolution1.ps}{5.0cm}{270}{0.4}{0.4}{-170}{220}
\caption{
Same as Figure \ref{all_mass_v598_pup_v1500_cyg_x55z02o10ne03},
but for V597 Pup.
We plot three model light curves of $M_{\rm WD}= 1.1$ ({\it thick
dash-dotted lines}), $1.05$ ({\it thick solid lines}), and
$1.0 ~M_\sun$ ({\it thin solid lines}) WDs 
for the envelope chemical composition of ``CO nova 2.''
We do not select the best model among these three
of 1.1, 1.05, and $1.0 ~M_\sun$ WDs, because the supersoft
X-ray duration is too short to be comparable with the
observation.  The X-ray count rate data
($0.3-10$ keV: {\it large open diamonds}) with error bars
 ({\it magenta bars}) are taken from 
an automatic analyzer of the {\it Swift} web page \citep{eva09}.
Observational V597 Pup data of $V$ ({\it filled squares})
magnitudes are taken from AFOEV and AAVSO.
}
\label{all_mass_v597_pup_v1500_cyg_x35z02c10o20}
\end{figure}

% Fig.21
%\placefigure{all_mass_v597_pup_v1500_cyg_x65z02o03ne03}

\begin{figure}
%\epsscale{1.0}
\epsscale{1.15}
\plotone{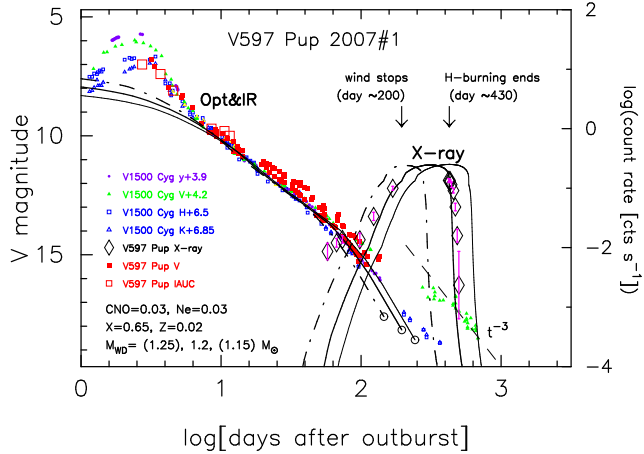}
%\plotone{f11bw.epsi}
%\plotone{all_mass_v597_pup_v1500_cyg_x65z02o03ne03_color.epsi}
%\plotfiddle{evolution1.ps}{5.0cm}{270}{0.4}{0.4}{-170}{220}
\caption{
Same as Figure \ref{all_mass_v597_pup_v1500_cyg_x35z02c10o20},
but we plot three model light curves of $M_{\rm WD}= 1.25$ ({\it thick
dash-dotted lines}), $1.2$ ({\it thick solid lines}), and
$1.15 ~M_\sun$ ({\it thin solid lines}) WDs 
for the envelope chemical composition of ``Ne nova 3.''
We select the $1.2 ~M_\sun$ WD as a best model among three
of 1.25, 1.2, and $1.15 ~M_\sun$ WDs and can reproduce
the supersoft X-ray data.
}
\label{all_mass_v597_pup_v1500_cyg_x65z02o03ne03}
\end{figure}

\subsection{V597 Pup 2007\#1}
%%%\label{subsection_v597_pup}

V597 Pup (Nova Puppis 2007 No.1) 
was discovered by Pereira on 2007 November 14.23 UT
% 2454418.5 + 0.23 = 2454418.73
% 2454418.73 - 2454416.0 = 2.73
(JD 2454418.73) at mag about 7.0 \citep{per07}.
%%\citet{kat02} reported a prediscovery magnitude of mag 5.5 on
%%September 18.465 UT.
This object was not detected \citep{per07}
on November 6.23, 7.22, 8.23, 10.23,
and 11.22 UT (limiting mag about 8), so we 
assume that $t_{\rm OB}=$~JD 2454416.0 (2007 November 11.5 UT)
is the outburst day.

A bright supersoft X-ray phase was reported by \citet{nes08c}.
% Jan. 8.02 and 17.18 UT, 2008
% 2454473.52 - 2454416.0 = 184.896
To search for the end epoch of supersoft X-ray phase, i.e.,
$t_{\rm H-burning}$, we use an automatic analyzer in the
{\it Swift} web page \citep{eva09}.  The X-ray (0.3 -- 10 keV)
count rates with the {\it Swift} XRT are plotted
in Figures \ref{all_mass_v597_pup_v1500_cyg_x35z02c10o20}
and \ref{all_mass_v597_pup_v1500_cyg_x65z02o03ne03} as well as
the optical light curves.
The X-ray data points show that the turnoff of supersoft
X-ray is $t_{\rm H-burning} \sim 430$ -- 460 days and 
the turn-on is probably $t_{\rm wind} \sim 150$ -- 170 days.
This is a rather rare case in which both the
turn-on and turnoff times of supersoft X-ray are specified. 

     The chemical composition of ejecta is not known,
so we first assumed a chemical composition of ``CO nova 2'' 
and made model light curves as shown in Figure 
\ref{all_mass_v597_pup_v1500_cyg_x35z02c10o20}.
The durations of supersoft X-ray phases are too short to be
comparable with the observation.  This is because the duration of
supersoft X-ray is shorter for a smaller hydrogen content of $X$.   

     Our reasonably-reproducing model light curve of a $1.2 ~M_\sun$ WD is
plotted in Figure \ref{all_mass_v597_pup_v1500_cyg_x65z02o03ne03}
for a chemical composition of ``Ne nova 3'', which
shows $t_{\rm break} = 74$, $t_{\rm wind} = 200$, and
$t_{\rm H-burning} = 430$ days after the outburst. 
We estimated the WD mass to be $1.2 \pm 0.05 ~M_\sun$ mainly from
the supersoft X-ray data.
Adopting a magnitude of 8.4 on day 2.7 as the maximum magnitude
along our model light curve of $1.2 ~M_\sun$ WD, we obtain
$t_2 = 8.7$ and $t_3 = 16.5$ days as listed in
Table \ref{fitting_t2_t3_tb_txonoff}.
We have estimated the ejecta mass of
$M_{\rm wind} \sim 1 \times 10^{-5} M_\sun$.
The optical light curve shape around the peak deviates largely from
our model light curves and is similar to that of V1500 Cyg. 
It is highly likely that V597 Pup is a superbright nova.
After the superbright phase, $V$ magnitudes are roughly
fitted with our model light curve until day $\sim 100$.

     We have also checked the accuracy of WD mass determination
from the ambiguity of chemical composition.
For chemical compositions of ``Ne nova 1'' and ``Ne nova 2,''
however, we could not obtain reasonable fits like those
as shown in Figure \ref{all_mass_v597_pup_v1500_cyg_x65z02o03ne03}
because supersoft X-ray durations of these models are too short to
fit with the supersoft X-ray observation like in Figure
\ref{all_mass_v597_pup_v1500_cyg_x35z02c10o20}. 
Therefore, we may conclude that the WD mass is
$M_{\rm WD}= 1.2 \pm 0.05 ~M_\sun$ and that the composition is
relatively hydrogen-rich like $X \sim 0.65$.

From the FF method, we obtain the distance modulus to V597 Pup, i.e.,
\begin{equation}
\left[ (m-M)_V \right]_{\rm FF} = m_{\rm w} - M_{\rm w} 
= 17.8 - 0.9 = 16.9,
\end{equation}
for the $1.15 ~M_\sun$ WD (``Ne nova 2'')
model.  Two other attendant models of $1.2 ~M_\sun$ and
$1.1 ~M_\sun$ WDs (not shown in the figure for ``Ne nova 2'' models),
give similar results of
$(m-M)_V = 17.5 - 0.7 = 16.8$ for $1.2 ~M_\sun$ or
$(m-M)_V = 18.3 - 1.2 = 17.1$ for $1.1 ~M_\sun$ WD.
From Figure \ref{all_mass_v597_pup_v1500_cyg_x35z02c10o20},
we also have 
$(m-M)_V = 17.3 - 0.5 = 16.8$ for $1.1 ~M_\sun$,
$(m-M)_V = 17.8 - 0.8 = 17.0$ for $1.05 ~M_\sun$, or
$(m-M)_V = 18.3 - 1.1 = 17.2$ for $1.0 ~M_\sun$ WD
with ``CO nova 2.''

      The LC method gives 
\begin{equation}
\left[5 \log(d/10) + A_V \right]_{\rm V1500~Cyg}
 - \left[5 \log(d/10) + A_V \right]_{\rm V597~Pup}  = -4.2.
\end{equation}
With  $(m-M)_V = 12.5$ from Equation
(\ref{distance_modulus_v1500_cyg}) for V1500 Cyg,
we obtain
\begin{equation}
\left[ ( m - M )_V \right]_{\rm LC}= 
\left[ 5 \log(d/10) + A_V \right]_{\rm V597~Pup} = 16.7.
\end{equation}

Kaler-Schmidt's law
% (Equation [\ref{kaler-schmidt-law}])
with $t_3 = 16.5$ days and Della Valle \& Livio's law
% (Equation [\ref{della-valle-livio-law}])
with $t_2 = 8.7$ days, and both with $m_{V, {\rm max}} = 8.4$
(at $t_0 = 2.7$ days) yield
\begin{equation}
\left[ \left( m - M \right)_{V, {\rm max}} \right]_{\rm MMRD}=
\left[ 5 \log(d/10) + A_V \right]_{\rm MMRD} = 17.1 ~(17.1),
\end{equation}
being consistent with the FF and LC results.
Then we have $(m-M)_V = 17.0 \pm 0.2$ 
%%% d=17.1, 14.3, 17.1, 17.1 kpc
and the distance to V597 Pup, $d = 16 \pm 2$ kpc,
for $A_V = 3.1 E(B-V) = 3.1 \times 0.3 = 0.93$ \citep{nes08c}.

     The orbital period of $P_{\rm orb}= 0.1112$~days 
(2.6687 hr) was derived by \citet{war09} from the orbital modulations
with a full amplitude of 0.2 mag (2008) -- 0.6 mag (2009).  Then,
we estimate the epoch when the companion emerges from
the nova envelope.  From Equation (\ref{warner_mass_formula}),
we have $M_2 = 0.22 ~M_\sun$, which corresponds to
$a = 1.1 ~R_\sun$ and $R_1^* =   0.58 ~R_\sun$ for $M_1 = 1.2 ~M_\sun$.
%% M1  M2  P-ORB(HOUR)
%% 1.2 0.2217 2.6687
%% A=   1.094    RoR1=  0.5770    RoR2=  0.2694    Ro
It is $t_{\rm emerge} \approx 52$ days
at $R_{\rm ph} \sim a$.

     \citet{war09} reported that a repetitive hump with an amplitude
of $\sim 0.2$ mag is seen but no obvious eclipse features in 2008
($\sim 125$ days after outburst).  Their Fourier
analysis of light curves provide an orbital period of 2.67 hr.
After that,
their 2009 observation revealed broad eclipses with $\sim 0.6$ mag
deep ($\sim 470 - 500$ days after outburst).
They interpreted that the drop in brightness (2.5 mag) between
these two observations in 2008 and 2009
was due largely or entirely to the dispersal and cooling of nova
ejecta far from the central binary and the eclipse and their 
related humps could well have been present even in the earlier
2008 light curves, that is, there are some
features resembling such diluted structures.
This interpretation is consistent with the emergence of 
companion on $\sim 50$ ($R_{\rm ph} = a$) or $\sim 60$ days
($R_{\rm ph} = a - R_2$) after outburst.  Moreover,
the WD photosphere is larger than that of the companion, i.e.,
$R_{\rm ph} \gtrsim R_2$, during the optically thick wind phase,
at least, until 130 days after outburst.  After the winds stop
on day $\sim 200$, the WD photosphere rapidly shrinks toward 
its original size, that is, it becomes a half by one month
\citep[see, e.g.,][]{hac03kb}.

% Fig.22
%\placefigure{light_curve_v1494_aql_v1974_cyg_x55z02o10ne03}

\begin{figure}
%\epsscale{1.0}
\epsscale{1.15}
\plotone{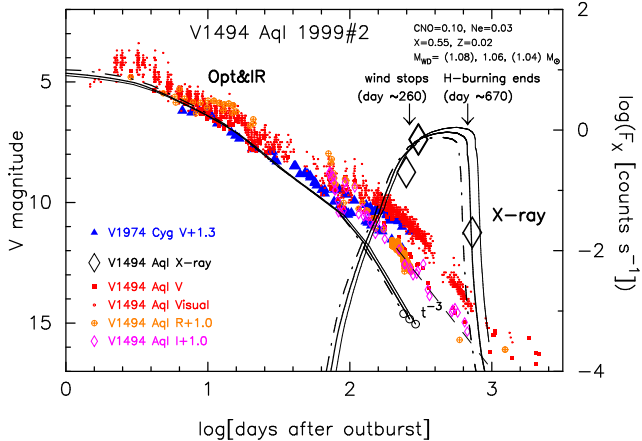}
%\plotone{f11bw.epsi}
%\plotone{light_curve_v1494_aql_v1974_cyg_x55z02o10ne03_color.epsi}
%\plotfiddle{evolution1.ps}{5.0cm}{270}{0.4}{0.4}{-170}{220}
\caption{
Optical and supersoft X-ray light curves of V1494 Aql.
Optical bands of visual ({\it small open circles}),
$V$ ({\it filled squares}), $R$ ({\it circles with a plus}), 
and  near IR band of $I$ ({\it small open diamonds})
magnitudes are shown, which are taken from AAVSO and VSNET.
The X-ray count rate data ({\it large open diamonds}) without error bars
are taken from \citet{roh09}.
We plot three model light curves of $M_{\rm WD}= 1.08$ ({\it thick
dash-dotted lines}), $1.06$ ({\it thick solid lines}), and
$1.04 ~M_\sun$ ({\it thin solid lines}) WDs 
for the envelope chemical composition of ``Ne nova 2.''
We select the $1.06 ~M_\sun$ WD as a best model among three
of 1.08, 1.06, and $1.04 ~M_\sun$ WDs and can reproduce
the supersoft X-ray data.
Observational $V$ magnitudes of V1974 Cyg are added for 
comparison, which are taken from \citet{cho93}.
}
\label{light_curve_v1494_aql_v1974_cyg_x55z02o10ne03}
\end{figure}

\subsection{V1494 Aql 1999\#2}
%%%\label{subsection_v1494_aql}

V1494 Aql (Nova Aquilae 1999 No.2)
was discovered by Pereira on 1999 December 1.785 UT
% 2451513.5 + 0.785 = 2451514.285
% 2451514.285 - 2451513.0 = 1.285
(JD 2451514.285) at mag about 6.0 \citep{per99}.
%Liller reported a prediscovery magnitude of mag 5.5 on
%September 18.465 UT.  
This object was not detected by Liller \citep{fuj99}
on November 25.035 UT (limiting mag about 10.5), so we 
assume that $t_{\rm OB}=$~JD 2451513.0 (1999 November 30.5  UT)
is the outburst day.
The star reached its maximum of $m_{\rm vis} \approx 3.9$ on
1999 December 3.4 UT (JD 2451515.9) about 3 days after the
outburst.  Then it declined by 2 mag in 6.6 days and by 3 mag
in 16 days \citep[e.g.,][]{kis00}.  The transition oscillation
began on day $\sim 20$ and ended on day $\sim 130$, during which
the star exhibited quasi-periodic oscillations with periods of
7 to 22 days and with an amplitude reaching 1.5 mag.
%%% (Fig. \ref{light_curve_v1494_aql_v1974_cyg_x55z02o10ne03}).

We plot the optical and supersoft X-ray light curves in
Figure \ref{light_curve_v1494_aql_v1974_cyg_x55z02o10ne03}.
A bright supersoft X-ray phase was detected with
{\it Chandra} on day 250 (2000 August 6 UT),
% 2451762.5 - 2451513.0 = 249.5
day 302 (2000 September 28 UT),
% 2451815.5 - 2451513.0 = 302.5
and day 305 (2000 October 1 UT)
% 2451818.5 - 2451513.0 = 305.5
\citep{dra03}.
\citet{nes07a} reported that the supersoft X-ray flux
% 2452241.5 - 2451513.0 = 728.5
had already declined by a factor of 40 on day $\sim 730$
(2001 November 28 UT).  Then the supersoft X-ray phase started
at least on day $\sim 250$ and ended before day $\sim 700$.

\citet{iij03} analyzed their spectra of V1494 Aql and estimated
the helium/hydrogen ratio to be He/H$= 0.13 \pm 0.01$ by number
ratio, which corresponds to $Y/X = 0.52 \pm 0.04$.
This helium/hydrogen ratio is close to our chemical composition
of ``Ne nova 2,''  in which $Y/X \approx 0.54$.  \citet{ark02}
derived He/H$=0.126$, N/H$= 0.04$ (or 0.0015), O/H$= 0.002$, and
Fe/H$= 5 \times 10^{-5}$ by number ratio.  This corresponds
to $X= 0.48$, $Y= 0.24$, $X_{\rm CNO}= 0.26$, and $Z= 0.02$,
(or to $X= 0.62$, $Y= 0.34$, $X_{\rm CNO}= 0.02$, and $Z= 0.02$).
We do not know which is better, so that we adopt an arithmetic
average of these two, i.e., $X= 0.55$, $Y= 0.29$, $X_{\rm CNO}= 0.14$,
and $Z= 0.02$, which is close to ``Ne nova 2.''  Therefore, we
assume ``Ne nova 2''as the envelope chemical composition of V1494 Aql.

     Our best-reproducing light curve is a $1.06 ~M_\sun$ WD model 
among 1.04, 1.06, and 1.08 $M_\sun$ WDs mainly from
the supersoft X-ray data
(Figure \ref{light_curve_v1494_aql_v1974_cyg_x55z02o10ne03}), i.e.,
%$X= 0.55$, $Y= 0.30$, $X_{\rm CNO}= 0.10$,
%$X_{\rm Ne}= 0.03$, and $Z=0.02$ 
$M_{\rm WD}= 1.06 \pm 0.02~M_\sun$.  For the $1.06 ~M_\sun$ WD model,
we derive $t_{\rm break} = 102$, $t_{\rm wind} = 260$, and
$t_{\rm H-burning} = 670$ days. 
Adopting a magnitude of $m_{V, {\rm max}}= 5.0$ on day 2.7
as the optical maximum along our model light
curve of $1.06 ~M_\sun$ WD, we obtain $t_2 = 11.7$ days and
$t_3 = 22.1$ days (Table \ref{fitting_t2_t3_tb_txonoff}).
The visual ($V$) magnitudes are not well fitted with our model
light curve, from day 10 to day 150, mainly because the light
curve has wavy structures and seems to be strongly contaminated
by emission lines.  Our model light curve almost follows 
the bottom line of $V$ magnitudes during the transition
oscillations, which is supported by Figure
\ref{all_mass_gk_per_v1500_cyg_x55z02o10ne03} for GK Per
\citep[see also Figure 2 of][]{hac07k}. 
The nova entered a nebular phase $\sim 100$ days after the
outburst and the visual magnitudes deviated much from 
our model light curve due mainly to strong emission lines
such as [\ion{O}{3}] \citep[e.g.,][]{iij03}.

We have also estimated the ejecta mass to be
$M_{\rm wind} \sim 1.1 \times 10^{-5} M_\sun$ (Table
\ref{physical_properties_of_novae}),
which is lost by the optically thick wind \citep[see,
e.g.,][for more detail]{kat94h, hac06kb}.
\citet{iij03} derived the ejecta mass of 
$M_{\rm ej}\sim (6.2 \pm 1.4) \times 10^{-6} 
 M_\sun (d/1.6{\rm ~kpc})^2$, being consistent with our
model estimation if we take the distance of $d = 2.2$ kpc
as derived below.

Based on submillimeter- and centimeter-band fluxes,
a single MERLIN image, and optical spectroscopy, \citet{eyr05}
concluded that optical spectroscopy indicates continued
mass ejection for over 195 days, with the material becoming
optically thin sometime between 195 and 285 days
after the outburst.  This is consistent with 
$t_{\rm wind}= 260$ days of our $1.06 ~M_\sun$ WD model.

\citet{maz00} reported that V1494 Aql had entered into 
a strong coronal phase at least 230 days after the outburst.
This is closely related to the epoch when the optically thick
winds stop, because the photosphere rapidly shrinks and the
photospheric temperature quickly increases to
$\sim 3 \times 10^{5}$~K or more.

     We have checked the dependency of the WD mass determination
on the hydrogen content of the WD envelope.
\citet{kam05} reported a different result of helium/hydrogen ratio
of the ejecta from \ion{He}{2}~4686/H$\beta$ ratio,
that is, He/H$= 0.24 \pm 0.06$ by number ratio,
which corresponds to $Y/X = 0.96 \pm 0.24$ and is close to
our ``CO nova 2'' (see Table \ref{chemical_composition}).
When we adopt a chemical composition of ``CO nova 2,''
we obtain a best-reproducing model of $0.92 ~M_\sun$ WD.  The other physical
parameters are also listed in Tables \ref{fitting_t2_t3_tb_txonoff}
and \ref{physical_properties_of_novae}.
On the other hand, if we assume a chemical composition of
``Ne nova 3,'' we obtain a best model of $1.13 ~M_\sun$ WD.
Finally we have 
\begin{equation}
M_{\rm WD}(X) \approx \left\{ 1.06 + 0.7 (X-0.55) \right\} ~M_\sun,
\end{equation}
for V1494 Aql.  Here, we use $M_{\rm WD}(X=0.65) = 1.13 ~M_\sun$,
$M_{\rm WD}(X=0.55) = 1.06 ~M_\sun$,
and $M_{\rm WD}(X=0.35) = 0.92 ~M_\sun$.  Then, we may conclude that
the WD mass is $M_{\rm WD}= 1.06 \pm 0.07 ~M_\sun$ for
a typical hydrogen content between $X=0.45$ -- 0.65.
If the hydrogen content is $X \sim 0.55$, then the WD has a mass of
$1.06 \pm 0.03 ~M_\sun$ and is probably a ONeMg WD.  On the other
hand, when the hydrogen content is as low as $X \sim 0.35$,
then the WD is less massive ($0.92 \pm 0.03 ~M_\sun$) and
probably a CO WD.

From the FF method, we obtain the distance modulus to V1494 Aql, i.e.,
\begin{equation}
\left[ (m-M)_V \right]_{\rm FF} = m_{\rm w} - M_{\rm w} 
= 14.9 - 1.5 = 13.4,
\end{equation}
for the $1.06 ~M_\sun$ WD (``Ne nova 2'') model.
Two other attendant models of $1.08 ~M_\sun$ and
$1.04 ~M_\sun$ WDs give similar results of
$(m-M)_V = 14.6 - 1.1 = 13.5$ for $1.08 ~M_\sun$ or
$(m-M)_V = 15.0 - 1.7 = 13.3$ for $1.04 ~M_\sun$ WD.

      The LC method gives 
\begin{equation}
\left[5 \log(d/10) + A_V \right]_{\rm V1974~Cyg}
 - \left[5 \log(d/10) + A_V \right]_{\rm V1494~Aql}  = -1.3.
\end{equation}
With $(m-V)_V = 12.3$ from Equation (\ref{distance_modulus_v1974_cyg})
for V1974 Cyg, we obtain 
\begin{equation}
\left[ ( m - M )_V \right]_{\rm LC}= 
\left[5 \log(d/10) + A_V \right]_{\rm V1494~Aql}  = 13.6.
\end{equation}

Kaler-Schmidt's law
% (Equation [\ref{kaler-schmidt-law}])
with $t_3 = 22.1$ days and Della Valle \& Livio's law
% (Equation [\ref{della-valle-livio-law}])
with $t_2 = 11.7$ days
and $m_{V,{\rm max}} = 5.0$ (at $t_0 = 2.7$ days) yield
\begin{equation}
\left[ \left( m - M \right)_{V, {\rm max}} \right]_{\rm MMRD}=
\left[ 5 \log(d/10) + A_V \right]_{\rm MMRD}= 13.4 ~(13.6),
\end{equation}
being consistent with the FF and LC results.
Then, we obtain the distance modulus of $(m-M)_V = 13.5 \pm 0.1$ and
%%% d= 2.06, 2.26, 2.06, 2.26 kpc
the distance of $d = 2.2 \pm 0.2$ kpc to V1494 Aql  
for $A_V = 3.1 E(B-V) = 3.1 \times 0.6 = 1.83$ \citep{iij03}. 
This value is a bit larger than the distance of
$1.6 \pm 0.2$ kpc estimated by \citet{iij03} from the MMRD
relation.

     The orbital period of $P_{\rm orb}= 0.1346$~days 
(3.23 hr) was derived by \citet{bar03} from the orbital modulations
with an eclipse depth of $\sim 0.5$ mag \citep[see also][]{ret00,
bos01}.  Then,
we estimate the epoch when the companion emerges from
the nova envelope.  From Equation (\ref{warner_mass_formula}),
we have $M_2 = 0.28 ~M_\sun$, which corresponds to
$a = 1.22 ~R_\sun$ and $R_1^* =   0.60 ~R_\sun$ 
for $M_1 = 1.06 ~M_\sun$.
%% M1  M2  P-ORB(HOUR)
%% 1.08 0.2815 3.23
%% A=   1.224    RoR1=  0.6095    RoR2=  0.3314    Ro
%%  t_merge= 66 days
%% 1.06 0.281 3.23
%% A=   1.218    RoR1=  0.6046    RoR2=  0.3312    Ro
%%   t_merge= 72 days
%% 0.92 0.28 3.23
%% A=   1.174    RoR1=  0.5683    RoR2=  0.3312    Ro
%%   t_merge= 95 days
These give $t_{\rm emerge} \approx 72$ days at $R_{\rm ph} \sim a$.
Various physical parameters are summarized in Tables
\ref{fitting_t2_t3_tb_txonoff} and \ref{physical_properties_of_novae}.

\citet{iij03} and \citet{kam05} reported that higher ionization
lines became prominent about 65 days after the optical maximum
(i.e., day $\sim 70$ after the outburst).  At that time,
the photospheric temperature of our $1.06 ~M_\sun$ WD model
has already increased to $T_{\rm ph} \sim 60000-70000$ K and
the higher ionization lines mentioned above are possibly related
to this high photospheric temperatures.  Another possibility 
is high temperature optically thin plasma that emits
hard X-ray mentioned below, because high temperature plasma
that emits hard X-ray emerges around $t_{\rm emerge} \sim 70$ days.

Hard X-ray was detected with {\it Chandra}
long before the supersoft X-ray appeared.
\citet{roh09} reported their six {\it Chandra} X-ray spectra.
The first three observations were taken with ACIS-I
on 134, 187, and 248 days after outburst. The count rates are 1.0, 0.69
and 0.53 counts s$^{-1}$, respectively. (We do not plot these
point in Figure \ref{light_curve_v1494_aql_v1974_cyg_x55z02o10ne03}
because the detector ACIS-I is different from that of
supersoft X-ray observation, HRC-S.)
They found no significant variations in any of
these observations. The spectra are dominated by emission lines
originating from N, O, and Ne.  They fit isothermal APEC models with
the spectra and found best fits with $k T = 0.65$~keV and
$N_{\rm H} = 3.3 \times 10^{21}$~cm$^2$ 
\citep[$E(B-V)= N_{\rm H}/4.8 \times 10^{21} = 0.68$ by ][]{nes07a}.
They reported that, in all cases the elemental
abundances of O and N had to be significantly increased.
They further reported that on day 248 a bright soft component
appeared in addition to the fading emission lines.
When the harder X-ray declined, that is, when the optically thick
wind stopped, the soft component of X-ray emerged.
This is very consistent with our collision model between
the wind and the companion as an origin of high temperature plasma
that emits hard X-ray.
In our model, the hard component of X-rays emerges around
$t_{\rm emerge} \sim 70$ days and attains its peak at
$t_{\rm break} \sim 100$ days and then declines toward
$t_{\rm wind} \sim 260$ days.

% Fig.23
%\placefigure{all_mass_v2467_cyg_v1974_cyg_x55z02o10ne03}

\begin{figure}
%\epsscale{1.0}
\epsscale{1.15}
\plotone{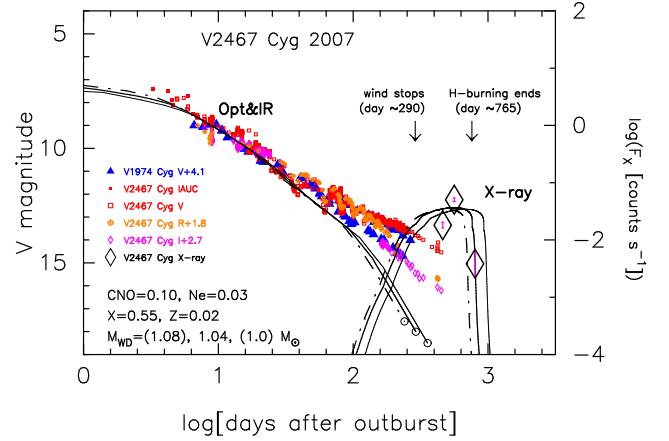}
%\plotone{f15bw.epsi}
%\plotone{all_mass_v2467_cyg_v1974_cyg_x55z02o10ne03_color.epsi}
%\plotfiddle{evolution1.ps}{5.0cm}{270}{0.4}{0.4}{-170}{220}
\caption{
Same as Figure \ref{light_curve_v1494_aql_v1974_cyg_x55z02o10ne03} but
for V2467 Cyg, i.e., optical and supersoft X-ray light curves
of V2467 Cyg.  Optical $V$ ({\it open red squares}), 
visual ({\it small red filled squares}),
$R$ ({\it ocher open circles with plus}),
and $I$ ({\it magenta open diamonds}) are taken from
IAU Circular 8821, \citet{tom07}, and AAVSO, while 
the supersoft X-ray data ({\it large black diamonds})  with error bars
 ({\it magenta bars}) are compiled 
from \citet{nes08b} and also taken from 
an automatic analyzer of the {\it Swift} web page \citep{eva09}.
We plot three model light curves of $M_{\rm WD}= 1.08$ ({\it thick
dash-dotted lines}), $1.04$ ({\it thick solid lines}), and
$1.0 ~M_\sun$ ({\it thin solid lines}) WDs for the envelope
chemical composition of ``Ne nova 2.'' 
We select the $1.04 ~M_\sun$ WD as a best model among three
of 1.08, 1.04, and $1.0 ~M_\sun$ WDs and can reproduce
the supersoft X-ray data.
%%{\it Dashed line}: the free-free light curve 
%%of $0.8~M_\sun$ WD is lifted up by 0.78 mag to fit
%%the upper boundaries of $V$, $R$, and $I$ band data
%%in the later transition phase in order to help the detection
%%of $t_{\rm break}$.
}
\label{all_mass_v2467_cyg_v1974_cyg_x55z02o10ne03}
\end{figure}

\subsection{V2467 Cyg 2007}
%%%\label{subsection_v2467_cyg}
V2467 Cyg (Nova Cygni 2007) was discovered by Tago on 2007 March
15.787 UT (JD 2454175.287) at 7.4 mag \citep{nak07b}.
% 2454174.5 + 0.787 = 2454175.287
% 2454175.287 - 2454172.0 = 3.287
Tago also reported that the nova was fainter than 12th magnitude 
three days ago (JD 2454172.296).  Here we assume the outburst
day of $t_{\rm OB}=$ JD 2454172.0 (2007 March 12.5 UT).  In Figure 
\ref{all_mass_v2467_cyg_v1974_cyg_x55z02o10ne03},
we plot optical and near infrared magnitudes taken
from IAU Circular 8821, \citet{tom07}, and AAVSO.
The nova reached its maximum of $m_{V, \rm max} \approx 7.4$
at $t_0 \approx$ JD 2454175.2 (March 16.7 UT).
Except for some early, low amplitude oscillations,
the light curve declined monotonically until day $\sim 30$.
Then V2467 Cyg started its transition phase until day $\sim 150$
in the form of a series of several oscillations
with a quasi-period of 20--22 days and a 0.5--1.1 mag amplitude.

\citet{nes08b} obtained two {\it Swift} observations.
The first observation was taken on 2008 June 15.17,
% 2454632.5 + 0.17 = 2454632.67
% 2454632.67 - 2454172.0 = 460.67
about 460 days after the outburst, and obtained
an X-ray count rate of 0.017 counts~s$^{-1}$.
% 2454737.5 + 0.3 = 2454737.8
% 2454737.8 - 2454172.0 = 565.8
About 100 days later (2008 September 23.3 UT $=$ 566 days after
the outburst), their second observation yields a count
rate of 0.048 counts~s$^{-1}$.  
The both X-ray spectra are soft with a blackbody temperature of
$kT=34$ eV and $kT=46$ eV, respectively. 
These two count rates are plotted
in Figure \ref{all_mass_v2467_cyg_v1974_cyg_x55z02o10ne03}.
To search for the end epoch of supersoft X-ray phase, i.e.,
$t_{\rm H-burning}$,
we use an automatic analyzer in the {\it Swift} web
page \citep{eva09}.  The three X-ray (0.3-10 keV) data points
with the {\it Swift} XRT are plotted
in Figure \ref{all_mass_v2467_cyg_v1974_cyg_x55z02o10ne03},
early two of which are the same data as those by \citet{nes08b}.
We have checked that the early two data points of the automatic
analyzer are very similar to those obtained by \citet{nes08b}.
The epoch of $t_{\rm H-burning}$ is around 760 days
after the outburst.

     Our free-free model light curves are plotted in
Figure \ref{all_mass_v2467_cyg_v1974_cyg_x55z02o10ne03}
for a chemical composition of ``Ne nova 2.''
The bottom line of $R$ and $I$ magnitudes are well fitted with
our model light curve until about day $\sim 100$.
On the other hand, $V$ magnitude is 
finally departing from the $R$ and $I$ magnitudes from
day $\sim 150$ due to strong emission line contributions.
The best-reproducing WD mass is $1.04 ~M_\sun$ among 1.08, 1.04, and 
$1.0 ~M_\sun$.  For the $1.04 ~M_\sun $ WD model, we obtain
$t_{\rm break} \approx 113$, $t_{\rm wind} \approx 290$, 
and $t_{\rm H-burning} \approx 765$ days.  We also obtain
``intrinsic``  $t_2 \approx 12$ and $t_3 \approx 23$ days
along our model light curve.
Here we assume that the model light curve reached
its maximum (7.9 mag) at $t_0 \sim 3.3$ days after the outburst
(at the first observational point).

\citet{swi08} reported the possible orbital period of V2467 Cyg,
$P_{\rm orb}= 0.1596$ days (3.82 hr),
from their unfiltered, $R$, and $I$ band CCD photometry.
%% M1  M2  P-ORB(HOUR)
%% 0.90  0.35  3.82
%% A=   1.331    RoR1=  0.6154    RoR2=  0.4004    Ro
%% 1.04  0.35  3.82
%% A=   1.379    RoR1=  0.6552    RoR2=  0.3995    Ro
%% 1.11 0.35 3.82
%% A=   1.402    RoR1=  0.6741    RoR2=  0.3991    Ro
Similar periods of 3.57 hr and 3.85 hr had been already observed
in 2007, several weeks after maximum, by \citet{tom07}.
We estimate the epoch when the companion emerges from
the nova envelope.   With the donor mass of $M_2 = 0.35 ~M_\sun$
estimated from Equation (\ref{warner_mass_formula}),
we obtain 
%%% A=   1.293    RoR1=  0.5854    RoR2=  0.4002    Ro
$a = 1.38 ~R_\sun$ and $R_1^* = 0.65 ~R_\sun$ for $M_1 = 1.04 ~M_\sun$,
giving $t_{\rm emerge} \approx 75$ days at $R_{\rm ph} \sim a$.

     We have checked the accuracy of WD mass determination 
from the ambiguity of chemical composition.
We obtain a best-reproducing model of $M_{\rm WD}= 0.90 ~M_\sun$ for 
$X=0.35$ (``CO nova 2''),  and $M_{\rm WD}= 1.11 ~M_\sun$ for 
$X=0.65$ (``Ne nova 3'').  Then, we have
\begin{equation}
M_{\rm WD}(X) \approx \left\{ 1.04 + 0.7 (X-0.55) \right\} ~M_\sun,
\end{equation}
for V2467 Cyg.  Thus, we may conclude that
the WD mass is $M_{\rm WD}= 1.04 \pm 0.07 ~M_\sun$ for
a typical hydrogen content between $X=0.45$ -- 0.65.
If the hydrogen content is $X \sim 0.55$, then the WD has a mass of
$1.04 \pm 0.03 ~M_\sun$ and is probably a ONeMg WD.  On the other
hand, when the hydrogen content is as low as $X \sim 0.35$,
then the WD is less massive ($0.90 \pm 0.03 ~M_\sun$) and
probably a CO WD.

From the FF method, we obtain the distance modulus to V2467 Cyg, i.e.,
\begin{equation}
\left[ (m-M)_V \right]_{\rm FF} = m_{\rm w} - M_{\rm w} 
= 18.0 - 1.7 = 16.3,
\end{equation}
for the $1.04 ~M_\sun$ WD (``Ne nova 2'') model.
Two other attendant models of $1.08 ~M_\sun$ and
$1.0 ~M_\sun$ WDs give similar results of
$(m-M)_V = 17.6 - 1.4 = 16.2$ for $1.08 ~M_\sun$ or
$(m-M)_V = 18.5 - 2.0 = 16.5$ for $1.0 ~M_\sun$ WD.

      The LC method gives 
\begin{equation}
\left[5 \log(d/10) + A_V \right]_{\rm V1974~Cyg}
 - \left[5 \log(d/10) + A_V \right]_{\rm V2467~Cyg}  = -4.1.
\end{equation}
With $(m-V)_V = 12.3$ from Equation (\ref{distance_modulus_v1974_cyg})
for V1974 Cyg, we have
\begin{equation}
\left[ ( m - M )_V \right]_{\rm LC} =
\left[5 \log(d/10) + A_V \right]_{\rm V2467~Cyg}  = 16.4.
\end{equation}

Kaler-Schmidt's law
% (Equation [\ref{kaler-schmidt-law}])
with $t_3 = 23.0$ days and Della Valle \& Livio's law
% (Equation [\ref{della-valle-livio-law}])
with $t_2 = 12.0$ days and $m_{V,{\rm max}} = 7.9$ at maximum
(at $t_0 = 3.3$ days) give
\begin{equation}
\left[ \left( m - M \right)_{V, {\rm max}} \right]_{\rm MMRD}=
\left[ 5 \log(d/10) + A_V \right]_{\rm MMRD} = 16.2 ~(16.5),
\label{v1668_cyg_v2467_cyg}
\end{equation}
being consistent with the FF and LC results.
Then, we obtain the distance modulus of $(m-M)_V = 16.4 \pm 0.2$ and
%%% d= 2.1, 2.2, 2.0, 2.3 kpc
the distance of $d = 2.2 \pm 0.2$ kpc to V2467 Cyg
for $A_V = 4.65$ \citep{maz07}.

% Fig.24
%\placefigure{all_mass_v5116_sgr_v1974_cyg_x65z02o03ne03}

\begin{figure}
%\epsscale{1.0}
\epsscale{1.15}
\plotone{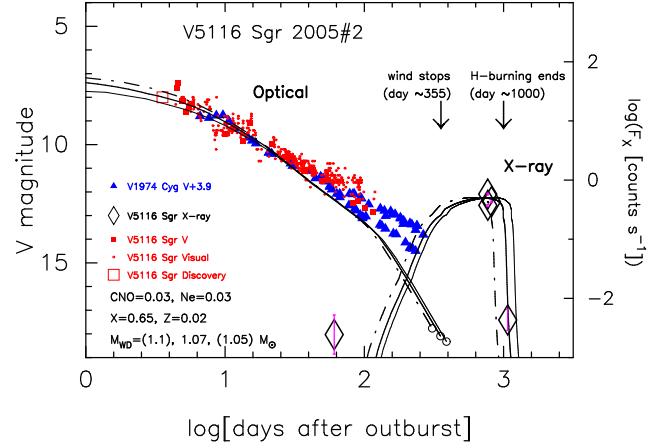}
%\plotone{f12bw.epsi}
%\plotone{all_mass_v5116_sgr_v1974_cyg_x65z02o03ne03_color.epsi}
%\plotfiddle{evolution1.ps}{5.0cm}{270}{0.4}{0.4}{-170}{220}
\caption{
Same as Figure \ref{light_curve_v1494_aql_v1974_cyg_x55z02o10ne03} but
for V5116 Sgr.  Optical $V$ ({\it filled red squares}), visual 
({\it small red open circles}), and supersoft X-ray 
({\it large diamonds}) light curves of V5116 Sgr.
The discovery magnitude is denoted by a large
open square \citep{lil05}.  The $V$ and visual data are taken
from AAVSO.  The X-ray flux data  with error bars
 ({\it magenta bars}) are taken from 
an automatic analyzer of the {\it Swift} web page \citep{eva09}.
We plot three model light curves of $M_{\rm WD}= 1.1$ ({\it thick
dash-dotted lines}), $1.07$ ({\it thick solid lines}), and
$1.05 ~M_\sun$ ({\it thin solid lines}) WDs 
for the envelope chemical composition of ``Ne nova 3.''
We select the $1.07 ~M_\sun$ WD as a best model among three
of 1.1, 1.07, and $1.05 ~M_\sun$ WDs and can reproduce 
the supersoft X-ray data.
}
\label{all_mass_v5116_sgr_v1974_cyg_x65z02o03ne03}
\end{figure}

\subsection{V5116 Sgr 2005\#2}
%%%\label{subsection_v5116_sgr}

V5116 Sgr (Nova Sagittarii 2005 No.2)
was discovered by Liller on July 4.049 UT
% 2453555.5 + 0.049 = 2453555.549
% 2453555.549 - 2453552.0 = 3.549
(JD 2453555.549) at mag about 8.0 \citep{lil05} and
became a supersoft X-ray source \citep{sal08}.
We have already published analysis of optical light curves for 
this nova \citep{hac07k} before the supersoft X-ray was detected
by \citet{sal07}.  This object was not detected on June 12
(limiting mag about 11.0), so \citet{hac07k} assumed
that $t_{\rm OB}=$JD 2453552.0 (July 1.5 UT) is the outburst day.
% 2453556.5 + 0.099 = 2453556.599
% 2453556.599 - 2453552.0 = 4.599
Their best-reproducing model is a $0.9 ~M_\sun$ WD for the chemical
composition of ``CO nova 2.''
Adopting 8.0 mag on JD 2453555.549 observed by
Liller \citep{lil05} as the maximum visual magnitude
for \object{V5116 Sgr}, Hachisu \& Kato derived $t_2 = 20$ days and
$t_3 = 33$ days.  From their best light curve of $0.9 ~M_\sun$ WD, 
Hachisu \& Kato predicted that the supersoft X-ray phase starts on
$t_{\rm wind} \sim 310$ days and ends on
$t_{\rm H-burning} \sim 770$ days after the outburst.

% 2453566.5 + 0.0 = 2453566.5
% 2453566.5 - 2453552.0 = 14.5

We plot the supersoft X-ray light curves as well as optical in
Figure \ref{all_mass_v5116_sgr_v1974_cyg_x65z02o03ne03}.
A bright supersoft X-ray phase was first detected with
{\it XMM Newton} \citep{sal07} on day 674 (2007 March 5 UT).
% 2454225.5 - 2453552.0 = 673.5
\citet{sal08} reported that the X-ray light curve shows abrupt
decreases and increases of the flux by a factor of $\sim 8$, which
is consistent with a periodicity of 2.97 hr, the orbital period
suggested by \citet{dob08}, and speculated that the X-ray light curve
may result from a partial coverage by an asymmetric accretion disk
in a high-inclination system.  \citet{nes07c} also reported
the supersoft X-ray detection with {\it Swift} on day 768 
(2007 August 7.742 UT), 
% 2454320.242 - 2453552.0 = 768.242
and \citet{nel07} observed a similar supersoft X-ray phase
on day 789
% 2454340.5 - 2453552.0 = 788.5
(2007 August 28 UT).
To search for the end epoch of supersoft X-ray phase, i.e.,
$t_{\rm H-burning}$, we use the automatic analyzer
in the {\it Swift} web page \citep{eva09}.
The X-ray (0.3-10 keV) count rate with the {\it Swift} XRT are plotted
in Figure \ref{all_mass_v5116_sgr_v1974_cyg_x65z02o03ne03}.
We have checked that the early several data points of the automatic
analyzer are very similar to those obtained by \citet{nes07c}.
The epoch of $t_{\rm H-burning}$ is somewhere between 820 and
1070 days after the outburst, being roughly
consistent with the prediction by \citet{hac07k}.

\citet{sal08} found, from their X-ray spectrum fit,
that oxygen/neon-rich WD atmosphere models provide a better
fit than carbon/oxygen-rich WD atmosphere models.  
Therefore, we adopt here the chemical composition of ``Ne nova 3''
and recalculated model light curves.
Our new best-reproducing model is a $1.07 ~M_\sun$ WD
(Figure \ref{all_mass_v5116_sgr_v1974_cyg_x65z02o03ne03}),
%%for the chemical composition of $X= 0.65$, $Y= 0.27$, 
%%$X_{\rm CNO}= 0.03$, $X_{\rm Ne}= 0.03$, and $Z= 0.02$.
%%for the chemical composition of $X= 0.55$, $Y= 0.30$, 
%%$X_{\rm CNO}= 0.10$, $X_{\rm Ne}= 0.03$, and $Z= 0.02$.
from which we obtain $t_{\rm break}= 131$, $t_{\rm wind}= 355$, 
$t_{\rm H-burning}= 1000$ days as listed 
in Table \ref{fitting_t2_t3_tb_txonoff}.
The estimated WD mass of $1.07 ~M_\sun$ is very
close or nearly equal to the upper limit mass
for CO WDs born in a binary, i.e.,
$M_{\rm CO} \lesssim 1.07 ~M_\sun$ \citep[e.g.,][]{ume99},
being consistent with the suggestion that the WD of V5116 Sgr
is an ONeMg white dwarf.  

Adopting a magnitude of $m_{V,{\rm max}}= 7.9$ at maximum ($t_0 =$
3.6 days) along our model light curve of the $1.07 ~M_\sun$ WD,
we obtain ``intrinsic'' $t_2 = 15.6$ and $t_3 = 28.5$ days
as listed in Table \ref{fitting_t2_t3_tb_txonoff}.
The $V$ magnitudes are roughly fitted with our model light
curve at least until day $\sim 100$.

     We have checked the accuracy of WD mass determination.
We obtain a best-reproducing model of $M_{\rm WD}= 1.0 ~M_\sun$ for 
$X=0.55$ (``Ne nova 2''),  and $M_{\rm WD}= 0.85 ~M_\sun$ for 
$X=0.35$ (``CO nova 2'').  Then, we have
\begin{equation}
M_{\rm WD}(X) \approx \left\{ 1.07 + 0.7 (X-0.65) \right\} ~M_\sun,
\end{equation}
for V5116 Cyg.  Only when the hydrogen content is $X \sim 0.65$,
the WD has a mass as large as $1.07 ~M_\sun$, which should be
consistent with an ONeMg WD.  On the other hand, when the hydrogen
content is as low as $X \sim 0.35$--0.55,
then the WD is less massive (0.85--$1.0 ~M_\sun$) and
probably a CO WD.

From the FF method, we obtain the distance modulus to V5116 Sgr, i.e.,
\begin{equation}
\left[ (m-M)_V \right]_{\rm FF} = m_{\rm w} - M_{\rm w} 
= 18.2 - 2.0 = 16.2,
\end{equation}
for the $1.0 ~M_\sun$ WD (``Ne nova 2'') model.
Two other attendant models of $1.05 ~M_\sun$ and
$0.95 ~M_\sun$ WDs (``Ne nova 2'') give similar results of
$(m-M)_V = 17.6 - 1.6 = 16.0$ for $1.05 ~M_\sun$ or
$(m-M)_V = 18.7 - 2.3 = 16.4$ for $0.95 ~M_\sun$ WD.

      The LC method gives 
\begin{equation}
\left[5 \log(d) + A_V \right]_{\rm V1974~Cyg}
 - \left[5 \log(d) + A_V \right]_{\rm V5116~Sgr}  = -3.9.
\end{equation}
With $(m-V)_V = 12.3$ from Equation (\ref{distance_modulus_v1974_cyg})
for V1974 Cyg, we obtain
\begin{equation}
\left[ ( m - M )_V \right]_{\rm LC}= 
\left[ 5 \log(d) + A_V \right]_{\rm V5116~Sgr} = 16.2.
\end{equation}

Kaler-Schmidt's law
% (Equation [\ref{kaler-schmidt-law}])
with $t_3 = 28.5$ days and Della Valle \& Livio's law
% (Equation [\ref{della-valle-livio-law}])
with $t_2 = 15.6$ days, and both with $m_{V, {\rm max}} = 7.9$
at maximum ($t_0 = 3.6$ days) yield
\begin{equation}
\left[ \left( m - M \right)_{V, {\rm max}} \right]_{\rm MMRD}=
\left[ 5 \log(d) + A_V \right]_{\rm MMRD} = 16.0 ~(16.2),
\end{equation}
being consistent with the FF and LC results.
%%% d= 12.0, 12.05, 11.0, 12.0 kpc
Then the distance modulus is $(m-M)_V = 16.2 \pm 0.2$
and the distance to V5116 Sgr is $d = 12 \pm 1$ kpc
for $A_V = 0.81$ \citep{bur08}.
These distance estimates are consistent with
$d = 11 \pm 3$ kpc discussed by \citet{sal08}.

     The orbital period of $P_{\rm orb}= 0.1238$~days 
(2.97 hr) was derived by \citet{dob08} from the orbital modulations
with a full amplitude of $0.3-0.4$ mag.  Then,
we estimate the epoch when the companion emerges from
the nova envelope.   From Equation (\ref{warner_mass_formula}),
we obtain $M_2 = 0.25 ~M_\sun$, which yields
$a = 1.15 ~R_\sun$ and $R_1^* = 0.58 ~R_\sun$
for $M_1 = 1.07 ~M_\sun$ and, as a result, 
$t_{\rm emerge} \approx 91$ days
at $R_{\rm ph} \sim a$.
%% M1  M2  P-ORB(HOUR)
%% 1.07 0.25 2.97
%% A=   1.146    RoR1=  0.5813    RoR2=  0.3011    Ro
%% 1.0 0.25 2.97
%% A=   1.125    RoR1=  0.5642    RoR2=  0.3011    Ro
%% 0.85 0.25 2.97
%% A=   1.078    RoR1=  0.5253    RoR2=  0.3015    Ro

% Fig.25
%\placefigure{all_mass_v574_pup_v1668_cyg_x65z02o03ne03}

\begin{figure}
%\epsscale{1.0}
\epsscale{1.15}
\plotone{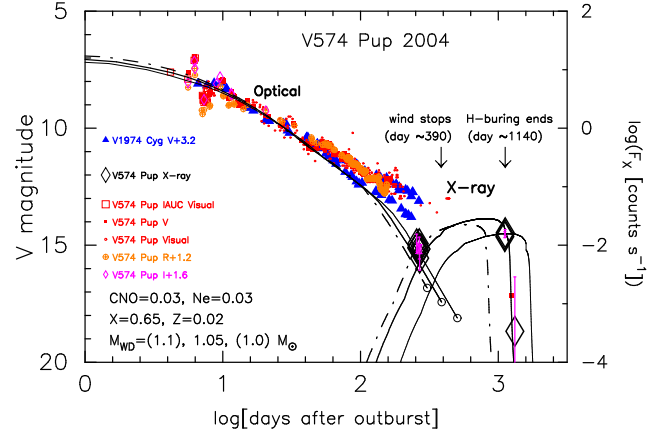}
%\plotone{f13bw.epsi}
%\plotone{all_mass_v574_pup_v1668_cyg_x65z02o03ne03_color.epsi}
%\plotfiddle{evolution1.ps}{5.0cm}{270}{0.4}{0.4}{-170}{220}
\caption{
Same as Figure \ref{light_curve_v1494_aql_v1974_cyg_x55z02o10ne03}
but for V574 Pup.
We plot three model light curves of $M_{\rm WD}= 1.1$ ({\it thick
dash-dotted lines}), $1.05$ ({\it thick solid lines}), and
$1.0 ~M_\sun$ ({\it thin solid lines}) WDs 
for the envelope chemical composition of ``Ne nova 3.''
We select the $1.05 ~M_\sun$ WD as a best model among three
of 1.1, 1.05, and $1.0 ~M_\sun$ WDs and can reproduce
the supersoft X-ray data.
Optical data are taken from AAVSO, 
Variable Star Observers League in Japan (VSOLJ),
and IAU Circulars 8443
and 8445.  The X-ray flux data ({\it large diamonds})  with error bars
 ({\it magenta bars}) are taken from 
\citet{nes07a} and from 
an automatic analyzer of the {\it Swift} web page \citep{eva09}. 
}
\label{all_mass_v574_pup_v1668_cyg_x65z02o03ne03}
\end{figure}

\subsection{V574 Pup 2004}
%%%\label{subsection_v574_pup}

V574 Pup (Nova Puppis 2004) was discovered independently by Tago
and Sakurai on 2004 November 20.672 UT 
% 2453329.5 + 0.672 = 2453330.172
% 2453330.172 - 2453326.0 = 4.172
(JD 2453330.172) at mag about 7.6 \citep{nak04}.
This object was not detected on November 12 and 16 with
limiting mag about 13.6 \citep{nak04}, so
we assume that $t_{\rm OB}=$JD 2453326.0 (November 16.5 UT)
is the outburst day.
The optical and near infrared light curves are plotted in
Figure \ref{all_mass_v574_pup_v1668_cyg_x65z02o03ne03}.
Although \citet{siv05} derived $m_{V, {\rm max}} \approx 8.0$
on JD 2453335.6 (November 26.1 UT), this is not the $V$ maximum
but the second peak as can be seen in 
Figure \ref{all_mass_v574_pup_v1668_cyg_x65z02o03ne03}.
We derive $m_{V, {\rm max}} \approx 7.0$ at $t_0 \approx$ JD~2453332.0
(November 22.5 UT), 6 days after the outburst based on the data
reported in IAU Circular 8445 \citep{sos04}.  After the $V$ magnitude
drops by about 2 mag at $t \approx$ JD~2453333.7 (November 24.2 UT),
it increases to 8.0 mag at $t \approx$ JD~2453335.6 (November 26.1 UT).
After that, the $V$ magnitude declines to about 9.0 mag and
stays for 20 days and then declines almost monotonically.  In this
way, V574 Pup shows an oscillatory feature in the very early phase.

     {\it Swift} XRT observations of V574 Pup showed that the nova
became a supersoft X-ray source before 
2005 August 17 \citep{nes07a}, that is, 274 days after the outburst.
\citet{nes07b} reported that
V574 Pup was still in a supersoft X-ray phase on
2007 December 7 and 13, that is, 1121 days after the outburst.
%The prediction of X-ray on/off day are roughly consistent with 
%$t_{\rm wind} \sim 370$ and $t_{\rm H-burning} \sim 1100$.
To search for the end epoch of supersoft X-ray phase,
we use the automatic analyzer in the {\it Swift} web
page \citep{eva09}.  The X-ray (0.3-10 keV) count rate with the
{\it Swift} XRT are plotted
in Figure \ref{all_mass_v574_pup_v1668_cyg_x65z02o03ne03}.
We added the last observational point from {\it Swift} web page.
The epoch of $t_{\rm H-burning}$ is somewhere between 1130 and
1320 days after the outburst.

\citet{rud06} suggested that the observational
properties of V574 Pup were intermediate between those typical
of explosions on a CO or an ONeMg WD.   Their {\it Spitzer}
observation shows that V574 Pup reveals the strong coronal lines
of Mg and Ne ions (but no \ion{Ne}{2} at 12.8 microns), the behavior
of which is suggestive of an explosion on the surface of an ONeMg WD.
Therefore, we have calculated light curves for 
three different chemical compositions, i.e.,
``Ne nova 3,''  ``Ne nova 2,'' and ``CO nova 2.''
However, only the case of ``Ne nova 3'' provides a reasonable
fit to the observation.  Our best-reproducing light curve is plotted in 
Figure \ref{all_mass_v574_pup_v1668_cyg_x65z02o03ne03} 
for the chemical composition of ``Ne nova 3.''
The WD mass is estimated to be $\sim 1.05 \pm 0.05 ~M_\sun$,
which suggests a possibility that the WD is an ONeMg WD, because
the estimated WD mass of $1.05 ~M_\sun$ is marginally
consistent with the theoretically obtained upper limit mass
for CO WDs born in binaries,
$M_{\rm CO} \lesssim 1.07 ~M_\sun$ \citep[e.g.,][]{ume99}.

For the other two chemical compositions, i.e.,  ``Ne nova 2,'' and
``CO nova 2,'' we do not find any reasonable light curves which 
simultaneously satisfy both the start and end of supersoft X-ray 
phase, that is, the duration of supersoft X-ray phase is too short
to be compatible with the observation, 
like as has already shown in Figure
\ref{all_mass_v597_pup_v1500_cyg_x35z02c10o20}.

     We determine $t_2$ and $t_3$ along our model light curve.
Assuming that our model light curve reached its maximum
at the discovery day ($t_0 = 4.172$ days after the outburst),
we obtain ``intrinsic'' $t_2 = 18.6$ and $t_3 = 35$ days.
The other timescales are obtained to be $t_{\rm break}= 144$,
$t_{\rm wind}= 390$, $t_{\rm H-burning}= 1140$ days
(Table \ref{fitting_t2_t3_tb_txonoff}).

From the FF method, we obtain the distance modulus to V574 Pup, i.e.,
\begin{equation}
\left[ (m-M)_V \right]_{\rm FF} = m_{\rm w} - M_{\rm w} 
= 17.4 - 2.0 = 15.4,
\end{equation}
for the $1.0 ~M_\sun$ WD (``Ne nova 2'') model, although the models
of ``Ne nova 2'' are not shown in the figure.
Two other attendant models of $1.04 ~M_\sun$ and
$0.95 ~M_\sun$ WDs (``Ne nova 2'') give similar results of
$(m-M)_V = 16.9 - 1.7 = 15.2$ for $1.04 ~M_\sun$ or
$(m-M)_V = 17.8 - 2.3 = 15.5$ for $0.95 ~M_\sun$ WD.

      The LC method gives 
\begin{equation}
\left[5 \log(d/10) + A_V \right]_{\rm V1974~Cyg}
 - \left[5 \log(d/10) + A_V \right]_{\rm V574~Pup}  = -3.2.
\end{equation}
With $(m-V)_V = 12.3$ from Equation (\ref{distance_modulus_v1974_cyg})
for V1974 Cyg, we have
\begin{equation}
\left[ ( m - M )_V \right]_{\rm LC}= 
\left[5 \log(d/10) + A_V \right]_{\rm V574~Pup}  = 15.5.
\end{equation}

Kaler-Schmidt's law
% (Equation [\ref{kaler-schmidt-law}])
with $t_3 = 35.0$ days or Della Valle \& Livio's law
% (Equation [\ref{della-valle-livio-law}])
with $t_2 = 18.6$ days and $m_{V,{\rm max}} = 7.6$ at maximum
($t_0 \approx 4.2$ days) give
\begin{equation}
\left[ \left( m - M \right)_{V, {\rm max}} \right]_{\rm MMRD}=
\left[ 5 \log(d/10) + A_V \right]_{\rm MMRD} = 15.5 ~(15.7),
\end{equation}
being consistent with the FF and LC results.
%%% d= 4.4, 4.6, 4.6, 5.0 kpc
Thus the distance modulus is $(m-M)_V = 15.5 \pm 0.2$ and 
the distance to V574 Pup is $d = 4.6 \pm 0.3$ kpc
for $A_V = 2.2$ \citep{bur08}.

% Fig.26
%\placefigure{all_mass_v458_vul_v1974_cyg_x55z02c10o10}

\begin{figure}
%\epsscale{1.0}
\epsscale{1.15}
\plotone{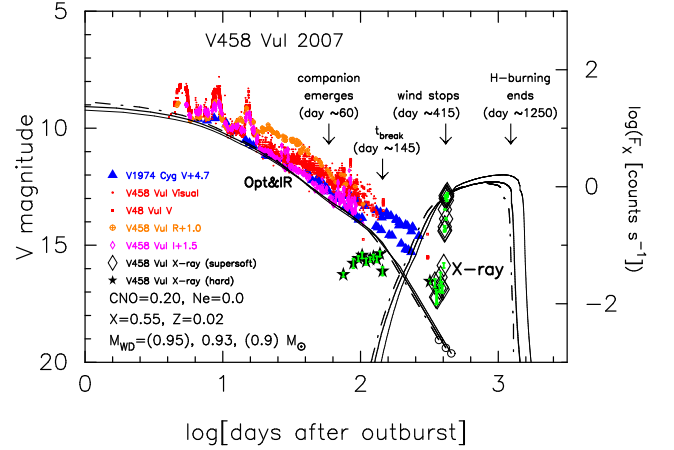}
%\plotone{f14bw.epsi}
%\plotone{all_mass_v458_vul_v1974_cyg_x55z02c10o10_color.epsi}
%\plotfiddle{evolution1.ps}{5.0cm}{270}{0.4}{0.4}{-170}{220}
\caption{
Same as Figure \ref{light_curve_v1494_aql_v1974_cyg_x55z02o10ne03} but
for V458 Vul.  Optical $V$ ({\it filled red squares}),
visual ({\it small red open circles}),
$R$ ({\it ocher circles with plus}),
and $I$ ({\it magenta open diamonds}) are taken from AAVSO and VSOLJ.
The X-ray flux data ({\it filled star marks} for hard and
{\it open large diamonds} for supersoft X-ray)
 with error bars
 ({\it green bars}) are taken from \citet{nes09}.   
We plot three model light curves of $M_{\rm WD}= 0.95$ ({\it thick
dash-dotted lines}), $0.93$ ({\it thick solid lines}), and
$0.9 ~M_\sun$ ({\it thin solid lines}) WDs for the chemical
composition of ``CO nova 4.''
We obtain $t_{\rm wind} = 370$, 416, and 456 days, 
for the 0.95, 0.93, and $0.9 ~M_\sun$ WDs, respectively.
Therefore we select the $0.93 ~M_\sun$ WD as a best model
among three of 0.95, 0.93, and $0.9 ~M_\sun$ WDs by fitting
our model value of $t_{\rm wind}$ 
with the observed turn-on of supersoft X-ray, i.e.,
$t_{\rm wind} \approx t_{\rm X-on} = 415 \pm 15$ days,
because supersoft X-ray is self-absorbed by wind itself.
We also indicate the epoch when the companion
emerges from the photosphere and the break point of optical
light curve in order to explain the flux of hard X-ray ({\it
filled star marks}).
}
\label{all_mass_v458_vul_v1974_cyg_x55z02c10o10}
\end{figure}

% Fig.27
%\placefigure{all_mass_v458_vul_v1974_cyg_x55z02o10ne03}

\begin{figure}
%\epsscale{1.0}
\epsscale{1.15}
\plotone{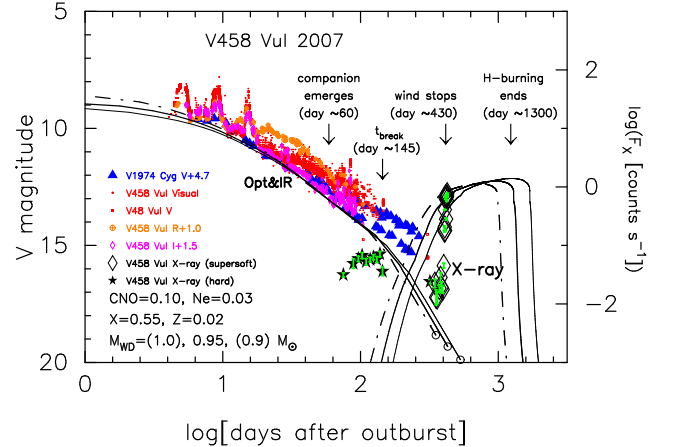}
%\plotone{f14bw.epsi}
%\plotone{all_mass_v458_vul_v1974_cyg_x55z02o10ne03_color.epsi}
%\plotfiddle{evolution1.ps}{5.0cm}{270}{0.4}{0.4}{-170}{220}
\caption{
Same as Figure \ref{all_mass_v458_vul_v1974_cyg_x55z02c10o10} but
for chemical composition of ``Ne nova 2.''
We plot three model light curves of $M_{\rm WD}= 1.0$ ({\it thick
dash-dotted lines}), $0.95$ ({\it thick solid lines}), and
$0.9 ~M_\sun$ ({\it thin solid lines}) WDs.
We obtain $t_{\rm wind} = 352$, 430, and 531 days, 
for the 1.0, 0.95, and $0.9 ~M_\sun$ WDs, respectively.
Therefore we select the $0.95 ~M_\sun$ WD as a best model
among three of 1.0, 0.95, and $0.9 ~M_\sun$ WDs. 
}
\label{all_mass_v458_vul_v1974_cyg_x55z02o10ne03}
\end{figure}

\subsection{V458 Vul 2007}
%%%\label{subsection_v458_vul}

V458 Vul (Nova Vulpeculae 2007) was discovered
by Abe on 2007 August 8.54 UT 
% 2454320.5 + 0.54 = 2454321.04
% 2454321.04 - 2454317.0 = 4.04
(JD 2454321.04) at mag about 9.5 \citep{nak07a}.
This object was not detected on July 23, 31, and August 4
with limiting mag about 11.5 \citep{nak07a}, so
we assume that $t_{\rm OB}=$JD 2454317.0 (August 4.5 UT)
is the outburst day.  The light curve is shown 
in Figure \ref{all_mass_v458_vul_v1974_cyg_x55z02c10o10}.
The $V$ magnitudes show a striking multi-peak (or multi-pulse)
structure in the early phase.
This kind of multi-peak structures were observed in HR Del 1967
and V723 Cas 1995, both of which are considered to be a nova
on a CO WD.  So we assume here that V458 Vul was also a nova
on a CO WD.

\citet{dra07} reported the supersoft X-ray detection of V458 Vul
with the {\it Swift} XRT on 2007 October 18.05 UT
% 2454391.5 + 0.05 = 2454391.55
% 2454391.55 - 2454317.0 = 74.55)
(75 days after the outburst).  \citet{tsu08}, however, concluded,
with the {\it Suzaku} observations, that the spectrum can be
described by a single temperature ($\sim 0.64$ keV) thin thermal
plasma model, suggesting shock-origin of the X-ray.
Recently, \citet{dra08} reported, with the {\it Swift} XRT
observations obtained from 2008 June 18 to September 16
(about $320-410$ days after the outburst), that
% 2454635.5 - 2454317.0 = 318.5
% 2454725.5 - 2454317.0 = 408.5
V458 Vul have entered a new phase characterized by a highly
variable supersoft X-ray component accompanied by partially
anti-correlated variations in the ultraviolet.  They also pointed
that an earlier report on entry into the supersoft phase \citep{dra07}
has proven premature.  \citet{nes09} reported a summary of the early
phase X-ray observation with {\it Swift} in which a supersoft X-ray
phase started about 400 days after the outburst.
Figure \ref{all_mass_v458_vul_v1974_cyg_x55z02c10o10} shows
the supersoft X-ray fluxes reported in \citet{nes09}.
The spectra show
hard ({\it star marks}) components similar to Tsujimoto et al.'s
until day $\sim 400$ and then turned into supersoft 
({\it open diamonds}).

The chemical composition of V458 Vul is not yet reported.  So
we here assume ``CO nova 4'' in Table \ref{chemical_composition},
which is similar to that of HR Del \citep{tyl78} and V723 Cas
\citep{iij06}.  Our model light curves are plotted in 
Figure \ref{all_mass_v458_vul_v1974_cyg_x55z02c10o10}.  This nova
has an oscillation phase in the early $V$, $R$, and $I$
light curves.  We thus fit our universal decline law with
the bottom line of each oscillation as shown in the figure.
The bottom line of $I$ magnitudes is nicely fitted with
our model light curve until day $\sim 100$.  On the other hand,
the other two, $V$ and $R$ magnitudes, are gradually departing
from the model light curve due to emission line contributions.

In the previous subsections, we have determined the WD masses 
mainly from X-ray light curve fitting because optical and near IR
model light curves (i.e., free-free emission model light curves) 
have a similar shape of light curves and are not sensitive to
the WD mass as explained in Figure 
\ref{v1668_cyg_vy_jhk_mag_x35z02c10o20}.  In the case of V458 Vul,
only the emergence of supersoft X-ray was clearly
detected but the supersoft X-ray phase does not end yet.
Therefore, we have to estimate the WD mass only from
the $t_{\rm X-on}$ time.

The supersoft X-ray flux replaced the hard X-ray component
and rose up around $t \sim 415$ days
as shown in Figure \ref{all_mass_v458_vul_v1974_cyg_x55z02c10o10}.
As shown in V382 Vel, we proposed a hypothesis that hard X-ray
component is a shock-origin by optically thick winds.  Therefore,
the substantial flux of hard X-ray indicates that optically thick
winds still blow during the hard X-ray phase.  On the other hand,
supersoft X-ray may be self-absorbed by wind itself.  Therefore,
the emergence of supersoft X-ray indicates that the optically 
thick wind has stopped or substantially weakened
\citep[e.g.,][]{hac03kb}.

Our supersoft X-ray model light curves show a gradual rise 
in Figure \ref{all_mass_v458_vul_v1974_cyg_x55z02c10o10}
because our flux is the photospheric flux and does not include
the effect of absorption by winds.  Here we suppose that
supersoft X-ray is absorbed during the optically thick wind
phase.  The sudden rise in the supersoft X-ray
flux just after the optically thick winds stopped matches our view
of dense optically thick winds \citep[see, e.g.,][for a sharp
increase/decrease supersoft X-ray flux of RX~J0513.9$-6951$]{hac03kb}.

The date of supersoft X-ray emergence is sharply determined to be
$t_{\rm wind} = 415 \pm 15$ days for V458 Vul.
We have calculated three models of 0.95, 0.93, and $0.9 ~M_\sun$ WDs
and obtained $t_{\rm wind} = 370$, 416, and 456 days, respectively.
Then the WD mass is estimated to be
$M_{\rm WD} \sim 0.93 \pm 0.03 ~M_\sun$ for ``CO nova 4''
from the emergence time of supersoft X-ray, i.e.,
$t_{\rm wind} \approx t_{\rm X-on} \sim 415$ days.  Here we regard
that the sharp increase in the supersoft X-ray flux 
corresponds to the end epoch of optically thick winds at about day 415.
For the best model of $M_{\rm WD}= 0.93 ~M_\sun$ among
$0.95$, $0.93$, and $0.9 ~M_\sun$ WDs,
we obtain $t_{\rm break} \approx 145$ days,
$t_{\rm wind} \approx 415$ days, 
and $t_{\rm H-burning} \approx 1250$ days 
(2011 January 5.5 UT) as listed 
in Table \ref{fitting_t2_t3_tb_txonoff}.

     We have checked the accuracy of WD mass determination especially
for CNO abundance.  We adopt other chemical composition
of ``Ne nova 2,'' same hydrogen content $X=0.55$ but different
CNO abundance $X_{\rm CNO}=0.1$ vs. 0.2, and
we obtain $t_{\rm wind} = 352$, 430, and 531 days
for $M_{\rm WD}= 1.0$, 0.95, and $0.9 ~M_\sun$ WDs, respectively.
Here we adopt the best-reproducing model of $M_{\rm WD}= 0.95 ~M_\sun$ among
these three WD masses
as shown in Figure \ref{all_mass_v458_vul_v1974_cyg_x55z02o10ne03}.
This result may suggest that
\begin{equation}
M_{\rm WD}(X_{\rm CNO}) \approx \left\{ 0.93 
- 0.2 (X_{\rm CNO}-0.2) \right\} ~M_\sun,
\end{equation}
for V458 Vul.  The slope of $-0.2$ for CNO abundance
is much smaller than that ($\sim 0.6$) for hydrogen content.
Therefore, if we regard the usual range of chemical composition as
$X_{\rm CNO}= 0.1 - 0.3$ and $X = 0.45 - 0.65$,
the ambiguity in WD mass determination
($\pm 0.02 ~M_\sun$) for CNO abundance is much smaller than
that ($\pm 0.06 ~M_\sun$) for hydrogen content.
Therefore it is likely that $M_{\rm WD}= 0.93 \pm 0.08 ~M_\sun$
for a broad range of the chemical composition 
$X_{\rm CNO}= 0.1 - 0.3$ ($X_{\rm CNO}= 0.2 \pm 0.1$)
and $X = 0.45 - 0.65$ ($X = 0.55 \pm 0.1$).

We have estimated $t_2$ and $t_3$ along our model light curve.
Placing our model light curve on
the bottom line of multi-pulse light curves as shown in
Figure \ref{all_mass_v458_vul_v1974_cyg_x55z02c10o10} and 
assuming that the $V$ magnitude attains its maximum of $\sim 9.4$ mag
at $t_0 \sim 4.4$ days, we obtain ``intrinsic'' $t_2 = 20.7$ and 
$t_3 = 38.7$ days.
The assumed value of our theoretical peak is $\sim 1.4$ mag
dimmer than that of the observed peak (8.0 mag).

From the FF method, we obtain the distance modulus to V458 Vul, i.e.,
\begin{equation}
\left[ (m-M)_V \right]_{\rm FF} = m_{\rm w} - M_{\rm w} 
= 19.3 - 2.3 = 17.0,
\end{equation}
for the $0.95 ~M_\sun$ WD (``Ne nova 2'') as shown in Figure
\ref{all_mass_v458_vul_v1974_cyg_x55z02o10ne03}.
Two other attendant models of $1.0 ~M_\sun$ and
$0.9 ~M_\sun$ WDs give similar results of
$(m-M)_V = 18.8 - 2.0 = 16.8$ for $1.0 ~M_\sun$ or
$(m-M)_V = 19.9 - 2.7 = 17.2$ for $0.9 ~M_\sun$ WD.
Here we do not use the model light curves of ``CO nova 4,''
because they are not calibrated yet.

      The LC method gives 
\begin{equation}
\left[5 \log(d/10) + A_V \right]_{\rm V1974~Cyg}
 - \left[5 \log(d/10) + A_V \right]_{\rm V458~Vul}  = -4.7.
\end{equation}
With $(m-V)_V = 12.3$ from Equation (\ref{distance_modulus_v1974_cyg})
for V1974 Cyg, we have
\begin{equation}
\left[ ( m - M )_V \right]_{\rm LC}= 
\left[5 \log(d/10) + A_V \right]_{\rm V458~Vul}  = 17.0.
\end{equation}

Kaler-Schmidt's law
% (Equation [\ref{kaler-schmidt-law}])
with $t_3 = 37.2$ days or Della Valle \& Livio's law
% (Equation [\ref{della-valle-livio-law}])
with $t_2 = 20.4$ days, and both with $m_{V,{\rm max}} = 9.3$
at maximum ($t_0 = 4.4$ days) yield
\begin{equation}
\left[ \left( m - M \right)_{V, {\rm max}} \right]_{\rm MMRD}=
\left[ 5 \log(d/10) + A_V \right]_{\rm MMRD} = 17.1 ~(17.3),
\end{equation}
being consistent with the FF and LC results.
%%% d= 10.7, 10.7, 11.1, 12.2 kpc
Thus we obtain the distance modulus of $(m-M)_V= 17.1 \pm 0.2$
and the distance of $d = 11 \pm 1$ kpc to V458 Vul
for $A_V = 3.1 E(B-V) = 3.1 \times 0.6 = 1.86$
\citep{lyn08, pog08, wes08}.
This distance is roughly consistent with $d= 13 \pm 3$ kpc
estimated by \citet{wes08}.

Finally, we introduce some characteristic features concerning
binary nature, that is, the emergence of the companion
from the WD photosphere.
     \citet{gor08} reported the orbital period of
$P_{\rm orb} =0.58946$ day with a full amplitude of 0.4 mag.
The orbital light curve shows a saw-tooth shape
similar to those in other supersoft X-ray
novae, e.g. CI Aql and V723 Cas.
%%%The epoch of Max = JDhel 2454461.479 (+/-0.030).
Because Equation (\ref{warner_mass_formula}) is not applicable
to such a long orbital period of $P_{\rm orb}= 14.15$ hr, we 
adopt a condition that the companion mass is smaller than
$0.74 ~M_\sun$.  This condition comes from 
$q = M_2/M_1 \lesssim 0.8$ ($M_1 = M_{\rm WD}= 0.93 M_\sun$)
required for thermally stable mass transfer in a binary
\citep[e.g.,][]{web83}.
If we assume $M_2 = 0.74 ~M_\sun$ as an upper limit of
the secondary mass, then we obtain 
%%  M1  M2  P-ORB(HOUR)
%%  0.93 0.74 14.15
%%  A=   3.509    RoR1=   1.400    RoR2=   1.261    Ro
%%  0.93 0.5 14.15
%%  A=   3.332    RoR1=   1.445    RoR2=   1.089    Ro
$a= 3.5 ~R_\sun$ and $R_1^* = 1.4 ~R_\sun$.
It is $t_{\rm emerge} \approx 62$ days at $R_{\rm ph} \sim a$.
Even if we take a smaller mass of $M_2 = 0.5 ~M_\sun$, 
we obtain similar values of $a= 3.3 ~R_\sun$,
$R_1^* = 1.45 ~R_\sun$, giving $t_{\rm emerge} \approx 65$ days
at $R_{\rm ph} \sim a$.

    \citet{tsu08} explained the early {\it Swift} X-ray 
data as shock-origin, which started about 70 days after the
outburst and peaked at about 140 days and then decreased
as can be seen in Figure \ref{all_mass_v458_vul_v1974_cyg_x55z02c10o10}. 
The shock may be formed by collision between two ejecta shells
\citep[e.g.,][]{fri87, cas04}
or by collision between nova winds and the companion
star \citep{hac05k, hac06kb, hac08kc}.

If these X-rays come from the shock between the optically thick
WD wind and the companion, the emergence of hard X-rays
should be coincident
with the emerge of the companion from the WD photosphere.
This is roughly consistent with $t_{\rm emerge} \approx 65$ days.
On the other hand, the break of free-free emission light curve
is caused by a more steeper decrease in the wind mass-loss rate after
the break point.  Therefore, the drop of X-ray flux at/near
140 days after the outburst is also consistent with our model of
$t_{\rm break} \approx 145$ days.  The hard X-ray lasted at least
until about $t_{\rm wind} \approx 410$ days, that is,
until when the optically thick wind stopped.

    It should be noted that, for the orbital period of V458 Vul,
R. Wesson et al. (private communication) proposed another orbital
period of 98.1 min.  If it is the case, the orbital period of 
$P_{\rm orb} = 0.0681$ day  (1.635 hr) is much shorter than
that obtained by \citet{gor08} and this period gives
$M_2 = 0.12 ~M_\sun$ from Equation (\ref{warner_mass_formula}), 
%% M1  M2  P-ORB(HOUR)
%% 0.93 0.12 1.635
%% A=  0.7132    RoR1=  0.3977    RoR2=  0.1587    Ro
$a= 0.71 ~R_\sun$, $R_1^* = 0.40 ~R_\sun$,
and $t_{\rm emerge} \approx 136$ days (at $R_{\rm ph} \sim a$).
Then, the emergence time of the companion star is too late, that is,
$t_{\rm emerge} \sim t_{\rm break} \sim 140$ days,
and our model on the origin of hard X-ray may not be viable.

% Fig.28
%\placefigure{supersoft_x_ray_time}

\begin{figure}
%\epsscale{0.75}
\epsscale{1.0}
%\epsscale{1.15}
\plotone{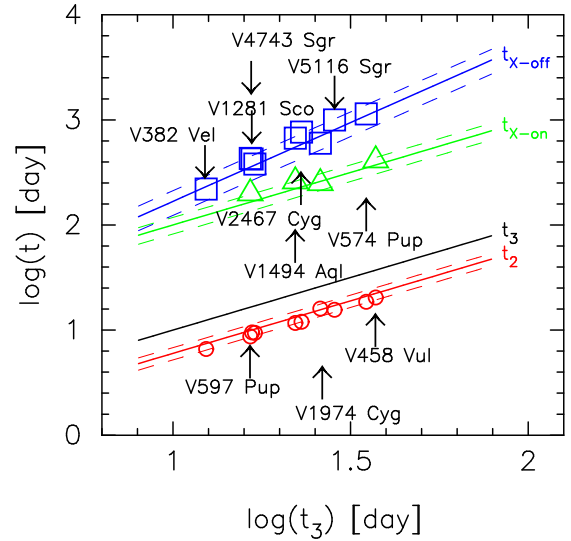}
%\plotone{f16bw.epsi}
%\plotone{supersoft_x_ray_time01_color.epsi}
%\plotfiddle{evolution1.ps}{5.0cm}{270}{0.4}{0.4}{-170}{220}
\caption{
Supersoft X-ray on/off and ``intrinsic'' $t_2$ time are plotted
against ``intrinsic'' $t_3$ time for 10 novae, V1974 Cyg 1992,
V382 Vel 1999, V1494 Aql 1999\#2, V4743 Sgr 2002,
V574 Pup 2004, V5116 Sgr 2005, V1281 Sco 2007,
V2467 Cyg 2007, V458 Vul 2007, and V597 Pup 2007\#1.
From top to bottom, $t_{\rm X-off}$, $t_{\rm X-on}$, and 
intrinsic $t_2$ in units of day.  Solid lines: central values
of Equations (\ref{t_3-t_h_burning}), (\ref{t_3-t_wind}),
and (\ref{t_2-t_3}).
Upper/lower dashed lines: upper/lower values
of Equations (\ref{t_3-t_h_burning}), (\ref{t_3-t_wind}),
and (\ref{t_2-t_3}),
respectively from top to bottom.
Each value of $t_{\rm X-off} \approx t_{\rm H-burning}$
({\it open squares}), $t_{\rm X-on} \approx t_{\rm wind}$
({\it open triangles}), intrinsic $t_2$ ({\it open circles}),
and $t_3$ is taken from Table \ref{fitting_t2_t3_tb_txonoff}.
}
\label{supersoft_x_ray_time}
\end{figure}

\subsection{Results on prediction formulae and $t_3$ time}
\label{t2_t3_txon_SSS}

     In the previous subsections, we have fitted our model light
curves with those of ten classical novae, in which a supersoft
X-ray phase was detected, and estimated various nova parameters.
Figure \ref{supersoft_x_ray_time} shows a summary of our results,
that is, $t_{\rm X-off}$, $t_{\rm X-on}$, and $t_2$ (from top
to bottom) against $t_3$ for the ten novae, excluding V598 Pup but
instead including V1974 Cyg, which was studied by \citet{hac06kb}.

This figure shows a very good
agreement between the data of ``intrinsic'' $(t_2,~t_3)$ and
Equation (\ref{t_2-t_3}).
This is not surprising because Equation (\ref{t_2-t_3}) can
be derived from the universal decline law 
with $F_\lambda \propto t^{-1.75}$ as shown in
Figures \ref{v1668_cyg_vy_jhk_mag_x35z02c10o20_model_scaling_law_logt}
and \ref{mass_v_uv_x_v1974_cyg_x55z02o10ne03_new_model}
\citep[see also][]{hac06kb},
and both $t_2$ and $t_3$ are calculated from our model light curve of
the universal decline law.

Figure \ref{supersoft_x_ray_time} also shows reasonable agreement
of the observed $t_{\rm X-on}$ and $t_{\rm X-off}$ against $t_3$
with Equations (\ref{t_3-t_wind}) and (\ref{t_3-t_h_burning}),
respectively.  Theoretically, $t_{\rm X-on}$ ($\approx$ the epoch
when optically thick winds stop) and $t_{\rm X-off}$ ($\approx$ 
the epoch when hydrogen-shell burning stops) are uniquely
determined if the WD mass and the chemical composition are given.
However, $t_2$ and $t_3$ times are theoretically not unique
even if both the WD mass and the chemical composition are given.
These timescales depend further on the initial envelope mass
as already explained in Section \ref{properties_of_light_curve}.
Therefore, we must obtain $t_2$ and $t_3$ times from each outburst
of various novae.  Good agreement between our prediction
formulae and the observation on $t_{\rm X-on}$ and $t_{\rm X-off}$
suggests that our nova models of optically thick winds and free-free
model light curves are reasonable for the entire description of nova
evolutions and, at the same time, the scatter in the initial nova
envelope mass is not so large from their average. In our modeling,
we estimated their average relations, $t_3$ vs. $t_{\rm wind}$ or
$t_3$ vs. $t_{\rm H-burning}$, from the model of V1668 Cyg.
Such a relatively small scatter of the 10 novae from these relations
strongly suggests that we reasonably predict the supersoft 
X-ray phases of novae from this diagram, i.e., from
Equations (\ref{t_3-t_wind}) and (\ref{t_3-t_h_burning}) if we can
estimate accurate ``intrinsic'' $t_2$ or $t_3$ time from early
optical light curves.

% Table 8
%\placetable{distance_moduli_of_novae}

\begin{deluxetable*}{llllcclllll}
\tabletypesize{\scriptsize}
\tablecaption{Distance Moduli of Novae
\label{distance_moduli_of_novae}}
\tablewidth{0pt}
\tablehead{
\colhead{object} &
\colhead{...} &
\colhead{$t_2$} &
\colhead{$t_3$} &
\colhead{$m_{V,{\rm max}}$} &
\colhead{$(m-M)_V$} &
\colhead{$(m-M)_V$} &
\colhead{$(m-M)_V$} &
\colhead{$A_V$} &
\colhead{distance} &
\colhead{ref.\tablenotemark{d}} \\
\colhead{} &
\colhead{} &
\colhead{(day)} &
\colhead{(day)} &
\colhead{} &
\colhead{FF\tablenotemark{a}} &
\colhead{LC\tablenotemark{b}} &
\colhead{MMRD\tablenotemark{c}} &
\colhead{} &
\colhead{(kpc)} &
\colhead{} 
}
\startdata
V598 Pup 2007\#2 & ... & -- & -- & -- & 11.7 & 12.0 & ~~--~~ (~~--~~) & 0.27 & $2.1 \pm 0.2$\tablenotemark{e} & 1 \\
V382 Vel 1999 & ... & 6.6 & 12.4 & 2.6 & 11.4 & 11.5 & 11.6 (11.4) & 0.62 & $1.5 \pm 0.1$ & 2 \\
V4743 Sgr 2002\#3 & ... & 9.4 & 17 & 5.0 & 13.8 & 13.6 & 13.7 (13.7) & 0.78 & $3.8 \pm 0.2$ & 3 \\
V1281 Sco 2007\#2 & ... & 9.5 & 16.7 & 9.2 & 17.8 & 17.5 & 17.9 (17.9) & 2.17 & $13 \pm 1$ & 4 \\
V597 Pup 2007\#1 & ... & 8.7 & 16.5 & 8.4 & 16.9 & 16.7 & 17.1 (17.1) & 0.93 & $16 \pm 2$ & 5 \\
V1494 Aql 1999\#2 & ... & 11.7 & 22.1 & 5.0 & 13.4 & 13.6 & 13.4 (13.6) & 1.83 & $2.2 \pm 0.2$ & 6 \\
V2467 Cyg 2007 & ... & 12.0 & 23.0 & 7.9 & 16.3 & 16.4 & 16.2 (16.5) & 4.65 & $2.2 \pm 0.2$ & 7 \\
V5116 Sgr 2005\#2 & ... & 15.6 & 28.5 & 7.9 & 16.2 & 16.2 & 16.0 (16.2) & 0.81 & $12 \pm 1$ & 8 \\
V574 Pup 2004 & ... & 18.6 & 35.0 & 7.6 & 15.4 & 15.5 & 15.5 (15.7) & 2.2 & $4.6 \pm 0.3$ & 8 \\
V458 Vul 2007 & ... & 20.7 & 38.7 & 9.4 & 17.0 & 17.0 & 17.2 (17.3) & 1.86 & $11 \pm 1$ & 9 \\
\\
V1974 Cyg 1992 & ... & 16 & 26 & 4.2 & -- & [12.3]\tablenotemark{f} & 12.4 (12.5)& 1.00 & 1.8 & 10 \\
V1668 Cyg 1978 & ... & 14.4 & 26 & 6.2 & -- & [14.3] & 14.4 (14.6) & 1.24 & 4.1 & 11 \\
V1500 Cyg 1975 & ... & 7.2 & 13 & 3.6 & -- & [12.5] & 12.6 (12.4) & 1.60 & 1.5\tablenotemark{g} & 12 \\
GK Per 1901 & ... & 8.1 & 14.5 & 0.5 & -- & [9.2] & 9.3 (9.3) & 0.93 & 0.455\tablenotemark{g} & 13
\enddata
\tablenotetext{a}{distance modulus estimated from direct fitting (FF)
with the calibrated free-free model light curve}
\tablenotetext{b}{distance modulus estimated from two light curve (LC)
fitting}
\tablenotetext{c}{distance modulus estimated from Schmidt-Kaler's law
of Equation (\ref{kaler-schmidt-law}) or Della Valle \& Livio's law
of Equation (\ref{della-valle-livio-law}) in parenthesis, i.e., from the
Maximum Magnitude-Rate of Decline (MMRD)}
\tablenotetext{d}{reference for quoted values of $A_V$ or $E(B-V)$,
where we assume that $A_V = 3.1 E(B-V)$:
1-\citet{rea08}, %%% 1-\citet{wes08},
2-\citet{sho03}, %%% 2-\citet{rea08},
3-\citet{van07}, %%% 3-\citet{nes08c},
4-\citet{rus07a}, %%% 4-\citet{maz07},
5-\citet{nes08c}, %%% 5-\citet{rus07a},
6-\citet{iij03}, %%% 6-\citet{bur08},
7-\citet{maz07}, %%% 7-\citet{van07},
8-\citet{bur08}, %%% 8-\citet{sho03},
9-\citet{wes08}, %%% 9-\citet{iij03},
10-\citet{cho93},
%%% 11-\citet{hac08kc},
11-\citet{sti81}, %%% 12-\citet{sti81},
12-\citet{dow00}, %%% 13-\citet{dow00},
13-\citet{wu89}   %%%14-\citet{wu89}
}
\tablenotetext{e}{error means a standard deviation of the four 
distance estimates, i.e., FF, LC, and two MMRDs}
\tablenotetext{f}{parenthesis means the distance estimated from
other method, see text for details}
\tablenotetext{g}{the distance is taken from \citet{dow00}}
\end{deluxetable*}

% Fig.29
%\placefigure{st_mariko}

\begin{figure}
%\epsscale{1.0}
\epsscale{1.15}
\plotone{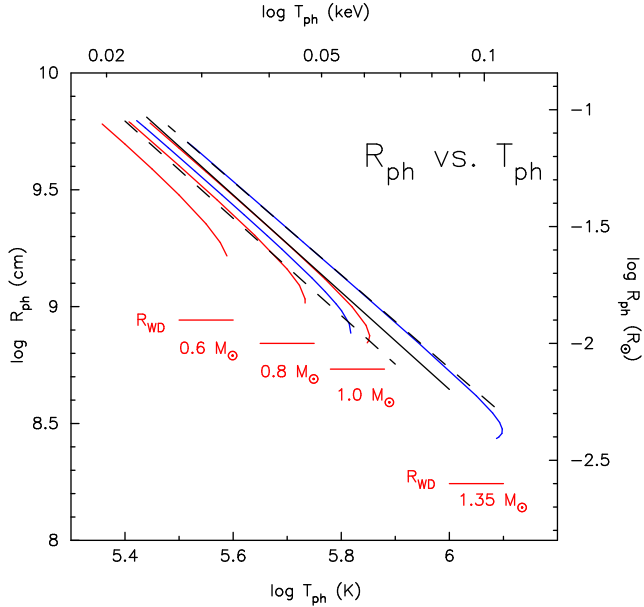}
%\plotone{f16bw.epsi}
%\plotone{st_mariko.epsi}
%\plotfiddle{evolution1.ps}{5.0cm}{270}{0.4}{0.4}{-170}{220}
\caption{
Photospheric radius against photospheric temperature for WDs with 
various masses only during the static phase (after the optically
thick winds stop).  Red solid lines denote those for 
0.6, 0.8, and 1.0 $M_\sun$ WDs with the chemical composition of
``CO nova 2.''  Blue solid lines denote those for 
1.0 and 1.35 $ M_\sun$ WDs with the chemical composition of ``Ne nova 2.''
Black thick solid line indicate an approximate formula of
Equation (\ref{photospheric_radius_vs_temperature}).
Its $\pm 0.1$ upper/lower bounds are indicated by two black dashed lines. 
Short horizontal lines indicate the WD radius 
(i.e., the bottom of a nuclear burning region) 
of 0.6, 0.8, 1.0 and 1.35 $ M_\odot$ WDs.
See text in Appendix B for more details.
%%% aurora/Xray.prediction/WDradius.T.wip
}
\label{st_mariko}
\end{figure}

\section{Conclusions}
\label{conclusions}
Our main conclusions consist of three parts: the first part is
for basic properties of our model light curves which may be
summarized as follows:
\begin{itemize}
\item[1.] We model optical and infrared light curves
with free-free emission, and UV 1455 \AA~ and supersoft X-ray light
curves with blackbody emission, based on the optically thick wind
model, as done in our previous paper \citep{hac06kb}.  
The model parameters governing the light curves are the WD mass
($M_{\rm WD}$),  chemical composition of the WD envelope
($X$, $Y$, $Z$, $X_{\rm CNO}$, and $X_{\rm Ne}$), 
and the initial envelope mass ($M_{\rm env,0}$).
\item[2.] Our free-free emission model light curves reasonably well
reproduce optical and near IR light curves of novae.
The decay timescale of our free-free emission model
light curves depends strongly on the WD mass but weakly
on the chemical composition. So the nova speed class is mainly
determined by the WD mass. 
\item[3.] The peak brightness of our model light curve depends 
also on the initial envelope mass ($M_{\rm env,0}$)
even if we fix the WD mass and the chemical composition of 
the envelope.  This indicates that nova speed class, i.e.,
$t_2$ and $t_3$ times, depends also on the initial envelope mass,
in other words, the mass accretion rate to the WD.
\item[4.] Our UV 1455 \AA~ model light curves
for different WD masses are homologous, that is, they are almost
overlapped to each other if we stretch them in the direction of
time ($t$) by a factor of $1/f_{\rm s}$ and properly normalize the flux.
All the free-free emission model light curves for different WD masses
are also homologous and overlapped with each other 
if we rescale them as $t'=t/f_{\rm s}$ in the time and 
$F'_\nu = f_{\rm s} F_\nu$ in the flux (see Figures 
\ref{v1668_cyg_vy_jhk_mag_x35z02c10o20_model_scaling_law_logt}
and \ref{mass_v_uv_x_v1974_cyg_x55z02o10ne03_new_model}),
that is, they are overlapped to each other in the
%%%$t/f_{\rm s}$--$f_{\rm s} F_\nu$ plane (i.e., in the
$t'$--$m'_V$ plane, where $m'_V = -2.5 \log F'_\nu +$ const.
The more massive the WD, the smaller the timescaling (or stretching)
factor of $f_{\rm s}$.  We call this ``a universal decline law.''
\item[5.]  Using the universal decline law, 
the apparent magnitudes of various WD mass models are restored by
$m_V = m'_V + 2.5 \log f_{\rm s}$ and
their absolute magnitudes are calculated from
$M_V = m'_V + 2.5 \log f_{\rm s} - (m-M)_V$.
Therefore, we can determine
the absolute magnitudes of free-free emission model light curves
by calibrating their absolute magnitudes with the known
distance moduli $(m-M)_V$ of V1668 Cyg for CO novae and V1974 Cyg
for neon novae.
\item[6.] Since the absolute magnitudes of our free-free emission model
light curves are now calibrated (Tables \ref{light_curves_of_novae_co}
and \ref{light_curves_of_novae_ne}), we can estimate the distance
moduli of individual novae by directly fitting them with observation.
We show that the distance moduli thus obtained for V1500 Cyg
are consistent with those calculated from the nebular expansion
parallax method.
This is a new method for obtaining the distance modulus to a nova.
\item[7.] It has been empirically proposed that the absolute 
$V$ magnitude 15 days after optical maximum, $M_V(15)$, is almost
the same among various novae.  We show that there is
only a small scatter of 0.1--0.3 mag for the absolute
brightnesses of our model light curves 15 days after maximum
(see Figures 
\ref{v1668_cyg_vy_jhk_mag_x35z02c10o20_model_real_scale_logt_no2}
and \ref{mass_v_uv_x_v1974_cyg_x55z02o10ne03_real_scale_model}).
This results theoretically support the above
empirical relation on $M_V(15)$.
\item[8.] We found that, when the timescales of two nova light curves
are similar, the absolute brightnesses of the two novae
are almost the same during the period in which the universal 
decline law is applied.  Therefore, we are also able to estimate
the distance modulus of a nova by
comparing it with the other nova the distance of which is known and
the timescale of which is similar.
We also show that the distance moduli thus obtained for GK Per
are consistent with those calculated from the nebular expansion
parallax method.
This is another new method for obtaining the distance modulus to a nova. 
\item[9.]  Our free-free model light curve reasonably follows the
observed light curve when free-free emission dominates the continuum
flux in novae.  Even when the nova has a transition oscillation or
a characteristic secondary peak, our free-free emission model light curve
can be fitted with the bottom line of observed nova light curves
(see Figures \ref{all_mass_gk_per_v1500_cyg_x55z02o10ne03},
\ref{all_mass_v1493_aql_v1500_cyg_x55z02o10ne03}, and
\ref{all_mass_v458_vul_v1974_cyg_x55z02c10o10}).  
\end{itemize}

The Second part contains timescales of novae,
our prediction formulae of supersoft X-ray on and off times, 
and our theoretical MMRD relations,
which are derived from the universal decline law:
\begin{itemize}
\item[10.]  We derive several relations between
the characteristic timescales of classical novae, i.e.,
$t_{\rm break}$, $t_{\rm wind}$, and $t_{\rm H-burning}$,
which depend mainly on the WD mass and weakly on the chemical
composition but almost independent of the initial envelope mass.
\item[11.] On the other hand, $t_2$ and $t_3$ depend sharply
on the initial envelope mass.
Adopting a typical (average) case of the initial envelope mass,
we obtain ``average'' relations between $t_3$ (or $t_2$) time and
the three timescales of $t_{\rm break}$, $t_{\rm wind}$, and
$t_{\rm H-burning}$.
These proposed relations are reasonably 
fitted with published data of novae.
We propose the relation between the ``intrinsic'' $t_3$ time and
$t_{\rm wind}$, i.e., Equation (\ref{t_3-t_wind}),
as a prediction formula 
for the start of a supersoft X-ray phase, 
and the relation between the ``intrinsic`` 
$t_3$ and $t_{\rm H-burning}$, i.e.,
Equation (\ref{t_3-t_h_burning}),
as a prediction formula for the end of it.  Here we propose
the ``intrinsic'' $t_3$ time which is calculated along
our free-free emission model light curve
fitted with the observation.
\item[12.] We obtain a relation between the maximum magnitude,
$m_{V, {\rm max}} = m'_{V, {\rm max}} + 2.5 \log f_{\rm s}$,
and the rate of decline, $\log t_3 = \log (t'_3 f_{\rm s})$, through
the stretching factor $f_{\rm s}$ of the free-free emission
model light curves, i.e.,
Equation (\ref{theoretical_apparent_MMRD_relation}).
A more massive WD corresponds to a smaller stretching factor of
$f_{\rm s}$ while the initial envelope mass determines
a set of $(t'_3, m'_{V, \rm max})$.   Calibrating the distance moduli
$(m-M)_V$ of V1668 Cyg and V1974 Cyg, we derive usual MMRD relations
between the absolute maximum magnitude $M_{V,{\rm max}}$
and $\log t_3$ by changing $f_{\rm s}$ for
a given set of $(t'_3, m'_{V, \rm max})$, i.e., 
Equation (\ref{theoretical_MMRD_relation}).
\item[13.] Our MMRD relations theoretically confirm that the main
parameter is the WD mass, $M_{\rm WD}$.  A more massive WD with
a smaller $f_{\rm s}$ corresponds to a faster nova, i.e., 
a shorter $t_3$ time.  However, there is a second parameter, i.e.,
the initial envelope mass $M_{\rm env,0}$, corresponding to
$(t'_3, m'_{V, \rm max})$.  The larger the envelope mass, the brighter
the maximum magnitude, and the shorter the rate of decline .
The main parameter $M_{\rm WD}$ produces a main trend of
MMRD relation and the second parameter $M_{\rm env,0}$
can reasonably explain the scatter around the main MMRD relation. 
\item[14.] Our MMRD relations are also in good agreement with 
Kaler-Schmidt's law \citep{sch57}
and Della Valle \& Livio's (1995) law at their central lines (values). 
\end{itemize}

The last part concerns light-curve fitting of ten classical novae
in which a supersoft X-ray phase was detected:

\begin{itemize}
\item[15.] The universal decline law reasonably reproduces
the observed light curves of ten classical novae, in which a supersoft
X-ray phase was detected.  The models reproducing
simultaneously the optical and supersoft X-ray
observations are WDs with
$1.28 \pm 0.04~M_\sun$ (V598 Pup),
$1.23 \pm 0.05~M_\sun$ (V382 Vel),
$1.15 \pm 0.06~M_\sun$ (V4743 Sgr),
$1.13 \pm 0.06~M_\sun$ (V1281 Sco),
$1.2 \pm 0.05~M_\sun$ (V597 Pup),
$1.06 \pm 0.07~M_\sun$ (V1494 Aql),
$1.04 \pm 0.07~M_\sun$ (V2467 Cyg),
$1.07 \pm 0.07~M_\sun$ (V5116 Sgr),
$1.05 \pm 0.05~M_\sun$ (V574 Pup),
$0.93 \pm 0.08~M_\sun$ (V458 Vul).
The first nine WDs are probably ONeMg WDs (ONe novae)
while the last one is probably a CO WD (CO nova). 
Our light curve fittings also suggest various physical values
of novae, such as the envelope mass of $M_{\rm env,0}$
at optical maximum, 
the ejecta mass of $M_{\rm wind}$, etc.  as tabulated in Table 
\ref{physical_properties_of_novae}.
\item[16.]  The newly proposed relationships of Equations 
(\ref{t_3-t_wind}) and (\ref{t_3-t_h_burning}) are
consistent with the emergence/decay epoch of the supersoft X-ray phase
of these novae including V1974 Cyg.
\item[17.] We estimate distances to the ten classical novae
by using our two new methods and show that they are consistent with
those derived from the MMRD relation based on the ``intrinsic'' $t_3$
time as tabulated in Table \ref{distance_moduli_of_novae}.
\item[18.] We propose a mechanism of early hard X-ray emission originated
from the shock between the optically thick winds and the companion star.
Our model predicts that the hard X-rays emerge when the companion
appears from the white dwarf photosphere and decay when the optically
thick winds stop.  Our estimates of emerging-time $t_{\rm emerge}$
of the companion from the nova photosphere are consistent
with the observed hard X-ray appearance in
V382 Vel, V1494 Aql, and V458 Vul; which may support
our idea of hard X-ray emission.
\end{itemize}


%% Included in this acknowledgments section are examples of the
%% AASTeX hypertext markup commands. Use \url without the optional [HREF]
%% argument when you want to print the url directly in the text. Otherwise,
%% use either \url or \anchor, with the HREF as the first argument and the
%% text to be printed in the second.

\acknowledgments
We thank AAVSO, AFOEV, VSNET, and VSOLJ for nova photometric data.
We are also grateful to the anonymous referee for useful comments
that helped to improve the paper.
This work made use of data supplied
by the UK Swift Science Data Centre at the University of Leicester.
This research has been supported in part by the
Grant-in-Aid for Scientific Research (20540227)
of the Japan Society for the Promotion of Science.



%%%\appendix

%%%%\section{Appendicial material}

\appendix
\section{Scaling Law of Brightness and Absolute Flux 
of Free-Free Emission Light Curves}
\label{scaling_law_of_brightness}
In this appendix, we present a formulation of scaling law for free-free
emission model light curves and derive 
Equation (\ref{simple_final_scaling_flux}).
We determine the flux of our free-free light curves
against that of $0.95 ~M_\sun$ WD ($\approx$ V1668 Cyg).
Now we explicitly express the proportionality constant in Equation
(\ref{wind-free-free-emission}) as
\begin{equation}
F_\nu^{\{M_{\rm WD}\}}(t) = C^{\{M_{\rm WD}\}} 
\left[ {{\dot M_{\rm wind}^2}
\over {v_{\rm ph}^2 R_{\rm ph}}} \right]^{\{M_{\rm WD}\}}_{(t)}.
\label{explicit-wind-free-free-emission}
\end{equation}
In free-free emission, $F_\nu$ is independent of frequency $\nu$,
then $C^{\{M_{\rm WD}\}}$ is also independent of frequency because
each term in the square bracket of the right-hand-side is also
independent of frequency.
The proportionality constant $C^{\{M_{\rm WD}\}}$ probably
depends not only on the WD mass but also weakly on the 
chemical composition and other physical conditions.
However, we here assume that $C^{\{M_{\rm WD}\}}$
depends only on $M_{\rm WD}$ when we fix the chemical composition.

Then, we obtained $F'_{\nu '} = f_{\rm s} F_\nu$ from Equation
(\ref{time-derivation-flux})
after time-stretching ($t' = t/f_{\rm s}$ and $\nu ' = f_{\rm s} \nu$).
Since the free-free flux is independent of the frequency $\nu$, i.e.,
$F'_{\nu '} = F'_\nu$, this simply means
\begin{equation}
F_{\nu}^{\{0.95~M_\sun\}}(t) 
= f_{\rm s} F_\nu^{\{M_{\rm WD}\}}(t/f_{\rm s}).
\label{scaled-wind-free-free-emission}
\end{equation}
Now, we define the apparent $V$ magnitudes as
\begin{equation}
m_V(t) \equiv -2.5 \log F_\nu^{\{M_{\rm WD}\}}(t) + H_V, 
\label{apparent_flux_def}
\end{equation}
and, for 0.95 $M_\sun$ WD ($\approx$ V1668 Cyg) with
$f_{\rm s} \approx 1$, we have
\begin{equation}
m''_V(t) \equiv -2.5 \log F_\nu^{\{0.95 ~M_\sun \} }(t)   
+ H_V,
\label{095_apparent_flux_def}
\end{equation}
where we introduce a constant $H_V$ in order to fit $m''_V$ of
$0.95 ~M_\sun$ WD model with the $V$ observation of V1668 Cyg,
and $H_V$ is the same among various WD masses.  
Using Equation (\ref{scaled-wind-free-free-emission}), we have
\begin{equation}
m''_V(t)  = m_V(t/f_{\rm s}) - 2.5 \log f_{\rm s}.
\label{final_scaling_flux}
\end{equation}

By the way, we have already defined the apparent $V$
magnitudes of our free-free emission model light curves as
\begin{equation}
m'_V(t/f_{\rm s}) \equiv -2.5 ~ \log ~ \left[ {{\dot M_{\rm wind}^2}
\over {v_{\rm ph}^2 R_{\rm ph}}} \right]^{\{M_{\rm WD}\}}_{(t/f_{\rm s})}
+ K_V,
\label{scaling_flux}
\end{equation}
in Equation (\ref{simple_scaling_flux}).
Here $K_V$ is a constant defined to fit
$m'_V$ of $0.95 ~M_\sun$ WD model with the $V$ observation
of V1668 Cyg, and common among all the WD mass models
(from $M_{\rm WD}= 0.55$ to 1.2 $M_\sun$ by 0.05 $M_\sun$ step)
for a chemical composition of ``CO nova 2.''
We saw that all the light curves are almost overlapped to each other
in Figure \ref{v1668_cyg_vy_jhk_mag_x35z02c10o20_model_scaling_law}.
This simply means that
\begin{equation}
\left[ {{\dot M_{\rm wind}^2}
\over {v_{\rm ph}^2 R_{\rm ph}}} \right]^{\{M_{\rm WD}\}}_{(t/f_{\rm s})}
\approx
\left[ {{\dot M_{\rm wind}^2}
\over {v_{\rm ph}^2 R_{\rm ph}}}
 \right]^{\{0.95~M_\sun\}}_{(t)}.
\label{overlapping-flux-free-free-emission}
\end{equation}
Substituting Equation (\ref{explicit-wind-free-free-emission}) into
Equation (\ref{scaled-wind-free-free-emission}) and
using Equation (\ref{overlapping-flux-free-free-emission}), we have
\begin{equation}
C^{\{0.95~M_\sun\}} \approx f_{\rm s} C^{\{M_{\rm WD}\}}.
\label{coefficient-free-free-emission}
\end{equation}
From Equation (\ref{overlapping-flux-free-free-emission})
together with Equations (\ref{095_apparent_flux_def})
and (\ref{scaling_flux}), we derive
\begin{equation}
m'_V(t/f_{\rm s}) - K_V
\approx 
m''_V(t) + 2.5 \log C^{\{0.95~M_\sun\}} - H_V.
\label{final-flux-free-free-emission}
\end{equation}
Therefore, we may regard that $m''_V(t)= m'_V(t/f_{\rm s})$,
or in other words,
\begin{equation}
m'_V(t/f_{\rm s}) = m_V(t/f_{\rm s}) - 2.5 \log f_{\rm s},
\end{equation}
if we define $K_V$ as
\begin{equation}
K_V =  H_V - 2.5 ~ \log ~ C^{\{0.95 ~M_\sun \} }.
\label{constant_relations}
\end{equation}
Thus, we derive Equation (\ref{simple_final_scaling_flux}).
Therefore, we plotted all the brightnesses of novae
in the $t' = t/f_{\rm s}$ vs. 
$m'_V = m_V - 2.5 \log f_{\rm s}$ diagram as
shown in Figure \ref{v1668_cyg_vy_jhk_mag_x35z02c10o20_model_scaling_law}.

It should be noted that the UV 1455\AA~  and supersoft X-ray
model light curves follow $F'_{\nu '} =  f_{\rm s} F_\nu$ 
after stretching but
do not satisfy $F_{\nu '} =  F_\nu$, so that these fluxes
do not obey Equations (\ref{scaled-wind-free-free-emission})
and (\ref{final_scaling_flux}),
because the blackbody emissivity depends on the frequency $\nu$ and,
as a result, the spectrum changes after time-stretching
($\nu ' = f_{\rm s} \nu$).   Therefore, 
we have to normalize the UV fluxes by different factors
to fit the peaks with the observational peak of V1668 Cyg,
as already mentioned in Section \ref{properties_of_light_curve}.

\section{An Empirical Formula of White Dwarf Photospheric Radius}
\label{static_photospheric_radius_of_wd}
The photospheric radius of a typical nova during a supersoft X-ray phase
is usually much larger than the Chandrasekhar radius.  
It is about ten times larger ($\sim 6 \times 10^9$ cm)
than the Chandrasekhar radius at the beginning of a supersoft 
X-ray phase and quickly shrinks with the photospheric temperature
being increasing.   Therefore, we here present a simple formula
on the radius of WD photosphere during a supersoft X-ray phase,
which may be used for estimating the X-ray emission area. 
The photospheric radius of the WD envelope in the supersoft X-ray phase 
is plotted in Figure \ref{st_mariko} against the photospheric
temperature for various WD masses with the chemical composition of
``CO nova 2'' and ``Ne nova 2.''  The left edge of each line corresponds
to the epoch when the wind stops (at $t_{\rm wind} \approx
t_{\rm X-on} =$ when the supersoft X-ray emerges).
As the photosphere shrinks, the photospheric temperature rises.
The rightmost point of each line 
denotes the epoch when hydrogen nuclear burning stops. 
We see that the photospheric radius $R_{\rm ph}$ is almost
inversely proportional to a square of the photospheric temperature
$T_{\rm ph}$, i.e., $R_{\rm ph} \propto 1 / T_{\rm ph}^2$,
because the total luminosity is almost constant,
$L_{\rm ph} =4 \pi R_{\rm ph}^2 \sigma T_{\rm ph}^4 \sim$~const.,
in the supersoft X-ray phase except the very later phase.

The radius of a typical fast classical nova ($1.0 ~M_\odot$ WD with 
a composition of ``CO nova 2'') can be approximated by the black solid
line as 
\begin{equation}
\log R_{\rm ph} {\rm (cm)} =-2.07 \left( \log T_{\rm ph} {\rm (K)} -5.7
\right) + 9.27,
\label{photospheric_radius_vs_temperature}
\end{equation}
for $5.45 < ~ \log T_{\rm ph}$ (K) $< 5.8$. 
The dashed lines denote the upper and lower bound values of
$\log R_{\rm ph} \pm 0.1$ which covers the most of the 
WD mass representative for classical novae, i.e.,
from $0.8 ~M_\odot$ with ``CO nova 2'' to $1.35 ~M_\odot$
with ``Ne nova 2.''
The slow novae are not covered in the region, i.e., $0.6-0.7 ~M_\odot$
with ``CO novae 2'' has smaller photospheric radius as shown
in Figure \ref{st_mariko}.  This figure also shows the WD radius,
i.e., the bottom of a hydrogen burning zone:
the Chandrasekhar radius for $M_{\rm WD} < 1.33 ~M_\sun$
and values taken from \citet{nom84} for $M_{\rm WD} \ge 1.33 ~M_\sun$.

\clearpage
% Table 1

%\clearpage
% Table 2

%\clearpage
% Table 3

%\clearpage
%Table 4

%\clearpage
% Table 5

%\clearpage
% Table 6

%\clearpage
% Table 7

%\clearpage
% Table 8

%\clearpage
% Fig.1

%\clearpage
% Fig.2

%\clearpage
% Fig.3

%\clearpage
% Fig.4

%\clearpage
% Fig.5

%\clearpage
% Fig.6

%\clearpage
% Fig.7

%\clearpage
% Fig.8

%\clearpage
% Fig.9

%\clearpage
% Fig.10

%\clearpage
% Fig.11

%\clearpage
% Fig.12

%\clearpage
% Fig.13

%\clearpage
% Fig.14

%\clearpage
% Fig.15

%\clearpage
% Fig.16

%\clearpage
% Fig.17

%\clearpage
% Fig.18

%\clearpage
% Fig.19

%\clearpage
% Fig.20

%\clearpage
% Fig.21

%\clearpage
% Fig.22

%\clearpage
% Fig.23

%\clearpage
% Fig.24

%\clearpage
% Fig.25

%\clearpage
% Fig.26

%\clearpage
% Fig.27

%\clearpage
% Fig.28

%\clearpage
% Fig.29

%% The following command ends your manuscript. LaTeX will ignore any text

\end{document}